%% file: main_v2.tex
\documentclass[twocolumn]{aastex63}

\newcommand{\sigacht}{$\sigma_{R_e/8}$}

\newcommand{\kms}{km\,s$^{-1}$}
\newcommand{\re}{$R_e$}
\newcommand{\vrms}{$V_\text{rms}$}
\newcommand{\obthree}{NGC~3923}
\newcommand{\obone}{NGC~1404}
\newcommand{\obsev}{NGC~7144}
\newcommand{\obsevsev}{NGC~7796}
\newcommand{\obfour}{NGC~4697}
\newcommand{\obfive}{NGC~5061}
\newcommand{\obm}{NGC~4552}
\newcommand{\obu}{UGC~1382}
\newcommand{\obic}{IC~1459}
\newcommand{\pystaff}{\textsc{PyStaff}}
\newcommand{\met}{[Z/H]}
\newcommand{\logten}{$\log$}
\newcommand{\Msun}{$M_\sun$}
\newcommand{\msun}{$M_\sun$}
\newcommand{\nablar}{$\nabla_r$}
\usepackage{amsmath}

\usepackage{ulem}

\received{May 9, 2021}
\revised{August 17, 2021}
\accepted{September 17, 2021}
\submitjournal{ApJ}

\shorttitle{Stellar Population Gradients of Early-type Galaxies}
\shortauthors{Feldmeier-Krause et al.}
\graphicspath{{./}{}}

\begin{document}

\title{Stellar Population and Elemental Abundance Gradients of Early-type Galaxies\footnote{This paper includes data gathered with the 6.5 meter Magellan Telescopes located at Las Campanas Observatory, Chile.}}

\correspondingauthor{A. Feldmeier-Krause}
\email{feldmeier@mpia.de}

\author[0000-0002-0160-7221]{A. Feldmeier-Krause}
\affiliation{Max-Planck-Institut f\"ur Astronomie, K\"onigstuhl 17, 69117, Heidelberg, Germany}

\author[0000-0001-8421-1005]{I. Lonoce}
\affiliation{Department of Astronomy $\&$ Astrophysics, The University of Chicago, 5640 South Ellis Avenue, Chicago, IL 60637, USA}

\author[0000-0003-3431-9135]{W. L. Freedman}
\affiliation{Department of Astronomy $\&$ Astrophysics, The University of Chicago, 5640 South Ellis Avenue, Chicago, IL 60637, USA}

\begin{abstract}
The evolution of galaxies is imprinted in their stellar populations. Several stellar population properties in massive early-type galaxies have been shown to correlate with intrinsic galaxy properties like the galaxy's central velocity dispersion, suggesting that stars formed in an initial collapse of gas ($z\sim$2). However, stellar populations change as a function of galaxy radius, and it is not clear how local gradients of individual galaxies are influenced by global galaxy properties and galaxy environment.  In this paper, we study the stellar populations of eight early-type galaxies as a function of radius. We use optical spectroscopy ($\sim$4000--8600\,\AA) and full-spectral fitting to measure stellar population age, metallicity, IMF slope, and nine elemental abundances (O, Mg, Si, Ca, Ti, C, N, Na, Fe) out to 1\,\re\space for each galaxy individually.  We find a wide range of properties, with ages ranging from 3--13 Gyr. Some galaxies have a radially constant, Salpeter-like  IMF, and other galaxies have a super-Salpeter IMF in the center, decreasing to a sub-Salpeter IMF at $\sim$0.5\,\re. We find a global correlation of the central [Z/H] to the central IMF and the radial gradient of the IMF  for the eight galaxies, but local correlations of the IMF slope to other stellar population parameters hold only for subsets of the galaxies in our sample. 
Some elemental abundances also correlate locally with each other within a galaxy, suggesting a common production channel. 
Those local correlations appear only in subsets of our galaxies indicating variations of the stellar content among different galaxies.


\end{abstract}


\section{Introduction}
\label{sec:intro}

Radial stellar population gradients in  early-type galaxies provide information on a  galaxy's formation and evolution. Age and metallicity help us to understand the star formation history; chemical abundances can constrain star formation time-scales and chemical enrichment history. The initial mass function (IMF) affects the chemical evolution.  All of these  parameters   are required to fully understand the evolution of  a galaxy.

Massive early-type galaxies are thought to have formed in two main stages \citep[e.g.][]{2009ApJ...699L.178N,2010ApJ...725.2312O}: in an early ($z\approx$2) collapse of gas, stars formed in situ in a compact core. 
The second stage is ongoing to the present:  satellite galaxies are accreted and deliver stars to the galaxy that were formed ex situ. The fraction of accreted stars is higher for more massive galaxies, and at large galaxy radius, due to the longer dynamical timescales \citep[e.g.][]{2015ApJ...807...11G}. 
A small fraction \citep[ca. 10\%,][]{2010MNRAS.404.1775T} of early-type galaxies show signs of recent star formation activity, with a young sub-population of stars that formed in situ. \cite{2010MNRAS.404.1775T} suggest that  rejuvenation may be triggered by galaxy interaction and merger activity in the past few gigayears.  
Since most early-type galaxies show only little recent star formation, their stellar mass is dominated by old stars, and we can learn about galaxy assembly and early star formation from their stellar populations. 

Important clues come from chemical elemental abundances, as different chemical elements have different origins. 
Hydrogen, deuterium and most helium were formed in primordial nucleosynthesis, but most other elements are produced by stellar nucleosynthesis or explosive burning in the late stages of stellar evolution \citep[see the review by][and references therein]{2019A&ARv..27....3M}. Stars more massive than 8\,\Msun\space produce most of the $\alpha$ elements (e. g. O, Ne, Mg, Si, Ca, Ti). 
They end their life as core-collapse supernovae (SNe), and  release elements into the ISM soon after the beginning of star formation. Massive  stars also produce iron peak elements (e.g. Fe, Ni), but a larger portion is produced  by type Ia SNe. Type  Ia SNe are thermonuclear explosions of C-O white dwarfs that accrete mass,  they produce Fe peak elements, but also Si, Ar, S, Ca. There is a delay between the explosion of massive stars and SNe Ia,  in the range of about 40-50 Myr to few Gyr.  Sodium is produced in both massive and intermediate-mass stars.  Carbon and Nitrogen are partly produced in massive (\textgreater 8\,\msun) stars, but a large portion ( 70\% of N, and $\sim$30-40\% of C, \citealt{2020A&A...639A..37R}) is produced   in intermediate-mass (2-8\,\msun) asymptotic giant branch (AGB) stars  and SNe Ia, leading also to a later enrichment compared to $\alpha$ elements.  The relative elemental abundances are thus caused by the delay between the production channels, and depend on the star formation history and the stellar initial mass function of a galaxy. However, other factors  influence the element production of a star, for example  the star's total metallicity and stellar rotation. 

Different studies revealed non-solar abundance ratios in early-type galaxies, and found increasing values as function of the galaxy mass or velocity dispersion \citep[e.g.][]{2000AJ....119.1645T,2005ApJ...621..673T,2012MNRAS.421.1908J,2014ApJ...780...33C,2014ApJ...783...20W}. This may be caused by a larger fraction of massive stars, i.e. a more top-heavy IMF,  in more massive galaxies. Other explanations are more rapid star formation  or quicker quenching in more massive galaxies, such that the star formation  has ended before type Ia SNe enrich the material with Fe. 

The IMF of massive early-type galaxies has been found to be bottom-heavy, i.e. dwarf-rich with a larger fraction of low-mass stars compared to the Milky Way \citep[e.g.][]{2003MNRAS.339L..12C,2010ApJ...709.1195T,2010ApJ...717..803G,2010Natur.468..940V,2012ApJ...760...71C,2013MNRAS.433.3017L}. In some galaxies the IMF slope varies as a function of radius, being more bottom-heavy in the center \citep[e.g.][]{2015MNRAS.447.1033M,2017ApJ...841...68V,2018MNRAS.478.4084S,2019MNRAS.489.4090L,2021A&A...649A..93B}, while in other galaxies the IMF slope appears to be constant \citep[e.g.][]{2017MNRAS.465..192Z,2017MNRAS.468.1594A,2018MNRAS.478.4464A,2018MNRAS.475.1073V}.

Several possible global and local correlations of the IMF slope with other structural and stellar population parameters have been suggested, for example the galaxy's velocity dispersion \citep{2012Natur.484..485C,2014MNRAS.438.1483S}, total metallicity \met\space \citep{2015ApJ...806L..31M,2018MNRAS.477.3954P,2019MNRAS.485.5256Z}, total mass density \citep{2015MNRAS.452L..21S}, [Mg/Fe] \citep{2012ApJ...760...71C}, [Na/Fe] \citep{2017ApJ...841...68V}, [$\alpha$/Fe], or age \citep{2014ApJ...792L..37M}. Some of these correlations have not been found in other studies e.g. age by \cite{2015ApJ...806L..31M} and \cite{2019MNRAS.485.5256Z}, [Mg/Fe] by  \cite{2015ApJ...806L..31M} and \cite{2017ApJ...841...68V}, [Na/Fe] and \met\space by \cite{2019MNRAS.489.4090L}, and local $\sigma$ by \cite{2015ApJ...806L..31M} and  \cite{2021A&A...645L...1B}. In summary,   there is still no consensus about which correlations hold, either globally or locally, and if any correlation signifies a causal relation or just  that two parameters vary within galaxies in a similar way
\citep{2019A&A...626A.124M,2020ARA&A..58..577S}.

In this paper we study stellar population gradients of eight early-type galaxies (\obone, \obthree, \obm, \obfour, \obfive, \obsev, \obsevsev, \obic) and one low-surface brightness spiral galaxy (\obu). We measure single stellar population light-weighted age, \met, the IMF slope in the mass range 0.08--1.0\,\msun, and nine elemental abundances (O, Mg, Si, Ca, Ti, C, N, Na, Fe). 
We investigate radial gradients for these parameters and test whether there are local correlations for the galaxies, and global correlations for the entire sample. 

This paper is organised as follows: We describe the long-slit spectroscopic data in Section \ref{sec:data}. We present our stellar population analysis and results in Section \ref{sec:pystaff}. 
In Section \ref{sec:correlation} we test whether local and global correlations of various stellar population parameters hold for our data set, and in Section \ref{sec:elprod} we compare relative elemental abundances, and how they vary as function of \met.  We discuss our results in Section \ref{sec:discussion}.

\begin{deluxetable*}{l  c r  r  r r  r  r r r r r r}
\tablecaption{Galaxy Sample \label{tab:sample}}
\tablecolumns{12}
\tablewidth{0pt}
\tabletypesize{\scriptsize}
\tablehead{
\colhead{Galaxy} &\colhead{Type} & \colhead{Distance} &
\colhead{\re}&\colhead{Scale}  & \colhead{$e$} &
\colhead{Major Axis PA}  &\colhead{$\mu_e$}& \colhead{\sigacht}& \colhead{$M_{B}$} & B-V
& \colhead{Env} & \colhead{$N_g$}\\
\nocolhead{} & \nocolhead{} & \colhead{(Mpc)} & \colhead{(arcsec)} & \colhead{(kpc/arcmin)} & \nocolhead{} & \colhead{(\degr)} & \colhead{(mag arcsec$^{-2}$)}  & \colhead{(\kms)}& \colhead{(mag)} & \colhead{(mag)}& \nocolhead{}}
\startdata
\obm & S0/E0 & 15.8 & 31.2 & 4.60 & \nodata  & \nodata  & 20.02 & 250 & -20.5&	0.94&C &714\\
\obone & E1& 18.6&24.6&5.41&0.195& 163.6 &19.84&230&-20.7&0.94&C&120\\
\obthree & E4-5 &20.9&  86.4&6.08 &	0.271 & 48.0&22.06&246&-21.1&	0.91&G&25\\
\obfour & E6 &11.6 &68.4& 3.38 &	0.308 & 66.1 &21.35&165&-20.3&0.87&G &19  \\
\obfive & E0 &26.0& 39.0&7.57 &	0.255& 	113.2 &21.14&188&-21.4&0.83&G&14\\
\obsev & E0  &24.9& 36.0& 7.25 &	0.044 &50.0&21.79&163&-20.4&0.88&G&3\\
\obsevsev & cD & 49.9&24.0& 14.52 &	0.141 & 170.4&21.53&254&-21.0&	0.93&F&2\\
\obic & E3-4  &26.9&55.8&7.83 &	0.264&45.0& 21.26&296&-21.4&0.95&G&16\\ 
\hline
\obu & S& 80.0& 16.0  & 23.27& \nodata &47.4 &\nodata &167&-20.2&...&F&...\\
\enddata
\tablecomments{Hubble type from RC3 \citep{1994AJ....108.2128C}; central velocity dispersion  \sigacht, integrated colors B-V from Hyperleda 
\footnote{\url{http://leda.univ-lyon1.fr/}};
\obm\space effective radius \re\space from \cite{2006MNRAS.366.1126C}, \obm\space surface brightness at 1\,\re\space in g-band $\mu_e$ (g) from \cite{2010ApJS..191....1C}; Remaining distance, effective radius \re, ellipticity $e$, major axis position angle PA, surface brightness at 1\,\re\space in V-band $\mu_e$ (V), and environment (C=cluster, G=group, or F=field) from Carnegie--Irvine Galaxy Survey  
\citep{2011ApJS..197...21H};  Number of galaxies in group $N_g$ from \cite{2017ApJ...843...16K}. \obu\space data from \cite{2016ApJ...826..210H}, see Appendix \ref{sec:ugc} for details. } 
\end{deluxetable*}

\begin{deluxetable*}{lllrrr }
\tablecaption{Summary of  Observations \label{tab:observations}}
\tablecolumns{6}
\tablewidth{0pt}
\tabletypesize{\scriptsize}
\tablehead{
\colhead{Galaxy} &
\colhead{Date} &
\colhead{Grating} &
\colhead{Slit Width} &
\colhead{Exposure Time} &
\colhead{PA}\\
\nocolhead{}&\nocolhead{}&\nocolhead{}&\colhead{(arcsec)}&\colhead{(minutes)}&\colhead{E of N (\degr)}
}
\startdata
  \obm& 2018 May 11&  600$\ell$/9\fdg71 &2.5&60&0  \\
(M~89) & 2018 May 11&  600$\ell$/10\fdg46 &2.5&60&0  \\
& 2019 April 28&  600$\ell$/17\fdg11 &2.5&60&0  \\
\hline
\obone & 2019 Dec. 2&  600$\ell$/9\fdg7   &2.5&60&151\\
 & 2019 Dec. 2&  600$\ell$/10\fdg45 &2.5&60&151    \\
 & 2019 Dec. 3&  600$\ell$/17\fdg11 &2.5&40&151  \\
\hline
\obthree& 2015 May 19/20&  600$\ell$/9\fdg8 &2.5&40&48  \\
 & 2015 May 19/20&  600$\ell$/16\fdg8 &2.5&20&48  \\
 \hline
\obfour & 2019 April 27 &  600$\ell$/9\fdg71 &2.5&60&0  \\
 & 2019 April 27 &  600$\ell$/10\fdg46 &2.5&60&0  \\
 & 2019 April 27 &  600$\ell$/17\fdg11 &2.5&60&0  \\
\hline
\obfive & 2019 April 29 &  600$\ell$/9\fdg71 &2.5&60&0   \\
 & 2019 April 30 &  600$\ell$/10\fdg46 &2.5&80&0   \\
 & 2019 April 30 &  600$\ell$/17\fdg11 &2.5&80&0  \\
\hline
\obsev & 2018 July 9&  600$\ell$/9\fdg95 &2.5&80&0  \\
 & 2018 July 9&  600$\ell$/10\fdg72 &2.5&80&0  \\
 & 2019 April 28&  600$\ell$/17\fdg11 &2.5&80&0 \\
\hline
\obsevsev & 2018 July 15&  600$\ell$/9\fdg95 &2.5&120&0  \\
 & 2018 July 15&  600$\ell$/10\fdg72 &2.5&120&0  \\
 \hline
 \obic & 2014 Sept. 19/20&  600$\ell$/9\fdg78 &1.5&30&0  \\
 & 2014 Sept. 19/20&  600$\ell$/16\fdg6 &1.5&120&0  \\
\obic & 2015 May 19&  600$\ell$/9\fdg78 &2.5&20&30  \\
 & 2015 May 19 &  600$\ell$/16\fdg6 &2.5&20&30  \\
 \hline
\enddata
\end{deluxetable*}

\section{Data}
\label{sec:data}
\subsection{Galaxy Sample and Observations}

We observed  nine galaxies with the  Inamori Magellan Areal Camera and Spectrograph  \citep[IMACS,][]{2006SPIE.6269E..0FD} on the Magellan Baade 6.5 m telescope. The data were collected over several observing runs from 2014--2019. For each run, nearby bright galaxies were selected that were observable on the allocated nights.
Most galaxies are elliptical or S0/E0, and most are located in group or cluster environments, only \obsevsev\space is  classified as a field galaxy. One galaxy, \obu, was believed to be an early-type galaxy, but was classified as a giant low-surface brightness galaxy (LSBG) by \cite{2016ApJ...826..210H}, with a   surrounding  disk and spiral arms. Our stellar population analysis of \obu\space is detailed in Appendix \ref{sec:ugc}.

The galaxies' central velocity dispersions  are in the range 160--300\,\kms,  and their physical sizes span \re$\sim$2.2\,kpc (\obone) to 8.8\,kpc (\obthree).  
We list details on the sample galaxies in Table \ref{tab:sample}, and summarise the observations in Table \ref{tab:observations}. 

For all galaxies we obtained optical longslit spectra in the ranges $\sim$4000-6600\,\AA\space (grating 600$\ell$/$\sim$9\fdg8), and for seven galaxies in addition  the Ca triplet (CaT) range $\sim$8100-8600\,\AA\space (grating 600$\ell$/$\sim$17\degr). For \obsevsev\space we do not have spectra in the CaT range, because IMACS showed a strong fringing signal,  and
 the spectroscopic flat spectra taken during the day were not sufficient to correct it. 
 
For some galaxies, we observed two slightly different settings at optical wavelengths (\obone, \obm, \obfour, \obfive, \obsev, \obsevsev), to ensure that the spectra cover the full wavelength range from $\sim$3700--6700\,\AA\space without gaps. However, for \obic\space and \obthree\space we observed only one optical grating angle, which causes three gaps, with each about 60\,\AA\space width,  in  the spectra, as IMACS consists of four chips along the wavelength direction. 
We have two sets of observations for \obic\space 
 at different position angles (PAs), 30\degr\,\space offset, and different slit widths. We analyzed these spectra separately.

\subsection{Data Reduction and Processing}
\label{sec:dr}

We used a combination of \textsc{iraf}\footnote{IRAF is distributed by the National Optical Astronomy Observatory, which is operated by the Association of Universities for Research in Astronomy (AURA) under a cooperative agreement with the National Science Foundation.} and custom-made \textsc{IDL} scripts for data reduction. Our reduction steps include bias subtraction, bad pixel detection and cosmic ray removal,  
distortion correction, and wavelength calibration to air wavelengths using He/Ne/Ar/Hg arc lamp exposures. 
For the CaT  spectra (grating 600$\ell$/$\sim$17\degr), we had an additional step in order to correct for  fringes. Following \cite{2021arXiv210702335L}, we divided the science frames by  spectroscopic flat field frames, which were taken close in time to the observations. This step was performed before distortion correction.  

The optical spectra of \obsev\space and \obsevsev\space show a wedge of additional light contamination  in the lower right part of the reddest chip (chip 1). This is probably caused by stray light from the Hall effect device on one of the grating tilt mechanisms (Dave Osip, priv. communication). We  decided not to use the spectral region affected by it, which limits our useful optical wavelength range to \textless6640\,\AA. 

The  IMACS slit length is 15\arcmin\space along two chips, i.e. about 7\arcmin\space  on a single chip.  This allowed us to use  the exposure itself to estimate the sky background, at a distance of $\gtrsim$2 \re\space from the galaxy center. We used the \textsc{iraf} tool \textsc{background}. For \obthree\space only, we used a slightly different approach (see \citealt{2020ApJ...902...12F} for details), which resulted in a similar S/N compared to using \textsc{background} subtraction. In short, we first extracted a sky spectrum at \textgreater3\,\re\space and subtracted it, and then we removed sky residuals using \textsc{pPXF}
 \citep{2004PASP..116..138C,2012ascl.soft10002C,2017MNRAS.466..798C}.  We combined individual exposures of a galaxy in the same setting before we extracted the spectra. 
We simultaneously extracted  spectra from the flat field exposures, processed in the same way as the science exposures, and corrected the extracted optical science spectra along the dispersion axis. The CaT spectra had already been flat-field corrected before sky subtraction, when we corrected for fringes. We extracted spectra in a finer radial sampling in the center ($\sim$0.01\,\re), but increased the extraction windows at larger radii to obtain  S/N\textgreater 50, typically up to  $\sim$0.25\,\re\space width. For most galaxies, the outermost bin extends to 1\,\re. The geometric mean of the  extraction window along the slit  is listed in Table \ref{tab:pyresult} as $x/$\re. When considering the slit width  we  obtain the geometric distance or radius from  the galaxy center  as $r/$\re.  
We derived the flux calibration from standard star observations, observed with the same settings as the science frames and processed in a similar way. The only exception are the data observed with a slit width of 1\farcs5\space (\obic\space observed 2014 Sept. 19/20), for which we only have standard star observations with slit width 2\farcs5. We compared the flux calibrated spectra with the \obic\space data that were observed with a slit width of 2\farcs5  (observed 2015 May 19). We found that the flux calibration for the bluest wavelength chip is not satisfactory, as it results in an overestimation of the flux towards shorter wavelengths ($\lesssim$3900\,\AA) for the slit width = 1\farcs5 data. For this reason, we do not use the spectra on the bluest chip (3600--4380\,\AA). 
We corrected the atmospheric transmission using the \textsc{ESO} tool \textsc{molecfit} \citep{2015A&A...576A..78K,2015A&A...576A..77S}. 
\begin{figure*}
\includegraphics[width=0.95\textwidth]{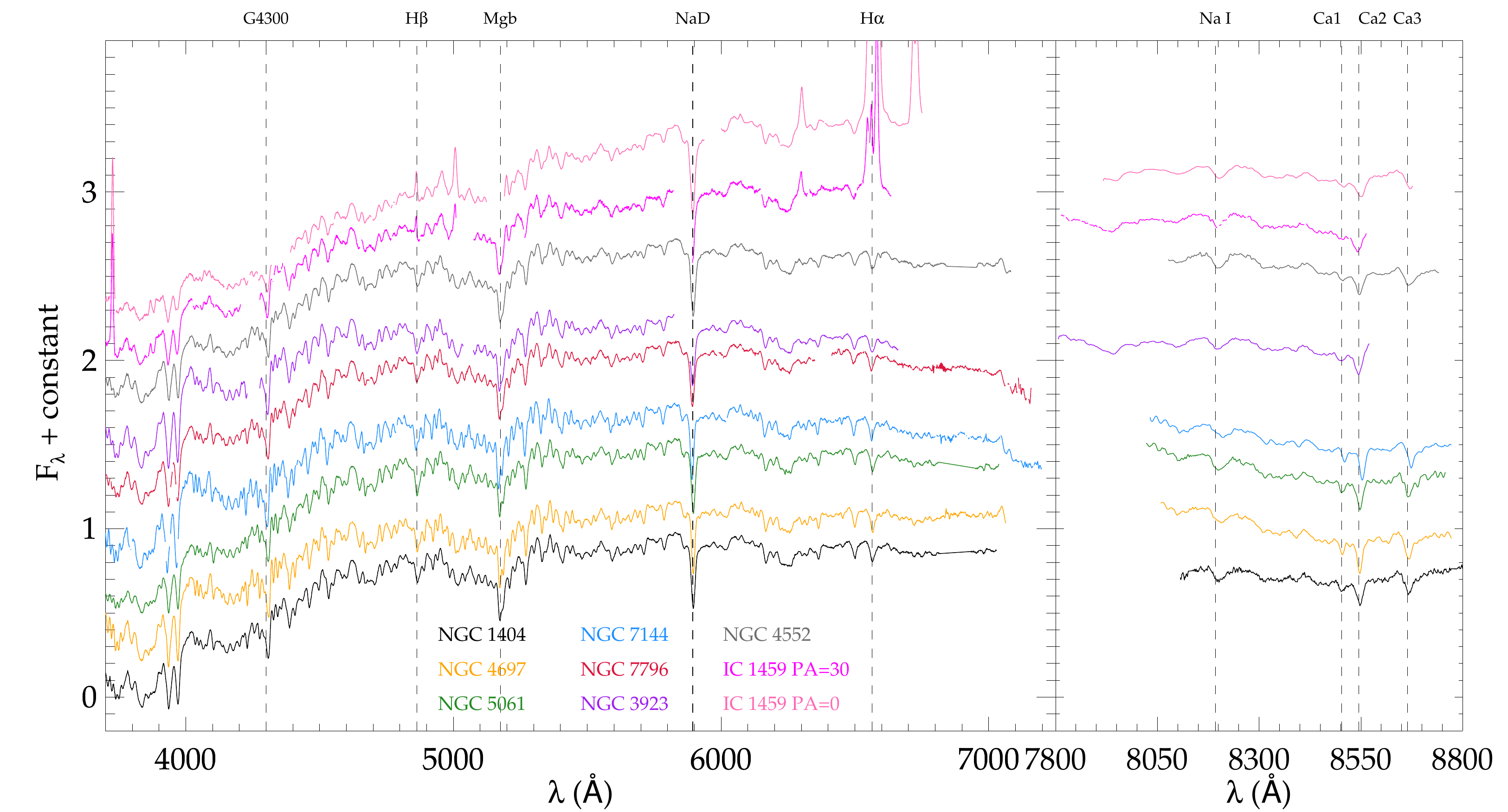}
 \caption{Central spectra of all galaxies in our sample, denoted with different colors. The spectra are normalised and  shifted to rest wavelength. Note the slightly different wavelength regions and the missing CaT region for \obsevsev\space (red). }
\label{fig:speccom}
\end{figure*}

In order to prepare the reduced spectra for our spectral fitting, we performed the following steps: We corrected the galaxy rotation as outlined in Appendix \ref{sec:rotcor}, we corrected all spectra for  Galactic extinction  using the map of \cite{2011ApJ...737..103S}, and converted the spectra  from air to vacuum wavelengths. 
We measured the spectral resolution of our data using arc line exposures. The full-width-half-maximum (FWHM) is usually $\sim$5.5\,\AA, with small variations between observing runs. Only the data observed with a slit width 1\farcs5 have a higher spectral resolution with FWHM $\sim$3\,\AA. For our analysis we use  \cite{2018ApJ...854..139C} template spectra, which  have a spectral resolution of $\sigma$=100\,\kms. We convolved our spectra to a spectral resolution of 100\,\kms\space  if the spectral resolution of the data was less than 100\,\kms.\footnote{For the spectral fitting in Section \ref{sec:pystaff}, we convolved the template spectra to the spectral resolution of the data in regions where it was \textgreater 100\,\kms.}  The convolution was necessary for the entire spectra observed with the 1\farcs5 slit, and for the CaT region (\textgreater8000\,\AA) of the 2\farcs5 slit  spectra. 
We  resampled all spectra to a 1.25\,\AA\space spaced wavelength grid, in order to decrease the time required for a fit. 
Finally, we  combined the spectra with different grating angles for each galaxy and radius  bin, with exception of \obic, where we kept the two sets at different PA and slit widths separate.  We show the central spectra of all galaxies in our sample in Figure \ref{fig:speccom}. In Appendix \ref{sec:snr} and Figure \ref{fig:snr}, we show the S/N of our data as a function of wavelength and radial extraction region, for each individual galaxy. In summary, the S/N lies usually in the range of 50 to 300/\AA.

\section{Spatially resolved stellar populations}

\label{sec:pystaff}

We applied full spectral fitting using the \textsc{python} code \pystaff\space \citep[Python Stellar Absorption Feature Fitting,][]{2018MNRAS.479.2443V} to measure 12 stellar population parameters. 
First we summarise the method, and then we present our results. 

\begin{figure}
\plotone{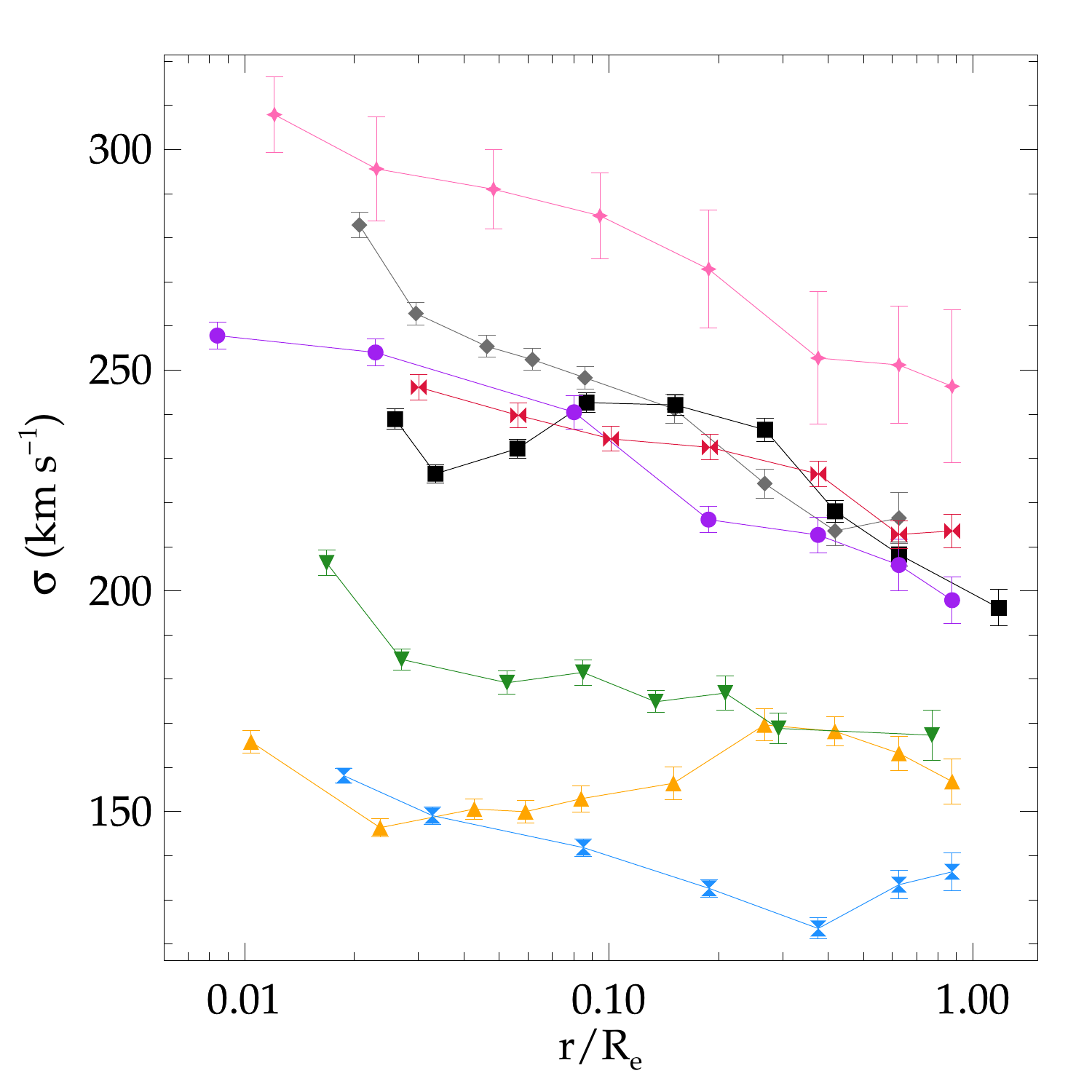}
\caption{Velocity dispersion $\sigma$ as a function of radius for all  galaxies, denoted by different colored symbols as in Figure~\ref{fig:speccom}.}
\label{fig:sigmaprof}
\end{figure}

\subsection{Method}
\subsubsection{Spectral Fitting}
The code \pystaff\space applies full spectral fitting to   find the best-fit single stellar population (SSP)  using the models by \cite{2018ApJ...854..139C}. 
We fit the stellar population parameters  age $t$, metallicity \met, nine elemental abundances ([Na/H], [C/H], [O/H], [Ca/H], [Fe/H], [N/H], [Ti/H], [Mg/H], [Si/H]), and the IMF slope. 
In addition,  we fit   the gas velocity and dispersion,  gas emission flux (see Appendix \ref{sec:gasfit} for details), and the stellar velocity and velocity dispersion $\sigma$.
 We show the stellar  velocity dispersion profiles, corrected for stellar rotation as described in Appendix \ref{sec:rotcor}, in  Figure \ref{fig:sigmaprof}.

We used the bounds [1, 14 Gyr] for age,  [-1.5, 0.3 dex] for \met,  [-0.45, 1.0 dex] for Na,   [-0.2, 0.3 dex] for C,  [0, 0.45 dex] for O,  [-0.45, 0.45 dex] for Ca, Fe, N, Ti, Mg, and Si, and [0.3, 3.5] for the IMF slope. 
We combined the two IMF slope parameters $x_1$ and $x_2$  of the SSP models (in the  initial stellar  mass ranges 0.08--0.5 $M_\sun$ and 0.5--1.0 $M_\sun$) to a single slope by forcing $x_1$ equal to $x_2$.  The IMF slope at \textgreater 1.0\,$M_\sun$ is fixed to the  \cite{1955ApJ...121..161S} slope.
\pystaff\space utilises the \textsc{emcee} package \citep{2013PASP..125..306F} to explore the large parameter space. We chose 100 walkers and 8000 steps.

We also ran fits with two  IMF slope parameters instead of one, for the two mass ranges 0.08--0.5\,\msun\space and 0.5--1.0\,\msun. However, the two parameters $x_1$ and $x_2$ are anti-correlated (see also \citealt{2020ApJ...902...12F,2021arXiv210702335L}), which led us to use only one slope for the full mass range 0.08--1.0\,\msun\space instead. The differences for the other parameters when using two instead of one IMF slope are within the 1$\sigma$ uncertainties.

We explored covariances of our fitting parameters. The probability distribution functions (PDFs) of our fits show some correlations, and the most common are a positive correlation between C and O, Fe and Mg, and an  anti-correlation of age and \met. These covariances are also shown in the upper right off-diagonal of Figure \ref{fig:corrlocal}. For each \pystaff\space stellar population fit, we computed the Spearman rank correlation coefficient $\rho$ of the PDFs of two stellar population parameters. We combined these measurements of $\rho$ to one mean $\rho$  per galaxy and parameter set, which we show as ellipses. A near circular shape and dotted lines mean no or a weak  covariance  ($\lvert\rho\rvert$\textless0.4),  a dot-dashed line a moderate (0.4$\leq\lvert\rho\rvert$\textless0.6), a dashed line a strong (0.6$\leq\lvert\rho\rvert$\textless0.8), a solid line  a very strong ($\lvert\rho\rvert\geq$0.8) covariance of two parameters.  
The covariances are included in the uncertainty estimates of our fitting results as listed in Table \ref{tab:pyresult}.

\subsubsection{Gradient Fitting}
\label{sec:radgradfit}
We measured radial gradients of the final single stellar population  parameters.
We used orthogonal distance regression (\textsc{scipy} package \textsc{odr}) to   fit a log-linear relationship for each galaxy and stellar population property $Y$ as a function of the geometric bin center or radius $r$, defined as $Y=\nabla_r\cdot$\logten$($r$/$\re$)+B_r$. 
The quantity $\nabla_r$ denotes the radial gradient per log($r$/\re); $B_r$ denotes the stellar population  value at 1\,\re\space (\logten($r/$\re)=0). 
We do not have an uncertainty for $r$, and used only the errors 
of the stellar population parameter in the fits. This means that the \textsc{odr} fits as a function of \logten($r$/\re) are equivalent to least-squares minimization. We fit the logarithmic age,  \logten\space  $t$.  We excluded data points located within the central 1\arcsec\space from the gradient fits, in order to avoid seeing effects.

Some stellar population gradients are not well described with a linear function, and we obtain  large reduced $\chi^2$ with our linear fits. 
We fit quadratic models and found them significantly better  than the linear models  for \met\space 
for four galaxies (\obm, \obone, \obfour, \obfive). For \obone\space and  \obfive\space the quadratic fit for [O/Fe] is better,  and for \obfive\space also [Na/Fe], [N/Fe], and the IMFs 
are better described with a quadratic model as a function of radius. Nevertheless, we use the linear model fits for an easier comparison among galaxies and parameters.

\begin{figure*}
\gridline{\fig{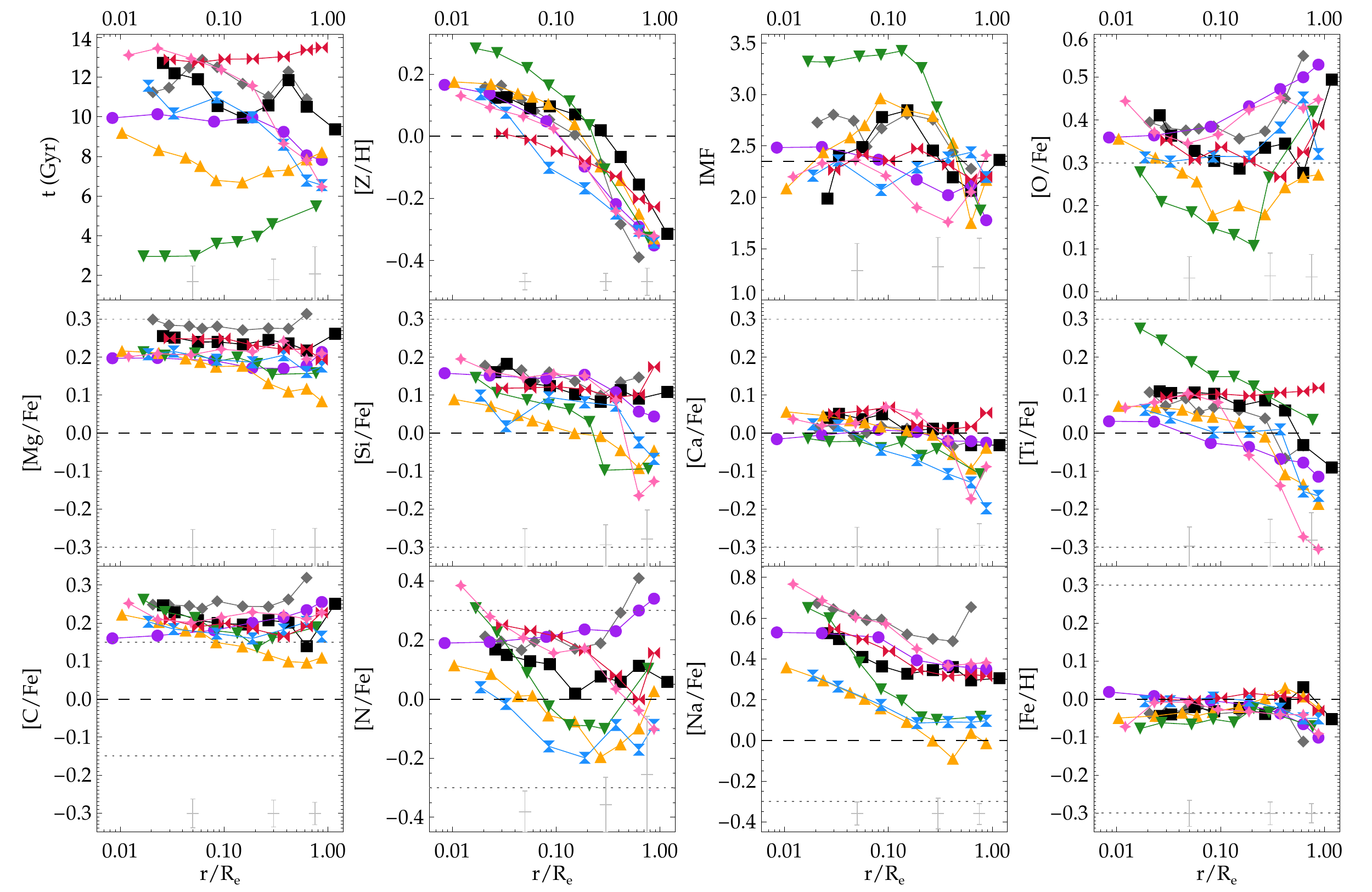}{0.9\textwidth}{}}
\gridline{\fig{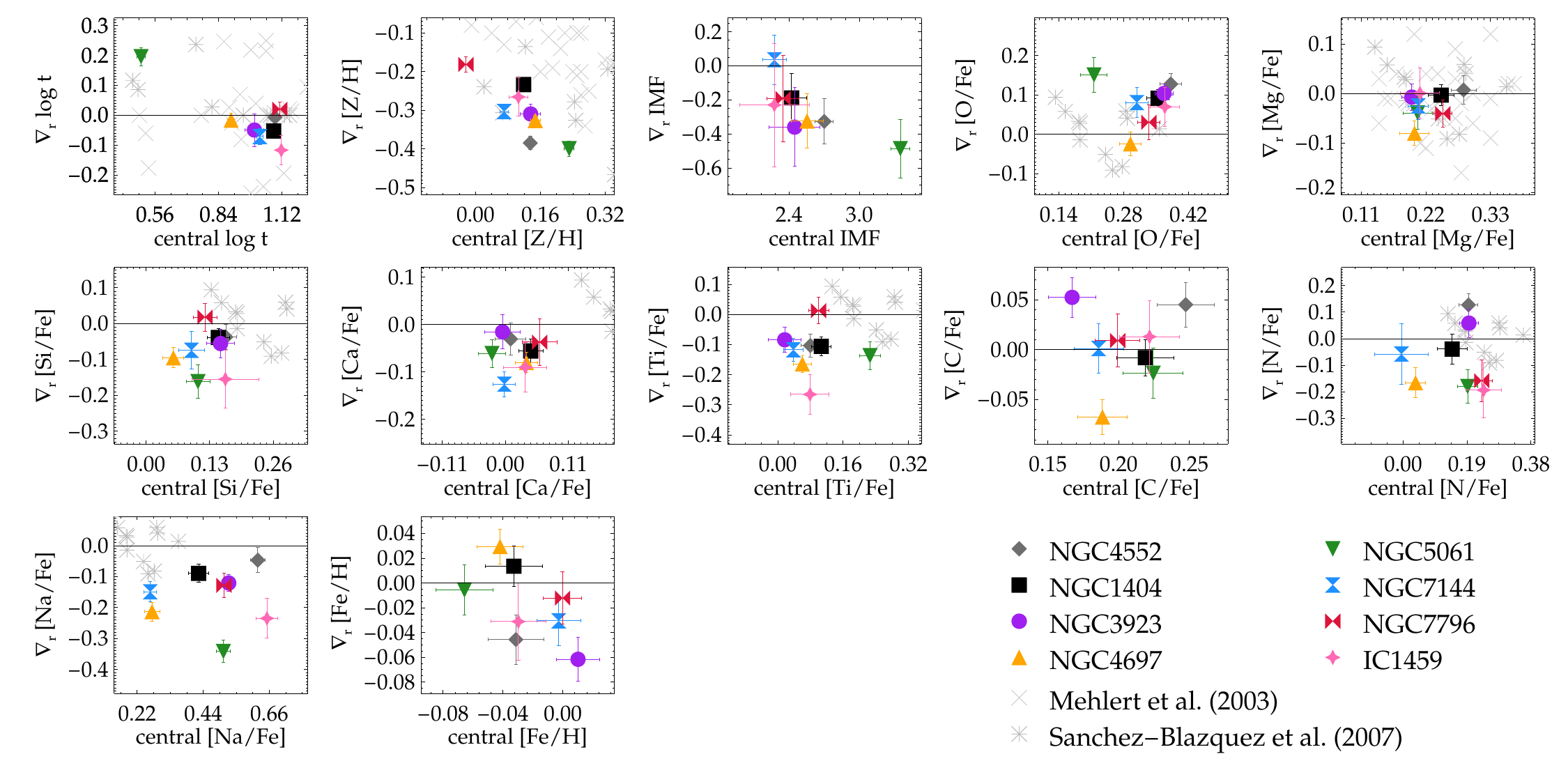}{0.9\textwidth}{}}
 \caption{Top: Stellar population profiles for each galaxy  as a function of $r$/\re. 
 Different colors and symbols denote different galaxies,  as outlined in the figure legend;  Gray plus symbols on the bottom of each panel are median uncertainties in three  radial ranges;  
Horizontal dashed lines are for solar metallicity and elemental abundances; Horizontal dotted lines for elemental abundances illustrate the approximate reference values of the \cite{2018ApJ...854..139C} model response functions, which were interpolated and extrapolated to measure  elemental abundances.
Bottom: Radial 
trends  $\nabla_r$  for each stellar population parameter and  galaxy, as a function of the error-weighted mean of the central ($r\leq$0.125\,\re) values. Horizontal lines denote constant value with radius.}
\label{fig:radgrad}
\end{figure*}

\subsection{Single Stellar Population Results}

In this section we present our results of the stellar populations and their radial gradients for our eight early-type galaxies.
For easier comparison with the literature, we converted the elemental abundances from [X/H], as they are parameterised in the models, to [X/Fe] by subtracting [Fe/H]. The exceptions are [Fe/H] and \met, which remain unchanged. 
We show the resulting radial stellar population profiles  in Figure \ref{fig:radgrad} (top), and list the results in Table \ref{tab:pyresult} for each galaxy.  In the panels of elemental abundances we show the reference values of the \cite{2018ApJ...854..139C} model response functions, which are used to derive elemental abundances (see Appendix \ref{sec:extrap})   as horizontal dotted lines.\footnote{The response function reference values are for [X/H], and not for [X/Fe] as in the plot. However, since [Fe/H]$\approx$0\,dex in all our fits, we neglected the difference to illustrate where the reference values of the response functions are. }
The radial gradients $\nabla_r$ are shown in Figure \ref{fig:radgrad} (bottom) as a function of the central value, computed as the error-weighted mean of the measurements located at $r\leq$0.125\,\re. 

The low-mass IMF slope  at \textless 1\,$M_\sun$ scatters around a median value of 2.44, with values ranging from 1.8 to 3.4.  \obfive\space has a significantly steeper low-mass IMF slope than  the \cite{1955ApJ...121..161S} IMF slope of 2.35, with values up to 3.4 in the central 0.1\,\re, but the low-mass IMF slope decrease at larger radii. 
 Some galaxies exhibit a constant low-mass IMF slope with radius, though there are also galaxies with a decreasing low-mass IMF with radius (\nablar IMF=$-0.4$/\logten($r$/\re)).

Most galaxies are older than 8\,Gyr, with some exceptions at larger radii. \obfour\space has an intermediate age (6.7-9.2 Gyr); \obfive\space is young, \textless6 Gyr. Our linear age  gradient ranges from --0.11 to +0.18 dex/\logten($r$/\re). \obfive\space is the only galaxy with a significantly increasing age with radius, all other galaxies have a decreasing or constant age.

All galaxies have a negative  \met\space gradient, with super-solar or roughly solar central values and decreasing to $\sim$--0.35\,dex. The galaxy with the steepest decreasing \met\space is \obfive\space ($-0.37$\,dex/\logten($r$/\re)).  \obsevsev\space ($-0.17$\,dex/\logten($r$/\re)) 
on the other hand has the most shallow decreasing \met, and  also  the lowest \met\space value in the center (0.01\,dex).  Interestingly, this galaxy is located in a low-density environment, whereas all of the other galaxies are either in clusters or groups (see Table \ref{tab:sample}).

Our measurements of [Fe/H] are close to solar (usually about --0.1 to 0.07\,dex) and either constant with radius or with a shallow slope ($-0.06\lesssim\lvert\nabla_r$[Fe/H]$\rvert\lesssim$ 0.03\,dex/\logten($r$/\re)); [O/Fe] spans 0.1 to 0.55\,dex; [Mg/Fe] is super-solar ($\sim$0.08 to 0.3\,dex); [Si/Fe] lies mostly within  --0.16 to 0.2\,dex; [Ca/Fe] is near solar  ($\sim$--0.2 to 0.1\,dex); [Na/Fe] is mostly super-solar  ($\sim$0.0 to 0.8\,dex); [C/Fe] is always super-solar ($\sim$0.1 to 0.3\,dex), [N/Fe] and [Ti/Fe]  have  a wide spread ($\sim$--0.2 to 0.4\,dex and --0.3 to 0.3\,dex).  
For most elements, the mean abundance measurement uncertainties are $\sim$0.05\,dex, and increase with radius, reflecting the decreasing S/N. The uncertainties of [N/Fe] are the largest, and they increase by a factor three from the center to the outer region, see gray colored symbols in the top panels of Figure \ref{fig:radgrad}.

We obtain the steepest radially decreasing  gradients for [Na/Fe], [N/Fe], and [Ti/Fe] ($ \nabla_r$[X/Fe]$ \lesssim-0.2$\,dex/\logten($r$/\re) for some galaxies, though the typical gradients are less steep ($\nabla_r\sim$--0.1 to --0.2\,dex/\logten($r$/\re). [Fe/H], [Mg/Fe], and [C/Fe] have  more shallow radial trends, with $-0.1$\textless$\nabla_r$\textless0.05 dex/\logten($r$/\re), see also Fig. \ref{fig:radgrad}, bottom. 
For a given element, the range of possible values in the sample increase with galaxy radius. This may indicate that the galaxies in our sample have different minor merger histories. The portion of stars brought to the galaxy via satellite accretion is larger at outer   radii than in the center, where the stellar populations are dominated by in situ formation.

We compare our results to stellar population measurements of the same and similar galaxies in the literature in Appendix \ref{sec:comparison}. In Appendix \ref{sec:sfts} we use the elemental abundances to constrain the star formation timescales. In Appendix \ref{sec:indiv} we summarise the results for each galaxy individually.

\section{Correlations of Stellar Population Parameters} 

\label{sec:correlation}

\subsection{Correlations of Central Stellar Population Parameters with Central Velocity Dispersion}\
\label{sec:globalsig}

Several studies found global correlations of galaxy overall  stellar population parameters (e.g. age, \met, [Mg/Fe]) with the velocity dispersion, using individual galaxies \citep[e.g.][]{2000AJ....120..165T,2010MNRAS.408...97K} or stacked spectra \citep[e.g.][]{2009ApJ...698.1590G,2010MNRAS.404.1775T,2014ApJ...783...20W,2014ApJ...780...33C}. 
 Here, we test if we can   reproduce  correlations  using our large set of stellar population parameters. 
  Such correlations are important to understand the assembly and evolution of galaxies.

\begin{figure*}
\gridline{\fig{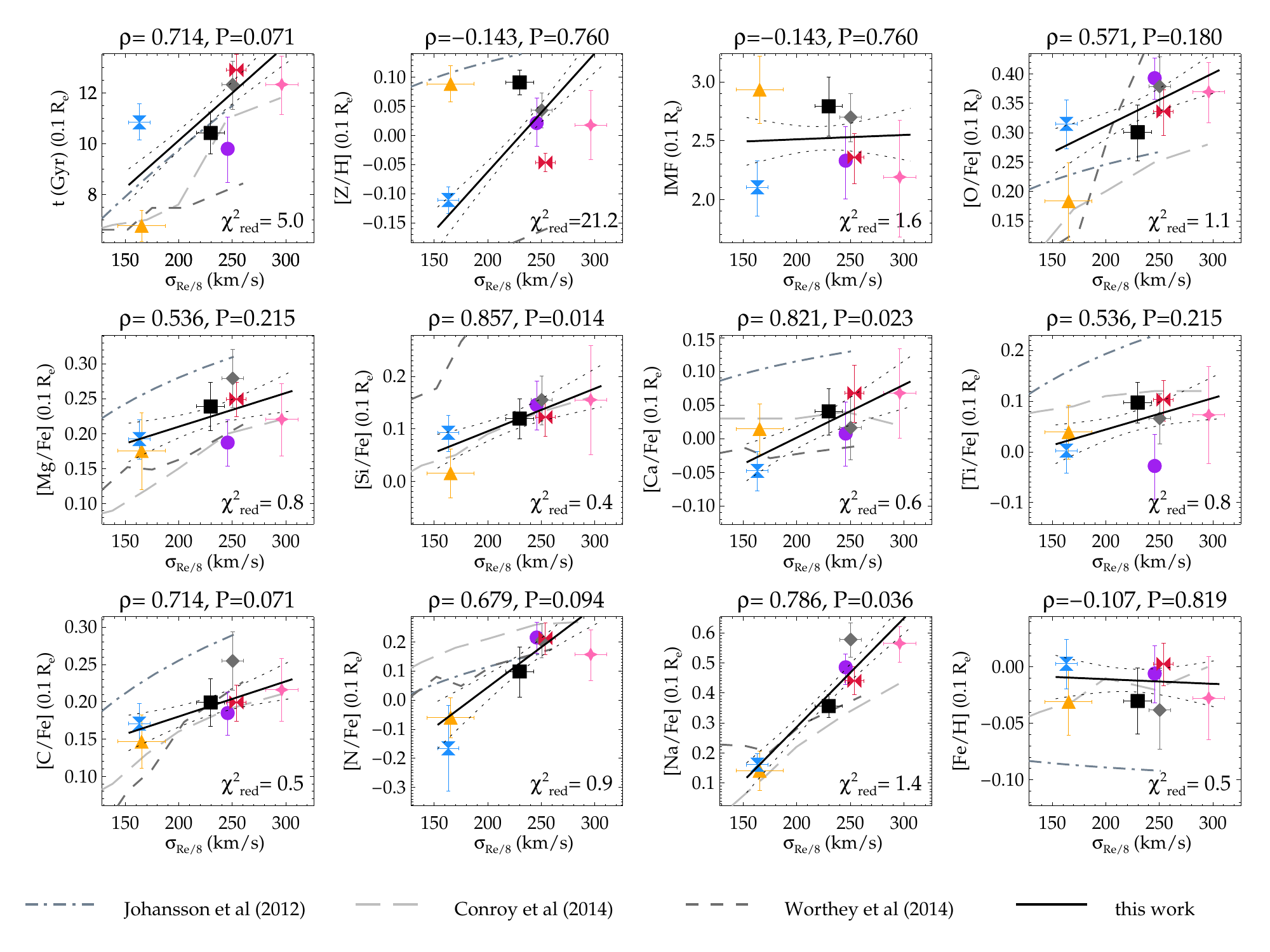}{0.95\textwidth}{}}
 \caption{Stellar population parameters (interpolated at 0.1\,\re) as a function of \sigacht\space   for seven old early-type galaxies, denoted by different colored symbols. 
   Spearman rank coefficient $\rho$ and  probability P  
 are shown at the top of each panel. The solid black line shows a linear fit, the gray dotted lines the 1-$\sigma$ uncertainties of the fit, the respective $\chi^2_\text{red}$ is denoted in the lower right corner of each panel. The dot-dashed, dashed, and long-dashed lines denote the relations found by \cite{2012MNRAS.421.1908J,2014ApJ...783...20W,2014ApJ...780...33C}.}
\label{fig:sigmallgrad01}
\end{figure*}

 \begin{table}
\centering
\tabletypesize{\scriptsize}
\caption{Linear fit to the stellar population parameters (interpolated at 0.1\,\re) as a function of \sigacht.} \label{tab:sigmacor1}
\begin{tabular}{l|ccr}
\tablewidth{0pt}
\hline
\hline
Parameter & Slope &Intercept& $\chi^2_\text{red}$\\
\hline
t& 0.038$\pm$ 0.007&   2.6$\pm$    1.5& 5.0\\
\met& 0.0020$\pm$ 0.0003&  -0.47$\pm$    0.06&21.2\\
IMF& 0.0004$\pm$ 0.0025&   2.4$\pm$    0.6& 1.6\\
\text{[O/Fe]}& 0.0009$\pm$ 0.0004&   0.13$\pm$    0.10& 1.1\\
\text{[Mg/Fe]}& 0.0005$\pm$ 0.0003&   0.11$\pm$    0.07& 0.8\\
\text{[Si/Fe]}& 0.0008$\pm$ 0.0004&  -0.07$\pm$    0.09& 0.4\\
\text{[Ca/Fe]}& 0.0008$\pm$ 0.0004&  -0.16$\pm$    0.08& 0.6\\
\text{[Ti/Fe]}& 0.0006$\pm$ 0.0005&  -0.08$\pm$    0.11& 0.8\\
\text{[C/Fe]}& 0.0005$\pm$ 0.0003&   0.09$\pm$    0.06& 0.5\\
\text{[N/Fe]}& 0.0028$\pm$ 0.0007&  -0.51$\pm$    0.17& 0.9\\
\text{[Na/Fe]}& 0.0036$\pm$ 0.0004&  -0.44$\pm$    0.10& 1.4\\
\text{[Fe/H]}&-0.0000$\pm$ 0.0002&  -0.00$\pm$    0.05& 0.5\\
\hline 
\hline
\end{tabular}
\end{table}

We investigated correlations with the velocity dispersion within \re/8, \sigacht, using the Hyperleda reference values from  Table \ref{tab:sample}. 
 We also computed  correlations with   the total B-band magnitude $M_B$,  and the dynamical mass \logten($M_{dyn}$) = 2 \logten($\sigma$)+\logten(\re)+3.1, with \re\space in pc \citep{2009ApJ...691L.138S}, using the  values for $M_B$,  and \re\space as listed in Table \ref{tab:sample}. Since  correlations with the latter parameters were usually less significant, we present only the correlations with \sigacht. 
 In our correlation analysis we do not consider \obfive, as it  has a more extended star formation history and therefore the SSP results may be biased (Appendix \ref{sec:sfh}). 
 We note that our sample size is relatively small (seven galaxies), and spans only a narrow range (\sigacht=163--296\,\kms, $M_B$=--20.3 to --21.4\,mag, \logten($M_{dyn}$) = 2 \logten($\sigma$)+\logten(\re)+3.1 = 11--12). 

We show the stellar population parameter values linearly interpolated to 0.1\,\re\space as a function of \sigacht\space in Figure \ref{fig:sigmallgrad01}.   We  performed linear fits, which are shown as solid black lines, the respective $\chi^2_\text{red}$ are noted in the lower right corners.  The linear fit  results are listed  in Table \ref{tab:sigmacor1}. Also shown in Figure \ref{fig:sigmallgrad01} are the   relations found by \cite{2012MNRAS.421.1908J,2014ApJ...783...20W,2014ApJ...780...33C}, 
  based on the global galaxy stellar populations. \cite{2012MNRAS.421.1908J} have a large data set  of $\sim$3800 individual galaxies, while \cite{2014ApJ...783...20W,2014ApJ...780...33C} derived their relations from stacked galaxy spectra. This explains why our data have a larger scatter. Moreover, these studies  used different methods and models. 
 Nevertheless, we see comparable trends with \sigacht.  
   
  For each parameter we 
 computed the Spearman rank correlation coefficient $\rho$ with \sigacht, denoted on the top of each panel. There are some  parameters with strong ($\lvert\rho\rvert\geq$0.6) or very strong ($\lvert\rho\rvert\geq$0.8)  correlations at this radius, in particular [Si/Fe], [Ca/Fe], [C/Fe], and [Na/Fe].  
 
We found no correlation of \met\space with \sigacht, which has been reported in the literature \citep[e.g.][]{2003A&A...407..423M,2010MNRAS.404.1775T,2015MNRAS.448.3484M},  for larger and/or more homogeneous samples. At a given value of \sigacht, our \met\space values span up to 0.2\,dex, which is within the range of scatter found in the literature for larger samples. 
 We  also tested whether the radial stellar population gradients $\nabla_r$ correlate with any mass proxies. No correlations exceed $\rho$=0.58, only \nablar [Mg/Fe] and \nablar [C/Fe] have moderate correlations ($\rho$=0.46 and 0.57, P=0.29 and  0.18) with \sigacht. 
\cite{2008MNRAS.389.1891R} and \cite{2019MNRAS.483.3420P} found no significant \nablar\space correlations, whereas  \cite{2005MNRAS.361L...6F,2009ApJ...691L.138S,2019MNRAS.489..608F,2020A&A...635A.129K} did, e.g. for \nablar\met. However, these studies have larger or more homogeneous samples, which allows them to detect correlations despite the intrinsic scatter among galaxies.  
 It is possible that we do not find these correlations   due to our small sample size, and narrow range of \sigacht.

\subsection{Global Correlations of Central Values and Gradients}

\label{sec:globcen}

\begin{figure*}
\gridline{\fig{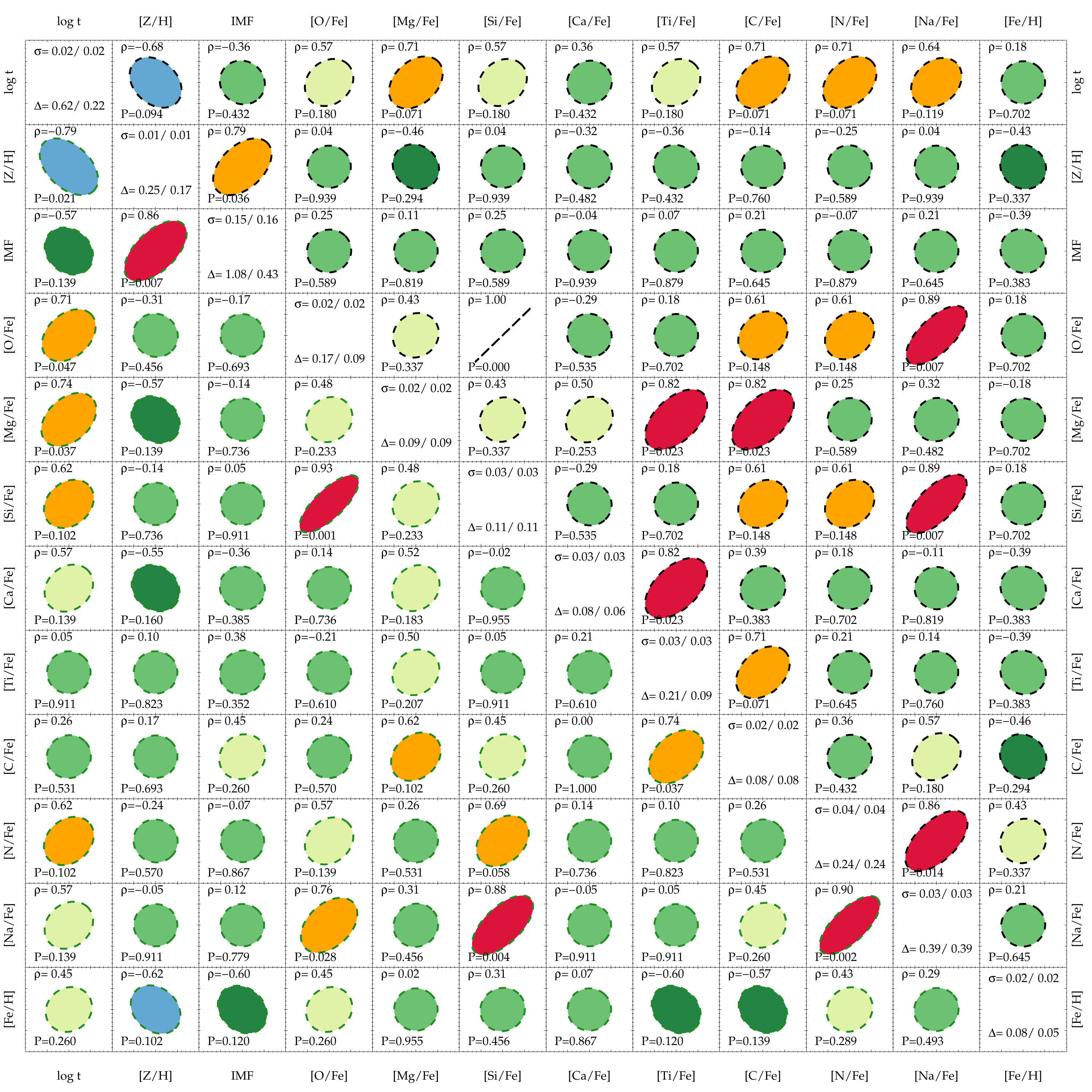}{0.99\textwidth}{}}
 \caption{Correlation matrix for  Central Stellar Population Parameters (error-weighted mean values at $r\leq$0.125\,\re). Larger eccentricities signify stronger correlations, i.e. greater  $\lvert\rho\rvert$. Upper right off-diagonal cells when excluding \obfive, lower left  when including \obfive.  Strong positive (negative) correlations with $\rho$\textgreater0.6 ($\rho$\textless--0.6) are marked with orange (light blue), very strong  positive (negative) correlations with $\rho$\textgreater0.8 ($\rho$\textless--0.8) with red (dark blue). Weak to moderate correlations (--0.6$\leq\rho\leq$0.6) are shown in different shades of green. $\rho$ and P are denoted in each panel, the numbers in the diagonal panels denote the mean uncertainty $\sigma$ and the range of values $\Delta$, with and without \obfive, respectively, for each parameter.}
\label{fig:valval}
\end{figure*}

\begin{figure*}
\gridline{\fig{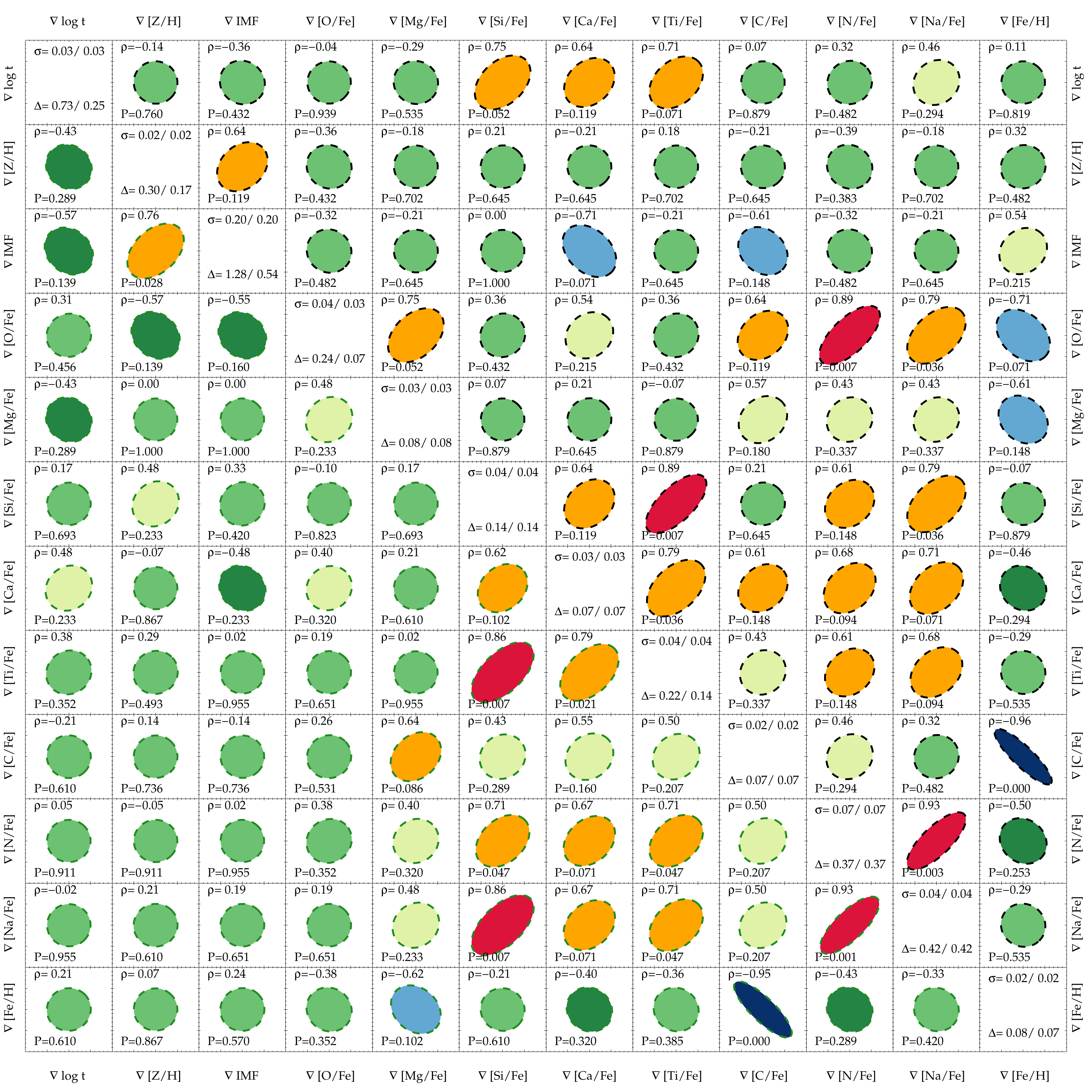}{0.99\textwidth}{}}
 \caption{Correlation matrix for  Radial Gradients \nablar\space of  Stellar Population Parameters, colors and symbols as in Figure \ref{fig:valval}.}
\label{fig:gradgrad}
\end{figure*}

\begin{figure*}
\gridline{\fig{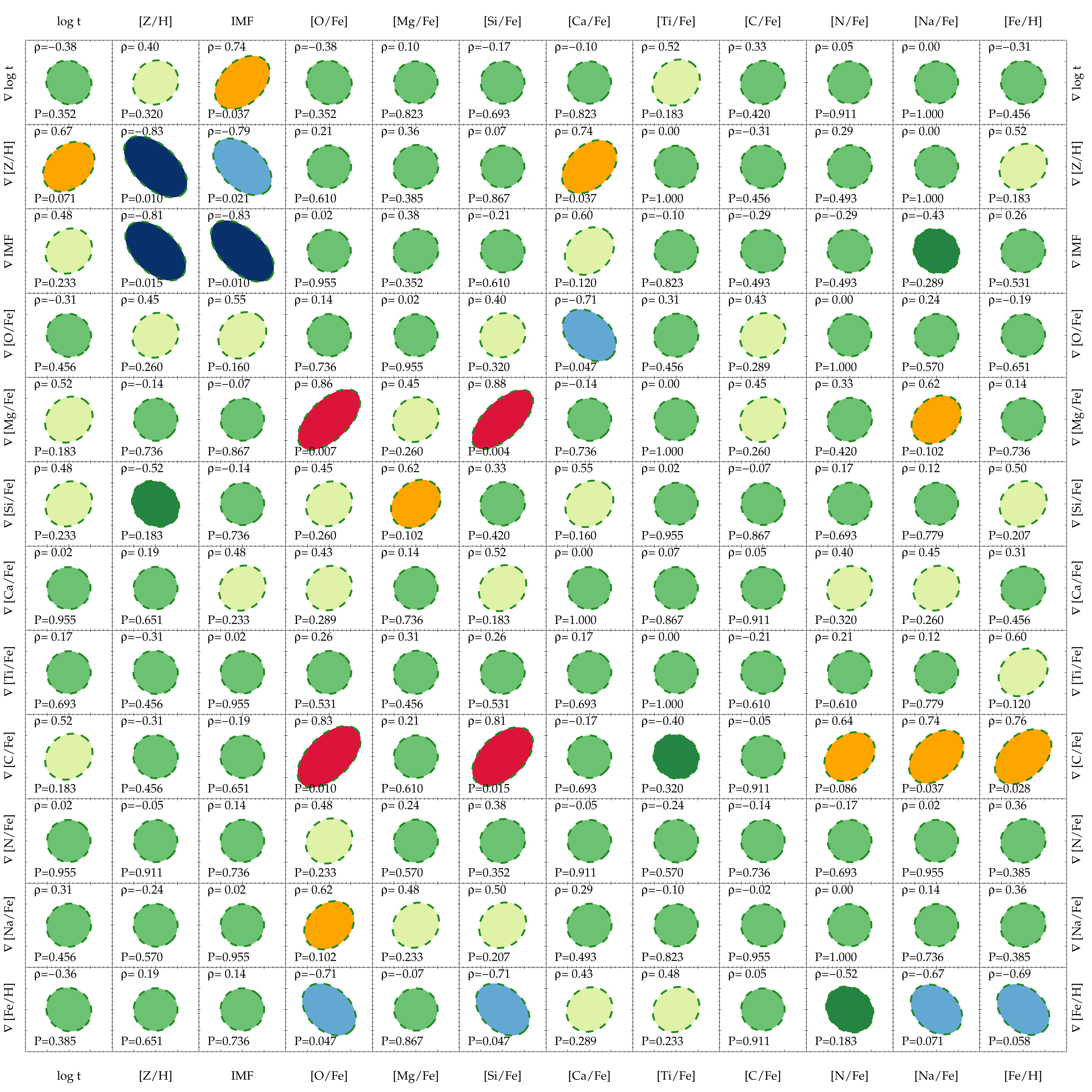}{0.99\textwidth}{}}
 \caption{Correlation matrix for Central Stellar Population Parameters (error-weighted mean values at $r\leq$0.125\,\re) with Radial gradients \nablar\space for all eight galaxies (including \obfive). }
\label{fig:valgrad}
\end{figure*}
\begin{figure*}
\gridline{\fig{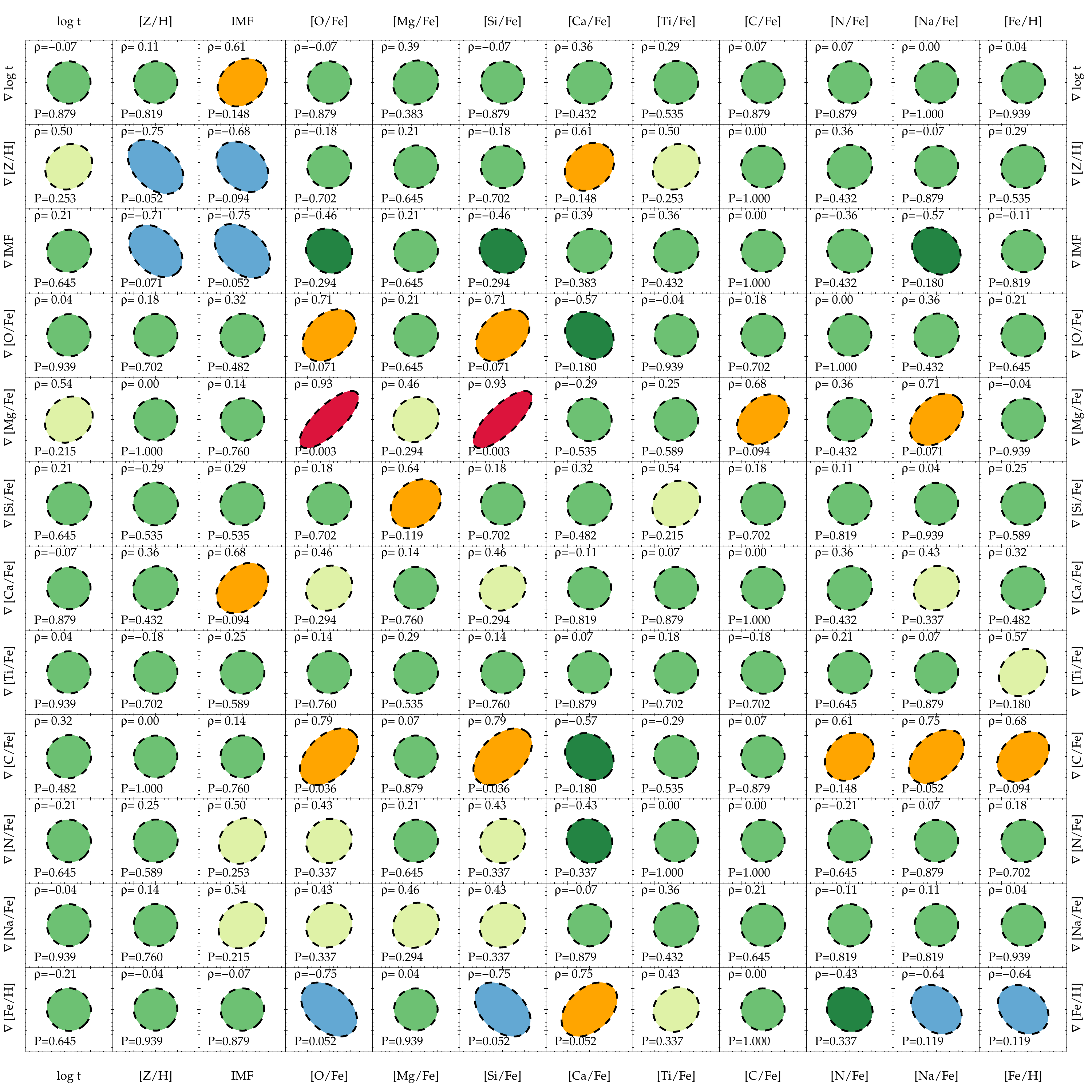}{0.99\textwidth}{}}
 \caption{Same as Figure \ref{fig:valgrad}, but  excluding \obfive.}
\label{fig:valgrad5}
\end{figure*}

We investigated whether there are any  correlations of the central stellar population parameters with each other, with the radial gradients $\nabla_r$, or among the gradients. We show the correlation matrixes in Figures \ref{fig:valval}, \ref{fig:gradgrad}, \ref{fig:valgrad}, and \ref{fig:valgrad5}.
The upper right off-diagonal of Figures \ref{fig:valval},  \ref{fig:gradgrad}  and all of Figure  \ref{fig:valgrad5} 
are when excluding \obfive; the lower left off-diagonal of  Figures \ref{fig:valval}, \ref{fig:gradgrad} and all of Figure  \ref{fig:valgrad} when including \obfive. 
Different colors and  the eccentricity of the ellipses denote the strength of the correlations; the Spearman rank correlation coefficient $\rho$ is also denoted on each panel. 

\begin{figure*}
\gridline{\fig{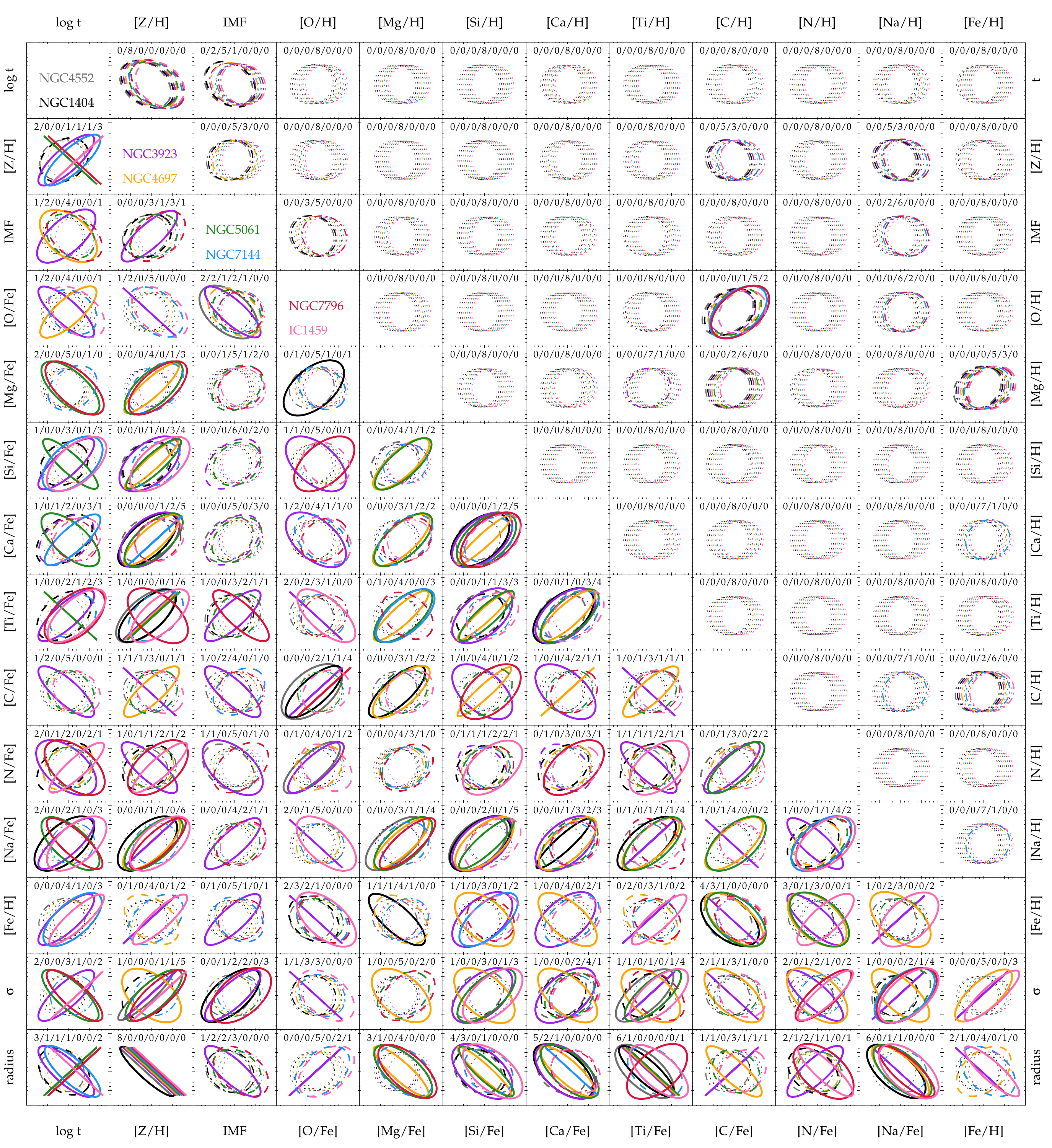}{0.99\textwidth}{}}
 \caption{Correlation matrix.  Upper-right off-diagonal displays the mean covariances   per galaxy  of fitting parameter PDFs obtained in the \pystaff\space fit;  Lower left off-diagonal  the  correlations of local values of Stellar Population Parameters for each individual galaxy.
 Different colors denote different galaxies as in Figure \ref{fig:speccom}, solid thick lines  very high correlations ($\lvert\rho\rvert\geq$0.8), dashed lines high  (0.6$\leq\lvert\rho\rvert$\textless0.8), dot-dashed lines  moderate (0.4$\leq\lvert\rho\rvert$\textless0.6), dotted lines no or weak correlations ($\lvert\rho\rvert$\textless0.4). In each panel we note the number of galaxies with $\rho\leq$-0.8/-0.8\textless$\rho\leq$-0.6/-0.6\textless$\rho\leq$-0.4/-0.4\textless$\rho$\textless0.4/0.4$\leq\rho$\textless0.6/0.6$\leq\rho$\textless0.8/$\rho\geq$0.8.   Each ellipse is slightly offset to the right to  make them easier visible.}
\label{fig:corrlocal}
\end{figure*}

We find a very strong positive correlation of the central \met\space and IMF slope. We caution that these parameters also have  positive covariances in several PDFs of our fits, though on average the covariance is weak to moderate (upper right off-diagonal of Figure \ref{fig:corrlocal}). We find  other strong  correlations  of central values for various combinations of elemental abundances, e.g. [Na/Fe] with [N/Fe] and [Si/Fe], or [O/Fe] with [Si/Fe]. 
However, these parameters have no strong covariances of their PDFs in our fits, with only [Na/H] and [O/H] showing  moderate covariances in some fits. 
Most of the correlations in Figure \ref{fig:valval}  are positive,  which is expected for elements that are, at least partially, produced by the same process, like $\alpha$ elements in massive stars;  O and Na, which are produced in both high-mass and to some extent in  intermediate-mass stars;  C and Ti, both produced in high mass stars;  or N and Na, which are  produced in intermediate-mass stars.

As shown in Figure \ref{fig:gradgrad}, we  find  several correlations  among stellar population gradients. As a consequence, [N/Fe] and [Na/Fe] not only correlate in the center, but also at larger radii. Also the radial gradients of various $\alpha$ elements correlate with one another (e.g. O and Mg, Si and Ti). 

Further, we obtain an  anti-correlation of the central \met\space with \nablar\met \space (Figures \ref{fig:valgrad} and \ref{fig:valgrad5}), 
which was also found by \cite{2007MNRAS.377..759S} and \cite{2008MNRAS.389.1891R}. It  is also shown in Figure \ref{fig:radgrad}. 
This means that the steepest \nablar \met\space is found in galaxies with high central \met, and  that at 1\,\re\space the range of  \met\space values in our data set is more narrow. 
Likewise, the central  \met\space anti-correlates with \nablar IMF such that galaxies with high central \met\space (and a more bottom-heavy IMF slope) tend to have a steeper decreasing \nablar IMF. Thus, the range of possible IMF slope values at 1\,\re\space is narrower (1.7-2.4) than at 0.01\,\re\space (2-3.3). 

There are  several further correlations of central stellar population parameters with another parameter's radial gradient. 
Some of the cases  have only a strong correlation when excluding \obfive\space (e.g. [O/Fe] with \nablar[O/Fe]),  and a weak correlation otherwise. They  require further confirmation, as they may be  caused by the low number of only seven objects, spanning a relatively narrow range of parameters.

 \begin{table}[ht]
\centering
\tabletypesize{\scriptsize}
\caption{Number of Galaxies with high Spearman Rank Correlation $\rho$ for Local Stellar Population Parameters, excluding  the central data points. We list only cases with at least five high   correlations ($\lvert\rho\rvert\geq$0.6)} \label{tab:localcor1}
\begin{tabular}{llccc}
\tablewidth{0pt}
\hline
\hline
Parameter & Parameter &sign& \multicolumn{2}{c}{Number of  galaxies with}\\
     &       & &$\lvert\rho\rvert\geq$0.8 & 0.6$\leq\lvert\rho\rvert$\textless 0.8\\
\hline
\decimals
r/\re          & \met     &neg.  & 8  &   0 \\
r/\re             & [Si/Fe]  &neg.    & 4  &   3 \\
r/\re             & [Ca/Fe]  &neg.    & 5  &   2 \\
r/\re             & [Ti/Fe]  &neg.    & 6  &   1 \\
r/\re             & [Na/Fe]  &neg.    & 6  &   0 \\
\hline
$\sigma$            & \met     &pos.  & 5  &   1 \\
$\sigma$            & [Ca/Fe]  &pos.    & 1  &   4 \\
$\sigma$            & [Ti/Fe]  &pos.    & 4  &   1 \\
$\sigma$            & [Na/Fe]  &pos.    & 4  &   1 \\
\hline
\logten(t)      & [Ti/Fe]    &pos. & 3  &   2 \\
\hline
\met            & [Si/Fe]    &pos.  & 4  &   3 \\
\met            & [Ca/Fe]    &pos. & 5  &   2 \\
\met            & [Ti/Fe]    &pos. & 6  &   1 \\
\met            & [Na/Fe]    &pos. & 6  &   0 \\
\hline

[Fe/H]            & [O/Fe]    &neg.  & 2  &   3 \\

[Fe/H]            & [C/Fe]   &neg.  & 4  &   3 \\

\hline

[O/Fe]            & [C/Fe] \tablenotemark{a}    &pos.  & 4  &   1 \\

\hline

[Mg/Fe]            & [Na/Fe]    &pos.  & 4  &   1 \\

\hline 

[Si/Fe]            & [Ca/Fe]    &pos.  & 5  &   2 \\

[Si/Fe]            & [Ti/Fe]    &pos.  & 3  &   3 \\

[Si/Fe]            & [Na/Fe]    &pos.  & 5  &   1 \\

\hline 

[Ca/Fe]            & [Ti/Fe]    &pos.  & 4  &   3 \\

[Ca/Fe]            & [Na/Fe]    &pos.  & 3  &   2 \\

\hline 
[Ti/Fe]            & [Na/Fe]    &pos.  & 4  &   1 \\

\hline

[N/Fe]            & [Na/Fe]    &pos.  & 2  &   4 \\

\hline 
\hline
\end{tabular}
\tablecomments{a: Some of the Parameter PDFs are correlated.}
\end{table}

\subsection{Correlations of Local Values for Individual Galaxies}
\label{sec:localsig}

Besides the correlations of our entire sample of galaxies,  we also studied whether the local parameters of the individual galaxies correlate.  We measured the Spearman rank correlation coefficients of our 12 stellar population parameters with each other,  the local velocity dispersion $\sigma$, and the radius $r$/\re. As for the radial gradient fitting, we omitted the most central measurements, as the spectrum is likely affected by seeing. 
We found several strong correlations ($\lvert \rho\rvert$\textgreater0.6) and even some very strong correlations ($\lvert \rho\rvert$\textgreater0.8).  However, except for  \met\space with $r$/\re, the correlations never hold for all eight galaxies. Often the correlation for at least one galaxy is very weak, or even in the opposite direction. We show the parameter combinations in the lower left off-diagonal of Figure \ref{fig:corrlocal}. Different colored ellipses denote different galaxies, solid lines a very strong ($\lvert \rho\rvert\geq$0.8), dashed lines a strong (0.6$\leq\lvert \rho\rvert$\textless0.8), dot-dashed lines a moderate (0.4$\leq\lvert \rho\rvert$\textless0.6), and dotted lines a weak or no correlation ($\lvert \rho\rvert$\textless0.4). On top of each panel we list the numbers of galaxies in different ranges of $\rho$ ($\rho\leq$-0.8/ -0.8\textless$\rho\leq$-0.6/ -0.6\textless$\rho$\textless-0.4/ -0.4\textless$\rho$\textless0.4/ 0.4\textless$\rho$\textless0.6/ 0.6$\leq\rho$\textless0.8/ $\rho\geq$0.8). 
 We also list the parameter combinations with at least five galaxies with a  strong local correlation in Table \ref{tab:localcor1}.

For a large number of galaxies we find high negative correlations with $r$/\re\space (bottom row in Figure \ref{fig:corrlocal}),  confirming our radial gradient measurements in Section \ref{sec:radgradfit}. Several \nablar\space  are negative for \met, [Si/Fe], [Ca/Fe], [Ti/Fe], and [Na/Fe], whereas e.g. the IMF slope, [N/Fe], or [O/Fe] have a non-monotonic radial behavior for some galaxies, leading to weaker correlations. 

The correlations with $\sigma$ (second row from bottom in Figure \ref{fig:corrlocal}) are mostly positive, but this may be simply caused by the radially decreasing $\sigma$ profile of most of our galaxies (see Figure \ref{fig:sigmaprof}). \obfour\space (shown in yellow), a galaxy with a  radially slightly increasing $\sigma$, has several very high negative correlations of stellar population parameters with $\sigma$, while the other galaxies have positive correlations. Also \cite{2021A&A...645L...1B} found that the positive IMF-$\sigma$ relation does not hold for a galaxy with radially increasing $\sigma$. We found that this is also the case for other stellar population parameters.

The positive correlations of several elemental abundances with \met\space (second column  in Figure \ref{fig:corrlocal}) are  expected, as increased elemental abundances contribute to the overall \met\space of a galaxy. There is also a  positive correlation of \met\space with the IMF slope for four galaxies. \cite{2021A&A...645L...1B} suggest that such correlations may be connected to the positive covariance of IMF and \met, however, in our spectral fitting the covariances are on average only  weak to moderate (0.14\textless$\rho$\textless0.46).  
Also  correlations of O and C (5 galaxies), IMF slope and O, Mg and C (4 galaxies each) may be to some extent connected to the covariances of these parameters in the PDFs of the spectral fits, though these tend to be weaker than our local correlations (see upper right off-diagonal  in Figure \ref{fig:corrlocal}). 
We note that Fe and C in our fits have a positive covariance, so the negative local correlation found for seven galaxies cannot be due to a covariance of  the  parameters'  PDFs.

Elements that are mostly  produced in massive  stars  (e.g. Mg, Si, Ca, Ti) correlate with each other and with Na in several galaxies.  Sodium also correlates with  nitrogen, which is produced in intermediate-mass stars. This suggests that sodium is produced  partially in high-mass, but also in   intermediate-mass stars. On the other hand, correlations of nitrogen with other elements produced dominantly in high-mass stars tend to be weaker. 
This may indicate a non-uniform IMF slope shape at the high-mass end (i.e. \textgreater 3 \msun) as a function of galaxy radius. But it may also be caused by  \met-dependent nitrogen yields (see Section \ref{sec:discussion}). We discuss the ratios of different elemental abundances and clues on the element production in Section \ref{sec:elprod}.

We found no strong local correlation with the low-mass IMF slope  that holds for more than four galaxies. The strongest cases are a negative correlation  of IMF slope with O, and a positive correlation with \met\space for four galaxies each. In both cases, the PDFs of these parameters  have covariances  in the same direction in some of our fits, though on average the covariances are only weak to moderate. 
\cite{2021A&A...645L...1B} suggest that the IMF-\met\space correlation may be, at least partially, caused by a covariance of these parameters in stellar population fits.  This correlation holds locally only for some  of our galaxies, but is very high  globally ($\rho\geq$0.79, Figure \ref{fig:valval}). This suggests that the global correlation is not  entirely caused by unresolved degeneracies in our fits, and these parameters may indeed correlate, though it is not clear if there is a causal relation.

 \section{Relative Elemental Abundances and Element Production}
 \label{sec:elprod}

 Here we compare  abundances of different elements, and how they may depend on \met. With a  focus  on oxygen, carbon, nitrogen, and sodium,  we discuss  some possible implication on  the production of these elements. 
 
\begin{figure}
\plotone{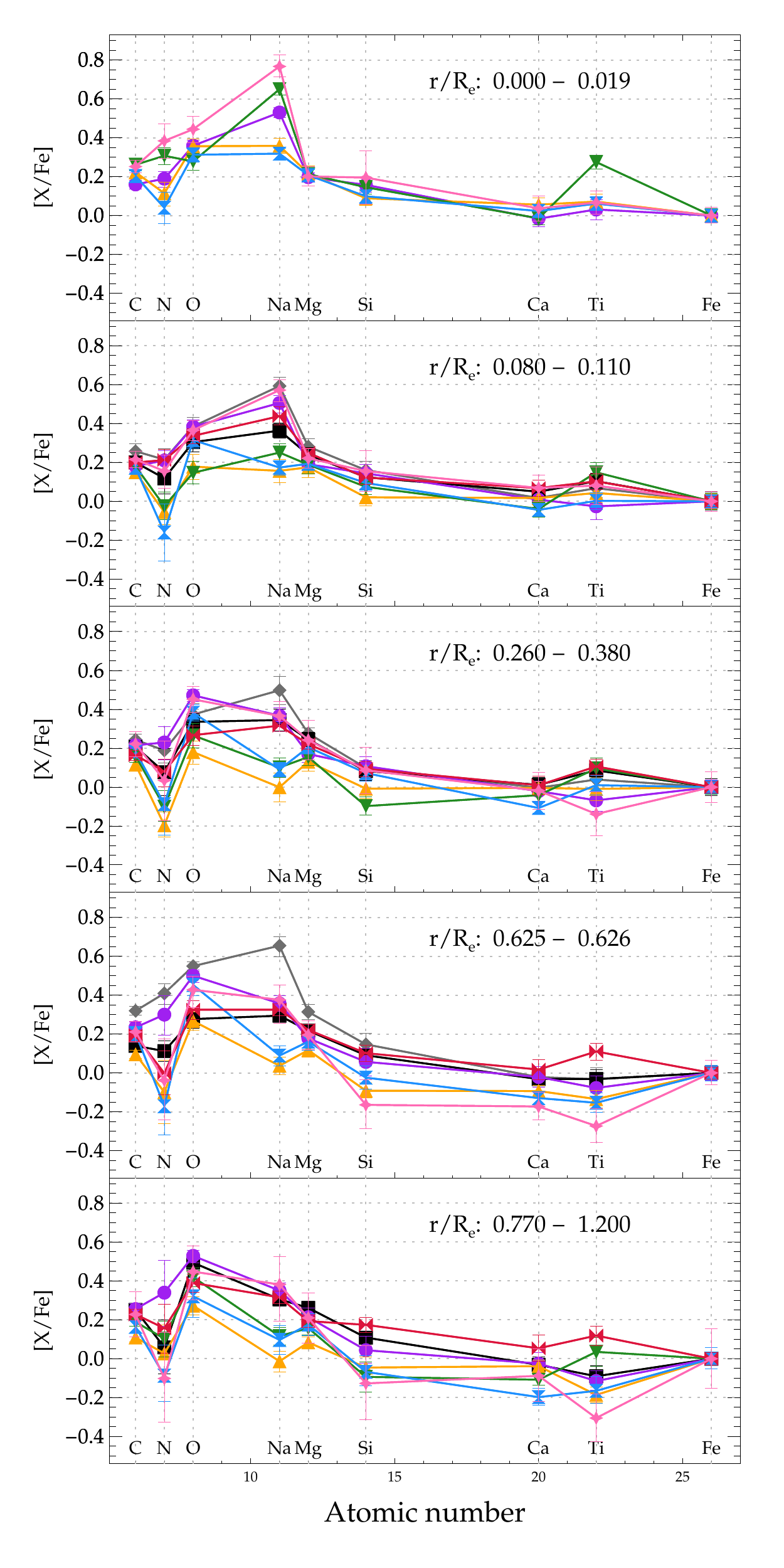}
\caption{Elemental abundances [X/Fe] as function of atomic number for five different radial bins (0--0.019 \re, 0.08--0.11 \re, 0.26--0.38 \re,  0.625--0.626 \re, and 0.77-1.2 \re).  Different colors and symbols denote different galaxies, as in previous Figures. }
\label{fig:atomic}
\end{figure}

 We found that $\alpha$ element abundances differ from element to element, as can be seen in Figure \ref{fig:atomic}, where we show the elemental abundances as a function of atomic number of X, different panels denote different radial ranges. Overall, lower mass elements tend to have higher abundances, i.e. [O/Fe]\textgreater [Mg/Fe]\textgreater[Si/Fe]\textgreater[Ca/Fe], though this does not always hold for [Ti/Fe].     

\begin{figure}
\plotone{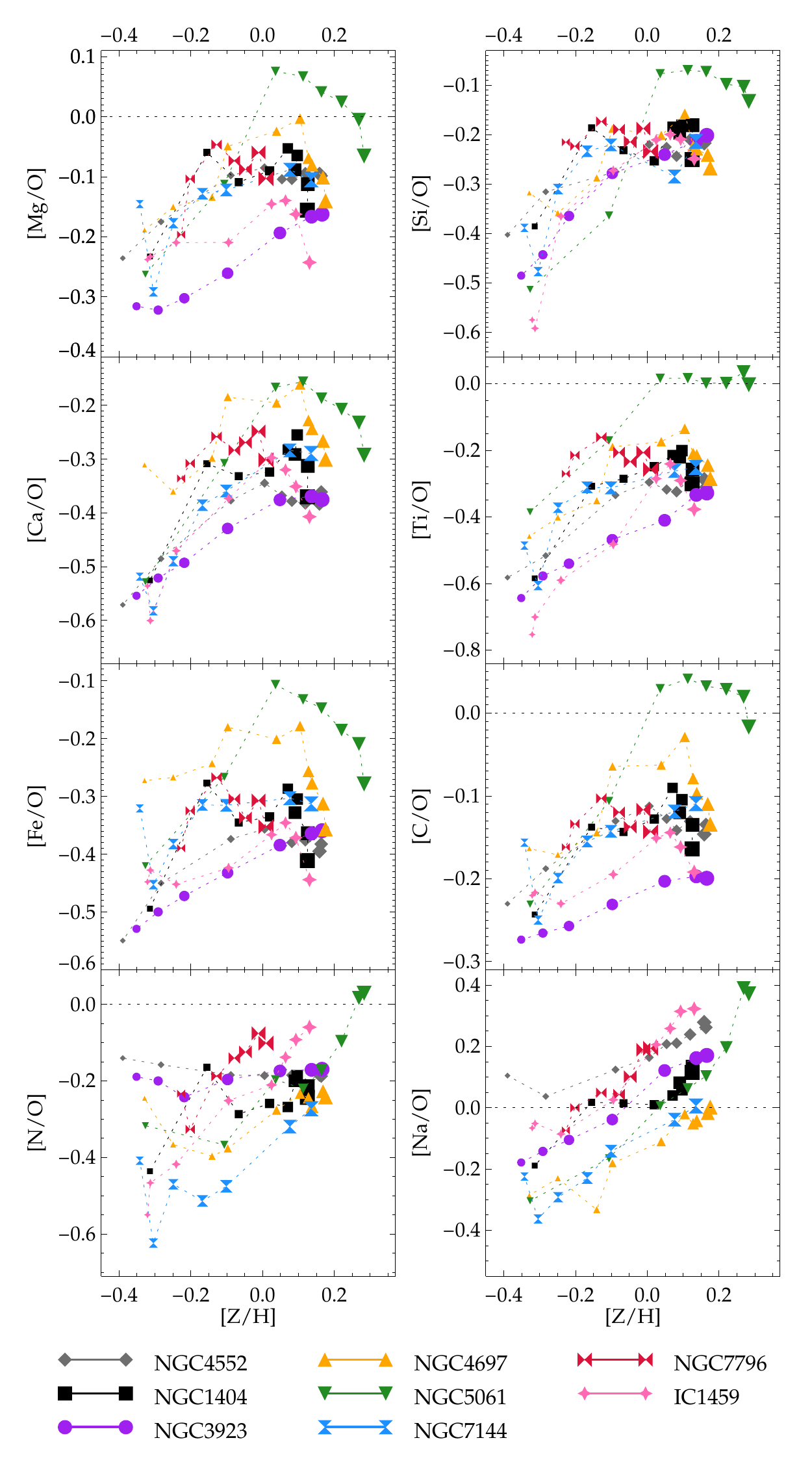}
\caption{Elemental abundances [X/O] as function of total metallicity \met\space  for eight galaxies, denoted by different  colors and symbols. Decreasing symbol sizes denote increasing galaxy radius. }
\label{fig:vsz}
\end{figure}

Oxygen is the most abundant element in the Universe (with the exception of hydrogen and helium, and followed by carbon and nitrogen), and thought to dominate \met. However, the measurement of oxygen is more complicated than e.g. Mg, as oxygen only appears in molecular transitions in combination with Ti in our spectra.  Oxygen  is also crucial to estimate C and N,  as those exist in CO, CN, and C$_2$ molecules. 
Oxygen is  to some extent also formed in intermediate-mass stars \citep[$\sim$10\%,][]{2019A&ARv..27....3M}. Therefore, it may not be surprising that O does deviate from  $\alpha$ elements such as Mg, meaning [$\alpha$/O]$\neq$0. Though, we find deviations as large as [Mg/O]=--0.3\,dex (see Figure \ref{fig:vsz}). Values ranging from --0.3 to +0.6 were also found in  Milky Way  bulge stars  \citep{2008AJ....136..367M}. 
Moreover, we found that the ratio of [X/O] seems to correlate with \met\space for several galaxies (see Figure \ref{fig:vsz}).  The general trends are similar for most [$\alpha$/O] and [C/O], probably since $\alpha$ elements and $\sim$70\% of carbon are produced in massive stars \citep{2020A&A...639A..37R}.  [X/O] is  increasing with \met\space at \met\textless0, then there is a flattening and sometimes a turnover towards higher \met, with decreasing [X/O] for higher \met. A possible explanation  are lower oxygen yields at increasing  \met, caused by higher mass loss \citep{1992A&A...264..105M,2002A&A...390..561M}.  At super-solar \met\space and in very massive stars,  carbon yields decrease even more than oxygen yields, causing the turnover of [C/O]  \citep{1998A&A...334..505P}.

Nitrogen is thought to be   primarily produced by intermediate-mass stars (4--8 $M_\sun)$, with  a small contribution of massive stars, that is more important at low \met. 
We find that [N/O] increases with \met\space for several galaxies, especially at \met$\gtrsim$-0.2\,dex. This may be since  nitrogen is produced as secondary element, from intermediate-mass stars that formed from gas that was already enriched with C and O \citep{2019A&ARv..27....3M}. At lower \met, studies of nitrogen and oxygen in the gas phase found a flattening of [N/O], which is attributed to the primary contribution from massive stars \citep{2019A&ARv..27....3M}. \met-dependent yields may also cause the weak local correlations of $\alpha$ elements to N, found in Section \ref{sec:localsig}.  Similar to  studies of N/O in the gas phase, we see indications of a flattening of [N/O] in individual galaxies at the outer radii, where \met\space decreases. However,  we caution that our nitrogen measurements at large radii have large uncertainties, exceeding 0.1\,dex. 

Most galaxies  have increasing [Na/O] with \met\space over the entire range. The correlation of [Na/O] with \met\space is even higher than of [Na/Fe] with \met\space (Section \ref{sec:localsig}), with a high Spearman rank correlation coefficient $\rho$\textgreater0.65 for all galaxies. The behavior of [Na/Fe] in the Milky Way bulge indicates that  Na may be produced in massive stars, and Na yields increase with [Fe/H] \citep{2016PASA...33...40M}. This may explain why there is no flattening of [Na/O] with \met, as we see for other [X/O] ratios.
N has only few high local correlations with other elements, with the exception of Na. Six galaxies have high local correlations of Na and N, which is mainly produced in intermediate-mass stars. This may suggest that Na is, to a larger  extent than $\alpha$ elements, formed in intermediate-mass stars.

To summarise,  different [X/O] ratios show different behaviors with \met\space at the high- and low-\met\space end. Most $\alpha$ elements behave in a similar way.  Oxygen and carbon production appear to be suppressed at high \met, but this is not the case for sodium. 

\section{Discussion}
\label{sec:discussion}

\subsection{Limitations of our Stellar Population Measurements}

One caveat in our study is that we assumed a single stellar population, although we see  signs of mixed stellar populations at larger radii (Appendix \ref{sec:sfh}). These outer stars were  possibly brought in from minor mergers, or formed later in a star formation event triggered by a galaxy merger, as part of  the second stage of galaxy assembly. Neglecting the star formation history may bias our SSP parameters at larger radii. For example,   the light-weighted age is biased to lower values. For \obfive, this bias seems to exist even over the entire galaxy, not just in  the outer region.  Including a second stellar population in the fit, as done by \cite{2017ApJ...841...68V}, is probably a better approximation, and we are planning to implement this in the future.  Although the assumption of a  SSP is an oversimplification, we found that most of our galaxies are dominated by an old, metal-rich  stellar population, at least in the inner regions (Appendix \ref{sec:sfh}),  and we can still  study differences among the   galaxies in our sample.

The SSP model spectra themselves are based on various assumptions: They are built from atomic and molecular line lists, which may be incomplete, and model atmospheres that assume local thermodynamic equilibrium (LTE).  However, we see no systematic residuals in our fits that  would indicate  a missing line, and  the relative error introduced by the LTE assumption is  negligible ($\ll$1\%), except for the Ca H\&K region \citep{2018ApJ...854..139C}, which we did not include in the fits.

In addition, our fitting method has certain restrictions: 
The elemental abundance measurements are obtained by applying linear multiplicative  response functions for each individual element \citep{2018MNRAS.475.1073V,2018MNRAS.479.2443V}. The response functions of the \cite{2018ApJ...854..139C}  models are  computed at reference values. In some of our fits, the resulting elemental abundances of [C/H] and [O/H] exceed the reference value, meaning our results are an extrapolation. 
The fitting code \pystaff\space  applies a Taylor expansion near the reference value. 
Compared to a linear extrapolation, our results for [O/H] that exceed the reference values of 0.3\,dex are  biased to lower values. The bias increases with increasing values of [O/H], reaching up to 0.1\,dex. The bias in [C/H] for values beyond the reference values of 0.15\,dex is smaller,  \textless0.05\,dex (see Appendix \ref{sec:extrap}).  
For other elements  (Mg, Si, Ca, Ti, N, Na, Fe), our results are mostly  within the reference values of the response functions, and for this reason unaffected by problems related to extrapolation.

\subsection{Variations of the IMF}

We measured the low-mass IMF slope (0.08-1\,\msun) and found some indications for global correlations, but few local correlations. 

Several studies have suggested a local correlation of the IMF with other SP parameters, but we can confirm these correlations only for subsets of our sample. For example, \cite{2018MNRAS.477.3954P} used stacked galaxy spectra and found local high positive correlations of the IMF slope with \met, [Na/Fe], $\sigma$, $r$/\re, and  age, while  we obtain these correlations for no more than four galaxies in our sample. 
Why do we find  these correlations   only for some galaxies? 
Concerning the IMF-\met\space correlation, \cite{2021A&A...645L...1B} suggest that it is, at least partially, due to a  degeneracy of these parameters in spectral fits. We also see a correlation of these parameters in the PDFs of some, but not all spectral fits. Thus, the strength of this correlation may be overestimated in other studies. 
The IMF-[Na/Fe] correlation of \cite{2018MNRAS.477.3954P}    may be a consequence of  local \met-[Na/Fe] correlations (Table \ref{tab:localcor1}); thus, breaking the  IMF-\met\space degeneracy in our fits means we will not detect an IMF-[Na/Fe] correlation.  
Some galaxies in our sample have a relatively flat IMF profile, and hence  no correlation with $r$/\re.
Lastly, the IMF-$\sigma$ correlation does not hold for galaxies with a radially increasing $\sigma$ \citep{2021A&A...645L...1B}; hence there is no simple causal relation of IMF and $\sigma$.

The lack of correlations between the IMF slope and elemental abundances may not be surprising. We measured the low-mass end of the IMF slope (\textless 1\,\msun), whereas the elements were mostly produced in more massive stars (\textgreater3\,\msun). The IMF  may be more complicated than a single slope at \textless 1\,\msun\space and the Salpeter slope at \textgreater1\,\msun, and require a non-parametric description \citep{2017ApJ...837..166C} over the entire stellar mass range.  
We may even see a variation  of the IMF in the intermediate-mass range in our Na and N abundances: At a given value of \met, [Na/Fe] and [N/Fe] vary by roughly 0.3--0.4\,dex from  galaxy to galaxy, while various $\alpha$/Fe have a range of only  $\sim$0.15\,dex. Also [C/Fe] varies by about 0.15\,dex, but the upper end of our [C/Fe] range may be biased to lower values by $\lesssim$0.05\,dex (Appendix \ref{sec:extrap}). One possible explanation is  that the IMF slope in the intermediate-mass range (3-8\,\msun), where most of N and some Na and C are produced,  varies among individual galaxies.

Instead of local, we see some global IMF correlations: 
Galaxies with higher central \met\space tend to have a higher  central IMF slope, steeper than Salpeter, but becoming more Salpeter-like with increasing radius  (see Section \ref{sec:globcen}). Galaxies with lower central \met\space  tend to have a radially flat, and Salpeter-to sub-Salpeter IMF slope. Whether there is causal relation for the correlation of central \met\space and IMF slope is unclear.  At r\textgreater 0.5\re\space the variation of possible IMF slopes is comparable to the measurement uncertainties, and therefore consistent with no IMF variations.

\subsection{Global $\sigma$ Correlations  and Possible Influence of Galaxy Properties}

Several studies have found global correlations of galaxy properties such as \sigacht\space with stellar populations parameters $t$, \met, and elemental abundances \citep[for a review see][]{2019A&ARv..27....3M}. 
These relations are fundamental to our  understanding of galaxy formation and evolution, and an important test bench for galaxy evolution models. 

We found some strong correlations with \sigacht\space for our sample of seven galaxies that are known in the literature, while some literature correlations are only weak. 
However, the galaxies are non-uniform concerning their star formation activity, AGN activity, and environment. Could these influence and weaken our SP-$\sigma$ correlations? 
\cite{2010MNRAS.404.1775T} showed that star formation activity influences the SP-$\sigma$ relationships. For this reason, we excluded \obfive\space from our correlation analysis.
Most galaxies in our sample are passive, though 
\obic\space is a LINER.   \cite{2007ApJ...671..243G} found that LINER galaxies follow similar $\sigma$ correlations for \met\space and abundances as passive galaxies, which led us to include \obic.  
The  galaxies in our sample are located in different environments (isolated, galaxy groups, and clusters). \cite{2010MNRAS.404.1775T} found that the correlations of the global galaxy age (considering only old galaxies), \met, and   [$\alpha$/Fe] with $\sigma$ are not affected. On the other hand, \cite{2019MNRAS.489..608F}, who measured  radial stellar population gradients, found some  variations between cluster and field/group environments, but only at a 1-2$\sigma$ level.
We found that the isolated galaxy \obsevsev\space  has  overall flatter  radial gradients and lower \met. The flatter gradients may be  caused by lower galaxy minor merger rates  in low-density environments \citep{2010ApJ...718.1158L}, but it is not clear if the lower \met\space in the center of \obsevsev\space compared to other galaxies with similar \sigacht\space in our sample is related to the environment. 

The weaker global correlation of e.g. \met\space and [Mg/Fe]  to \sigacht\space in our study compared to the literature may simply be caused by intrinsic scatter among the galaxies in our study. We summarise the properties of the individual galaxies in Appendix \ref{sec:indiv}. We found  that variations between the individual galaxies  of our small sample exceed global correlations. Also, elemental abundances cover a wider range with increasing radius compared to the center, as some galaxies have steep, others have shallow elemental abundance gradients. 
This emphasises the complexity of possible evolutionary paths of different early-type galaxies.

\section {Summary}
In this paper we analyzed long-slit spectroscopic data in the wavelength range 4000-8600\,\AA\space for eight early-type galaxies (\obone, \obthree, \obm, \obfour, \obfive, \obsev, \obsevsev, \obic). We used full spectral fitting and the single stellar population (SSP) models of \cite{2018ApJ...854..139C} to measure the following SSP  parameters: age $t$, metallicity \met, low-mass IMF slope in the mass range 0.08-1\,\msun, and the elemental abundances [O/Fe], [Mg/Fe], [Si/Fe], [Ca/Fe], [Ti/Fe], [C/Fe], [N/Fe], [Na/Fe],  [Fe/H]. Our measurements as a function of radius extend to $\sim$1\,\re, and we fit  radial gradients \nablar\space of each stellar population parameter as a function of \logten ($r$/\re).  We found a wide range of properties, with ages ranging from 3-13\,Gyr. All galaxies have radially decreasing \met. Some  galaxies have a  radially constant, Salpeter-like  IMF, and other galaxies have a super-Salpeter IMF in the center, decreasing to sub-Salpeter IMF at $\sim$0.5\,\re.

To understand how various SSP parameters depend on each other, we calculated the Spearman rank correlation coefficient $\rho$ of various parameters. Our data set allows us to test whether there are both  global and local correlations of stellar population parameters with each other, or other galaxy properties.

We found high global  correlations among central stellar population parameters (e.g. \met\space correlates  with the IMF slope;  [Si/Fe] with [Na/Fe]), radial gradients  (e.g. \nablar [Fe/H] anti-correlates with \nablar [C/Fe]; \nablar [N/Fe] correlates with \nablar [Na/Fe]), and of central values with gradients (e.g. \met\space anti-correlates with both  \nablar \met\space and \nablar IMF). 
Some correlations may, to some extent, be strengthened by covariances of the SP parameters in our spectral fitting. 
We also calculated the Spearman rank coefficient,  $\rho$, of our SSP parameters at 0.1\,\re\space with the central stellar velocity dispersion \sigacht\space for seven galaxies (excluding \obfive, which may have biased SSP parameters), and performed a linear fit. \sigacht\space is a proxy for the galaxy's central gravitational potential. Several of our linear relations are in  good agreement with   relations from the literature, though some parameters have a large scatter and hence only weak to moderate correlations. 
Our sample size is small (seven galaxies when excluding \obfive); therefore, our global correlations  require confirmation with a larger number of galaxies.

We tested if there are local correlations for each individual galaxy. The most common  is a very strong anti-correlation of \met\space with the galaxy's radius (i.e. negative \nablar\met); it holds for all galaxies. Further, there are strong positive correlations (i) of \met\space and some elemental abundances (Ca, Ti, Na) with local $\sigma$ for at least five  galaxies, (ii) of \met\space with several elemental abundances (e.g. Ca, Si, Ti, Na), and (iii) of several elemental abundances with each other (e.g. Ca with Si, Ti, Na).  Correlations of various elemental abundances can be expected if they are produced by the same process, e.g. $\alpha$ elements produced in explosions of massive stars. We caution that in a few cases (e.g. O with C), the parameters  also have correlated probability distribution functions, which may to some extent cause the high local correlations. 
While these correlations are interesting, they do not necessarily indicate a causal relation of parameters.


\acknowledgments

We would like to thank Barry Madore and the Las Campanas
staff who helped us to obtain the data, Sam Vaughan for sharing
the code \pystaff, and Charlie Conroy for sharing the SSP
models. We also thank the referee for very constructive comments
and suggestions.
We acknowledge the usage of the HyperLeda database (http://leda.univ-lyon1.fr, \citealt{2014A&A...570A..13M}).
This paper includes data gathered with the 6.5 meter Magellan Telescopes located at Las Campanas Observatory, Chile.
This research has made use of the NASA/IPAC Extragalactic Database (NED) which is operated by the Jet Propulsion Laboratory, California Institute of Technology, under contract with the National Aeronautics and Space Administration.
This research has made use of the SIMBAD database, operated at CDS, Strasbourg, France.
This research has made use of NASA's Astrophysics Data System Bibliographic Services.
This research made use of ds9, a tool for data visualization supported by the Chandra X-ray Science Center (CXC) and the High Energy Astrophysics Science Archive Center (HEASARC) with support from the JWST Mission office at the Space Telescope Science Institute for 3D visualization.

\facilities{Magellan (IMACS)}

\software{         EMCEE \citep{2013PASP..125..306F}, 
            L.A.Cosmic \citep{2012ascl.soft07005V}, 
            MOLECFIT \citep{2015A&A...576A..78K,2015A&A...576A..77S},
            pPXF \citep{2017MNRAS.466..798C}, 
            PYSTAFF \citep{2018MNRAS.479.2443V},
            Matplotlib \citep{2007CSE.....9...90H},
            NumPy \citep{2011CSE....13b..22V}, 
            IRAF \citep{1986SPIE..627..733T,1993ASPC...52..173T}, 
            corner \citep{2016JOSS....1...24F}
          }

\newpage

\appendix

\section{Correction for Galaxy Rotation}
\label{sec:rotcor}
Several galaxies of our sample have rotational velocities of $\sim$100\,\kms\space or varying rotation profiles \citep{1989ApJ...344..613F,1996A&A...306..363D,2000A&AS..145...71K,2004MNRAS.352..721E}, 
which we accounted for in the following ways: 
for \obone, \obfour, \obfive, and \obm\space we extracted spectra on both sides of the slit separately, measured the radial velocity shift  with \textsc{pPXF} \citep{2017MNRAS.466..798C}, and corrected the radial velocities before summing the two  spectra with the same distance to the galaxy center and from  the two sides of the slit. In this step we also fit Balmer gas emission flux and subtracted it. We used a different approach  for \obic, which has a counter-rotating stellar core. This galaxy has strong gas emission lines, but the gas emission does not co-rotate with the stars and is asymmetric about the nucleus \citep{2002ApJ...578..787C}. We separately extracted and analyzed the northern and southern spectra out to 0.5\,\re, to improve the emission line subtraction. For the outer radii \textgreater0.5\,\re, we combined the two regions, to ensure sufficiently high S/N. The central rotation is $\sim$100\,\kms. The rotation sense changes between 0.1 and 0.19\,\re, and  at 0.375\,\re, the rotation is $\sim$60\,\kms. Ignoring the rotation in the outer radial bins means that we will overestimate the velocity dispersion, as we actually measure the root-mean-square velocity, \vrms$=\sqrt{V^2+\sigma^2}$. This has only a minor effect on $\sigma$ ($\sim$5\,\kms), but we later correct for the rotation assuming a constant rotation velocity of 50\,\kms\space at r\textgreater0.5\,\re\space \citep{1988ApJ...327L..55F}.  

For the remaining galaxies (\obsev, \obsevsev, \obthree) we did not correct the stellar rotation of the spectra. We tested if this is acceptable by  extracting spectra on both the northern and southern side of the slit and cross-correlating them. We obtained only small shifts,   usually $\lesssim$ 0.5 pixel; that is,  \textless 10\,\kms\space for \obsev\space (see  \citealt{1989ApJ...344..613F} for rotation curve), \obsevsev\space (see  \citealt{1994A&A...292..381B}), and \obthree\space at our PA of observations. This means that, technically, for these galaxies our measurement of $\sigma$ is actually the root-mean-square velocity, \vrms. The rotation velocity of these galaxies is comparably low  and the profiles rather flat, and we simply correct \vrms\space to $\sigma$ by subtracting in quadrature the literature values for the rotation at the axis closest to our observations; that is  40\,\kms\space for \obsev\space  \citep{1989ApJ...344..613F,1993ApJ...403..567S}, 30\,\kms\space  for \obsevsev\space \citep{1994A&A...292..381B}, and $\sim$0\,\kms\space for \obthree\space \citep{1988MNRAS.235..813C}.

To summarise, we corrected the galaxy rotation where necessary.  We neglected rotation for galaxies with low and constant rotation velocity, allowing us   to increase the S/N of the spectra.

\begin{figure*}
\gridline{\fig{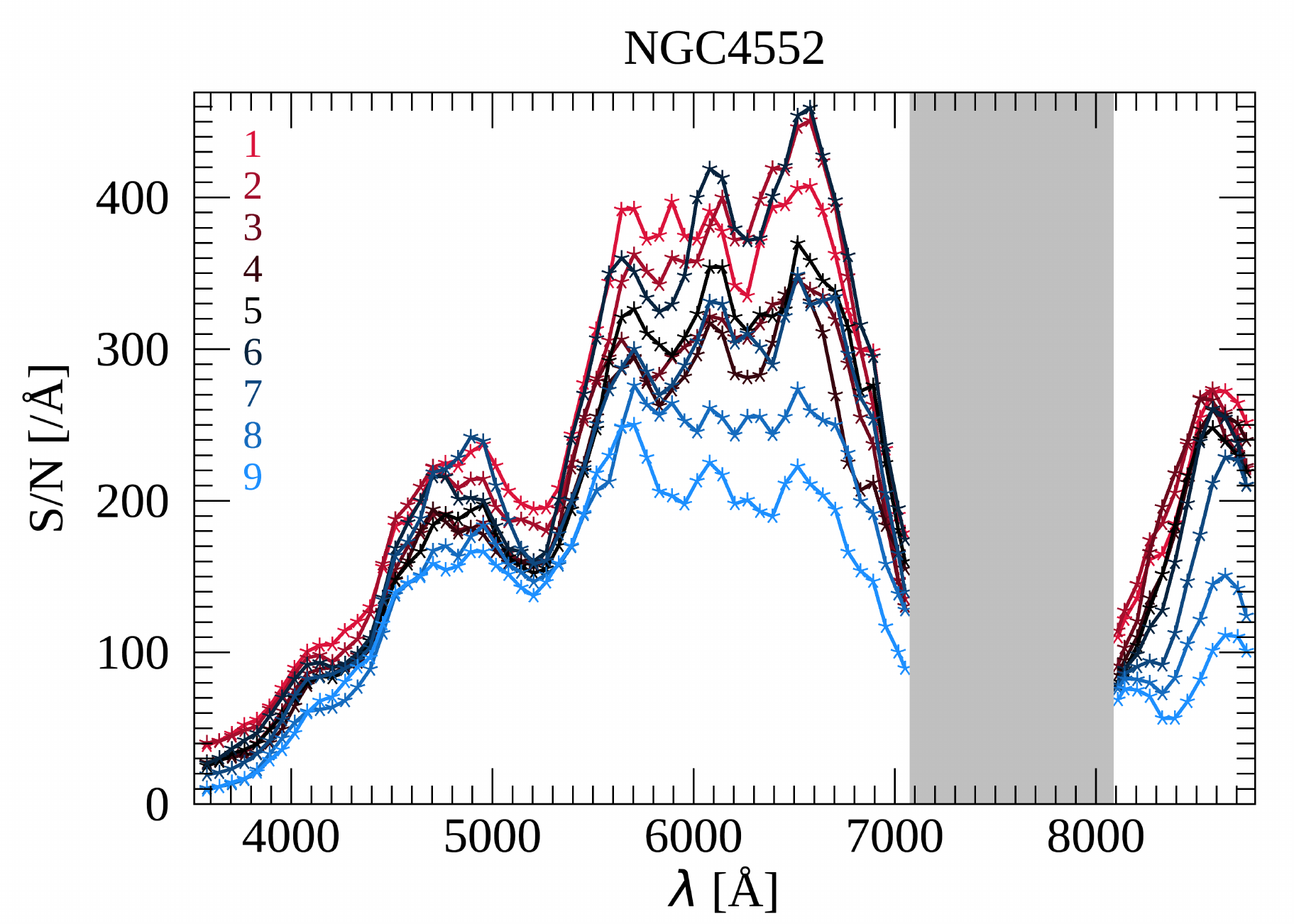}{0.3\textwidth}{(a)}
          \fig{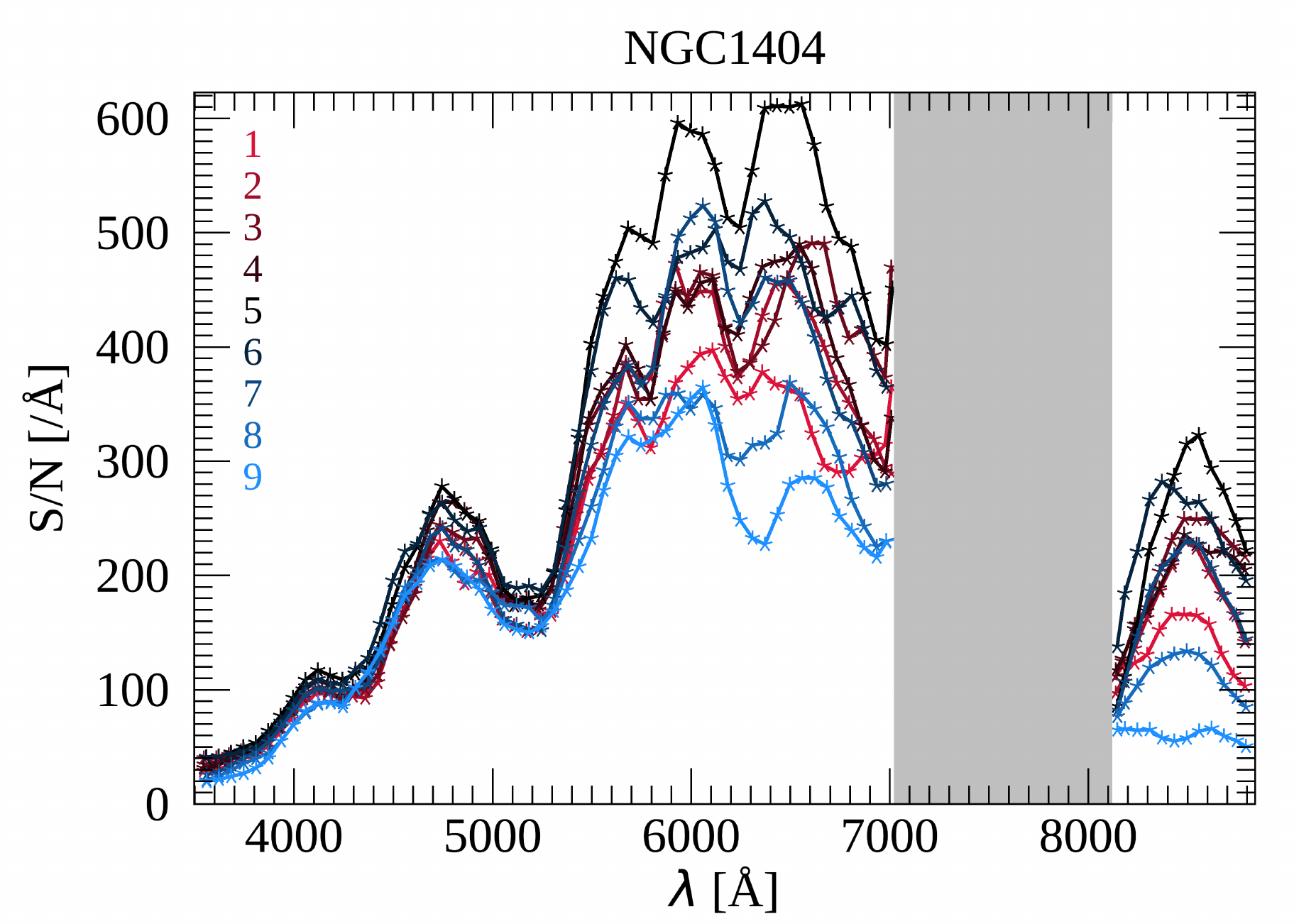}{0.3\textwidth}{(b)}
            \fig{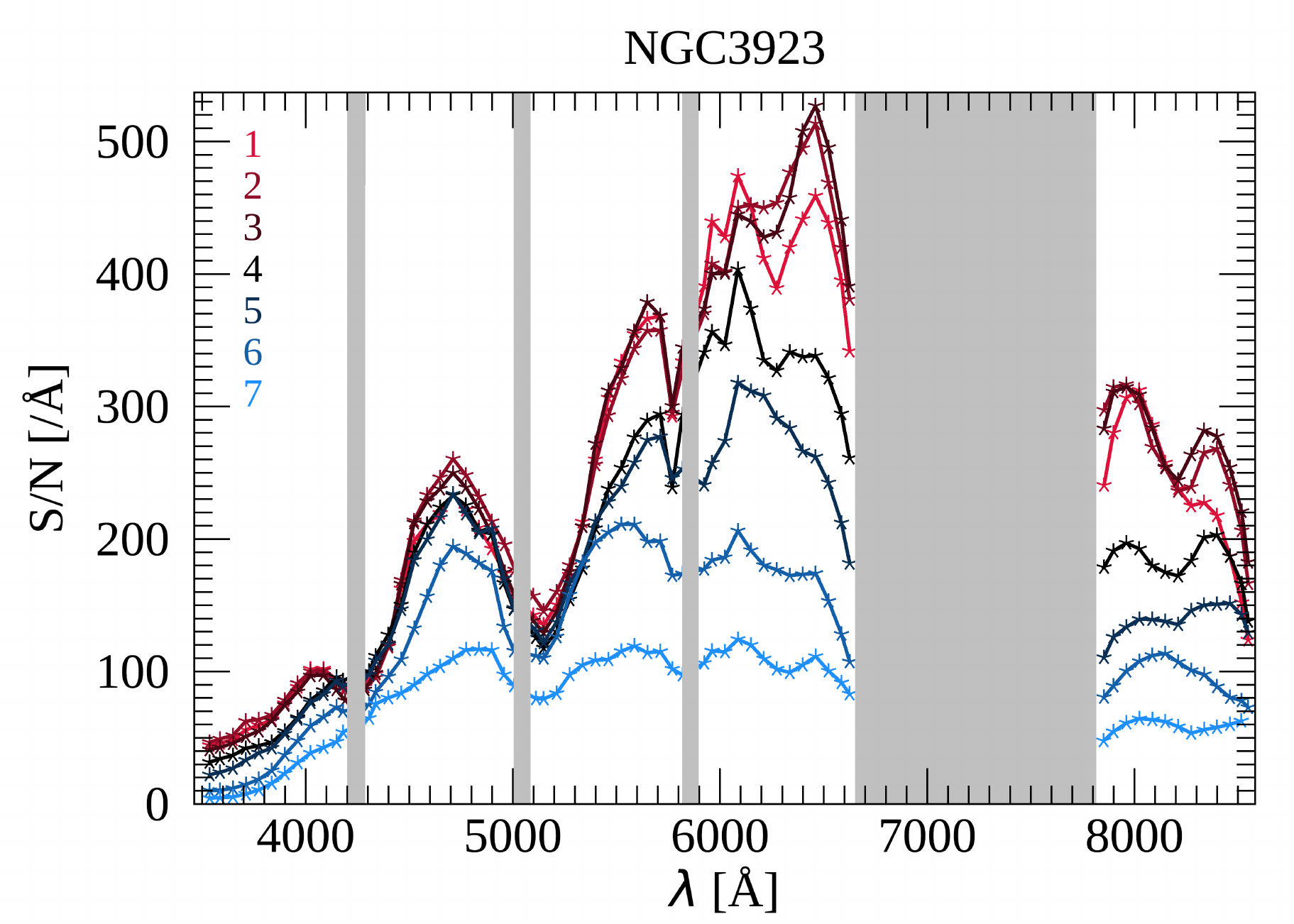}{0.3\textwidth}{(c)}
          }
\gridline{        \fig{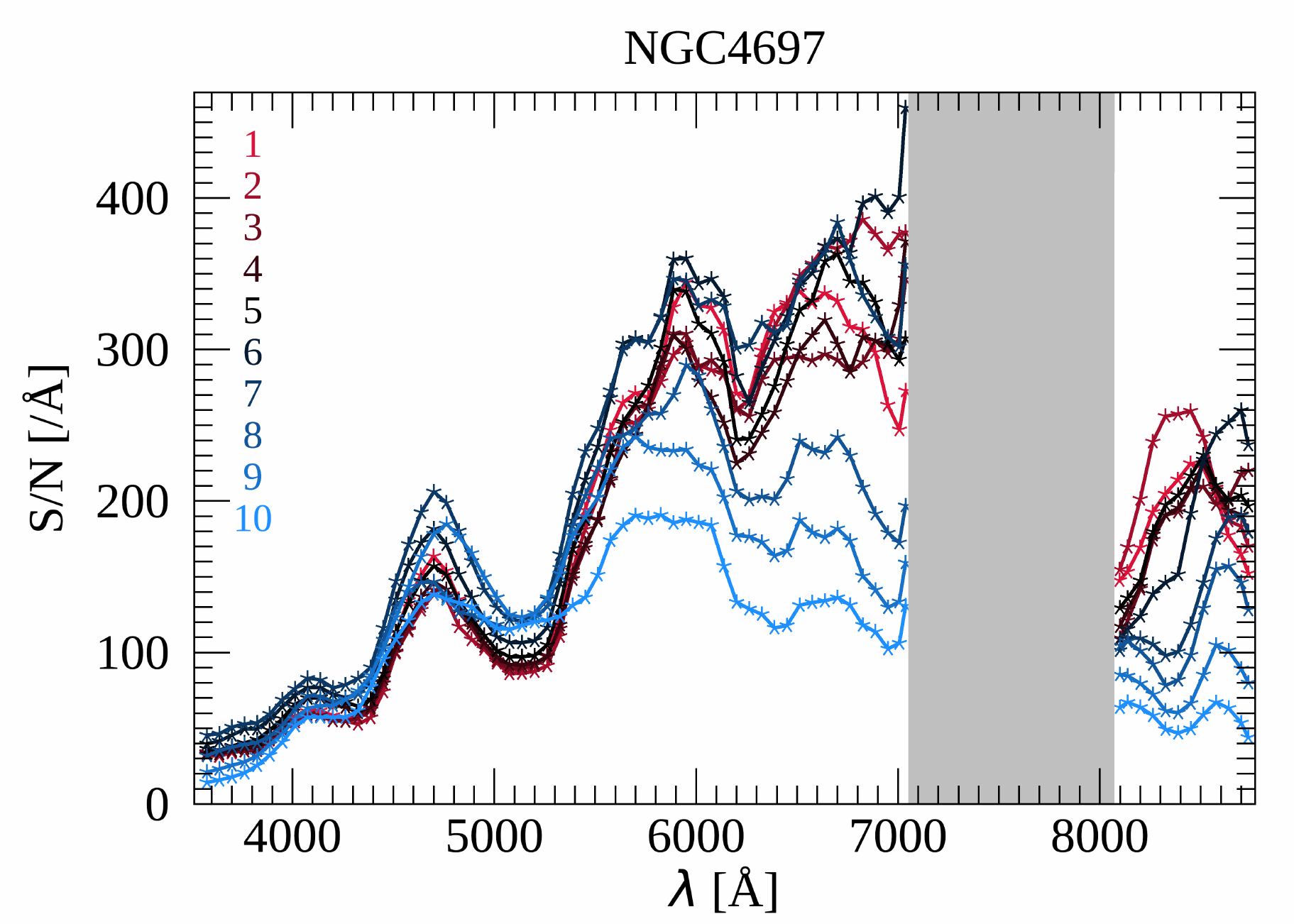}{0.33\textwidth}{(d)}
        \fig{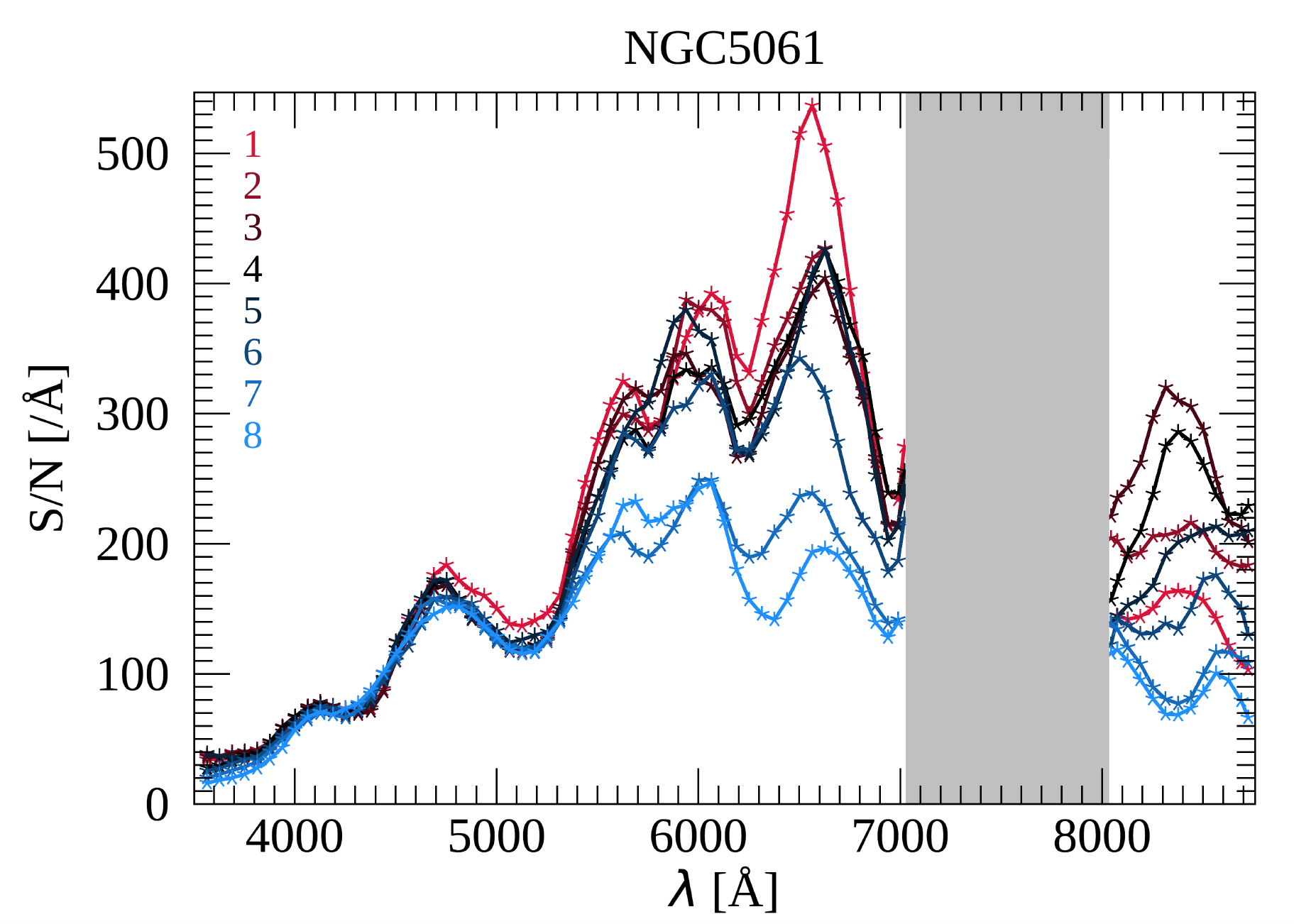}{0.33\textwidth}{(e)}
       	\fig{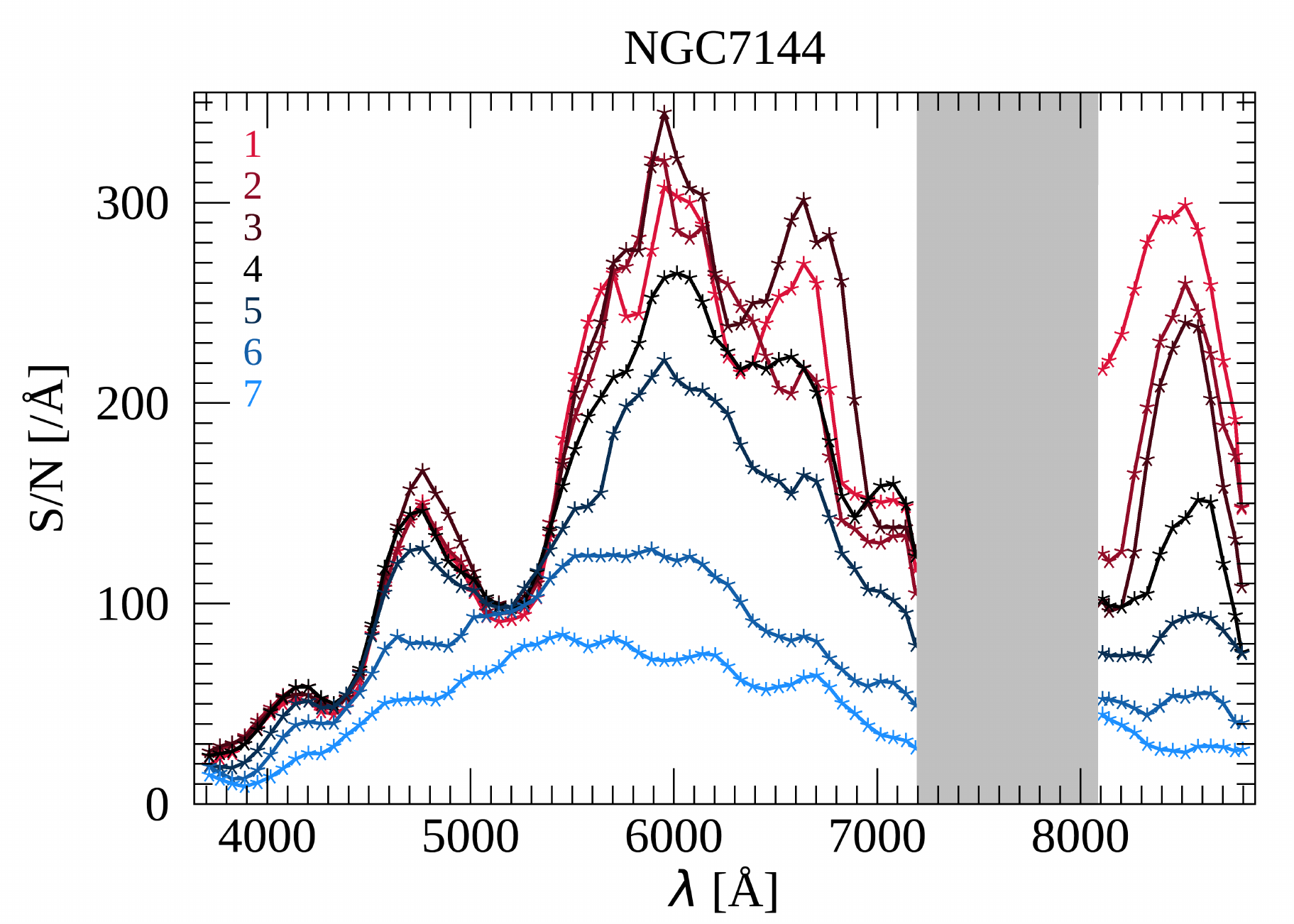}{0.33\textwidth}{(f)}}
\gridline{\fig{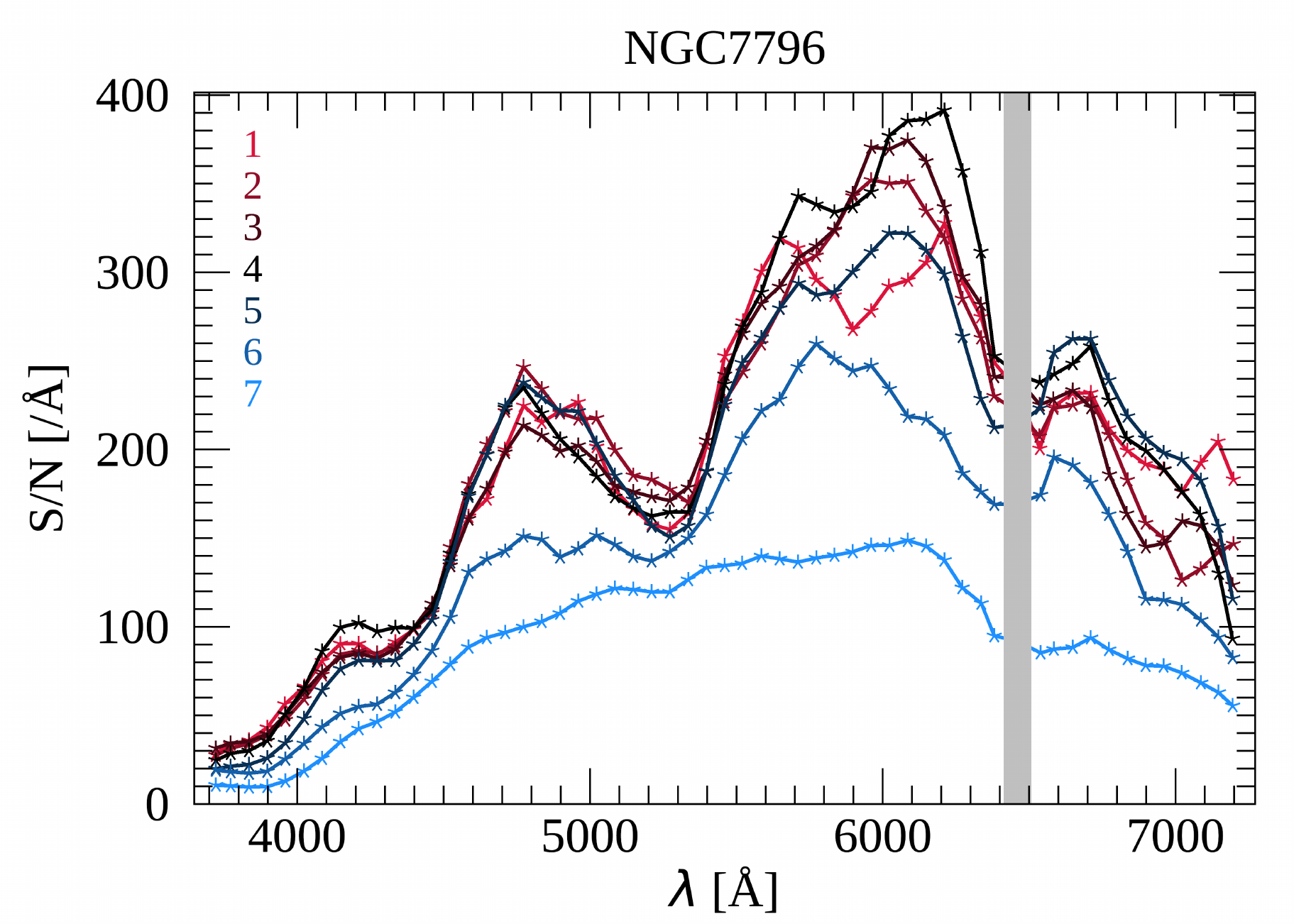}{0.33\textwidth}{(g)}
\fig{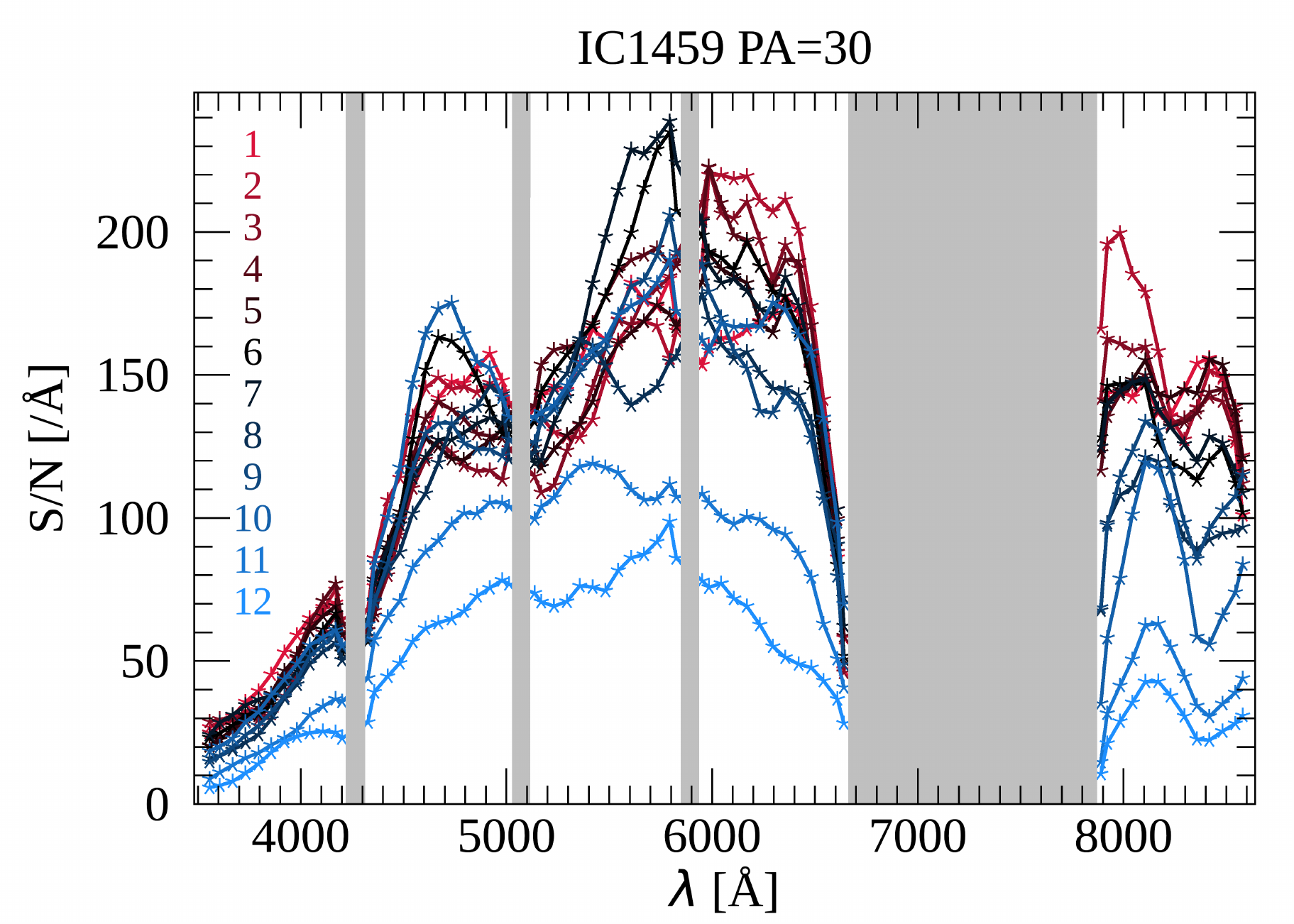}{0.33\textwidth}{(h)}
          \fig{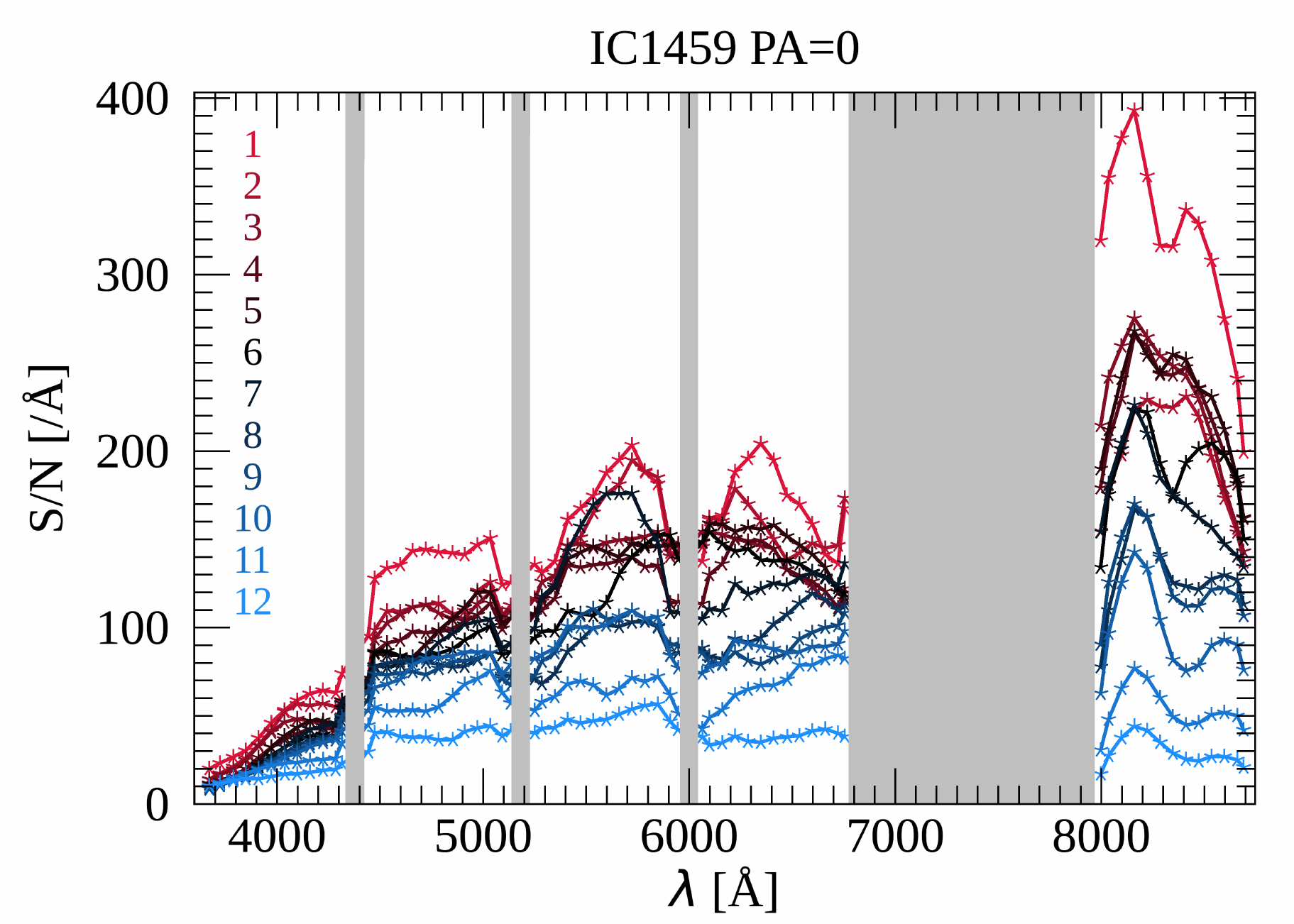}{0.33\textwidth}{(i)}
          }
\caption{Signal-to-noise ratio as function of wavelength for each galaxy. The different colors denote central spectra (red) to spectra at large radii (blue). 
Note the different ranges of the wavelength and S/N axes for different galaxies. Spectral regions without coverage are shown as gray shaded regions. 
\label{fig:snr}}
\end{figure*}

\section{Data Quality}
\label{sec:snr}
To estimate the S/N of our data, we used \textsc{DER\_SNR} \citep{2007STECF..42....4S}, on 100 pixel wide segments of  unconvolved and rebinned spectra, with steps of 50 pixels.  \textsc{DER\_SNR} is a robust algorithm that uses only the spectrum flux, and assumes that the noise is normally distributed and uncorrelated two pixels apart. While this assumption may not be strictly true for our  rebinned spectra (see Section \ref{sec:dr}), we can still learn how the S/N varies from galaxy to galaxy, for different radius  bins  for a single galaxy, and as a function of wavelength. We show the S/N in Figure \ref{fig:snr}. 

For all galaxies, we find an increase of S/N with wavelength from \textless4000\,\AA\space to $\sim$6000\,\AA, in some cases followed by a decrease to 7000\,\AA. The CaT region has usually similar or lower values of S/N, due to the lower exposure times, with the exception of \obic\space with PA=0\degr.  
All  galaxies have similar S/N, usually in the range of 50 to 300/\AA.

\section{Single Stellar Population (SSP) Fitting}
\subsection{Treatment of Gas Emission Lines and Systematic Uncertainties}
 \label{sec:gasfit}
\begin{figure*}
\gridline{\fig{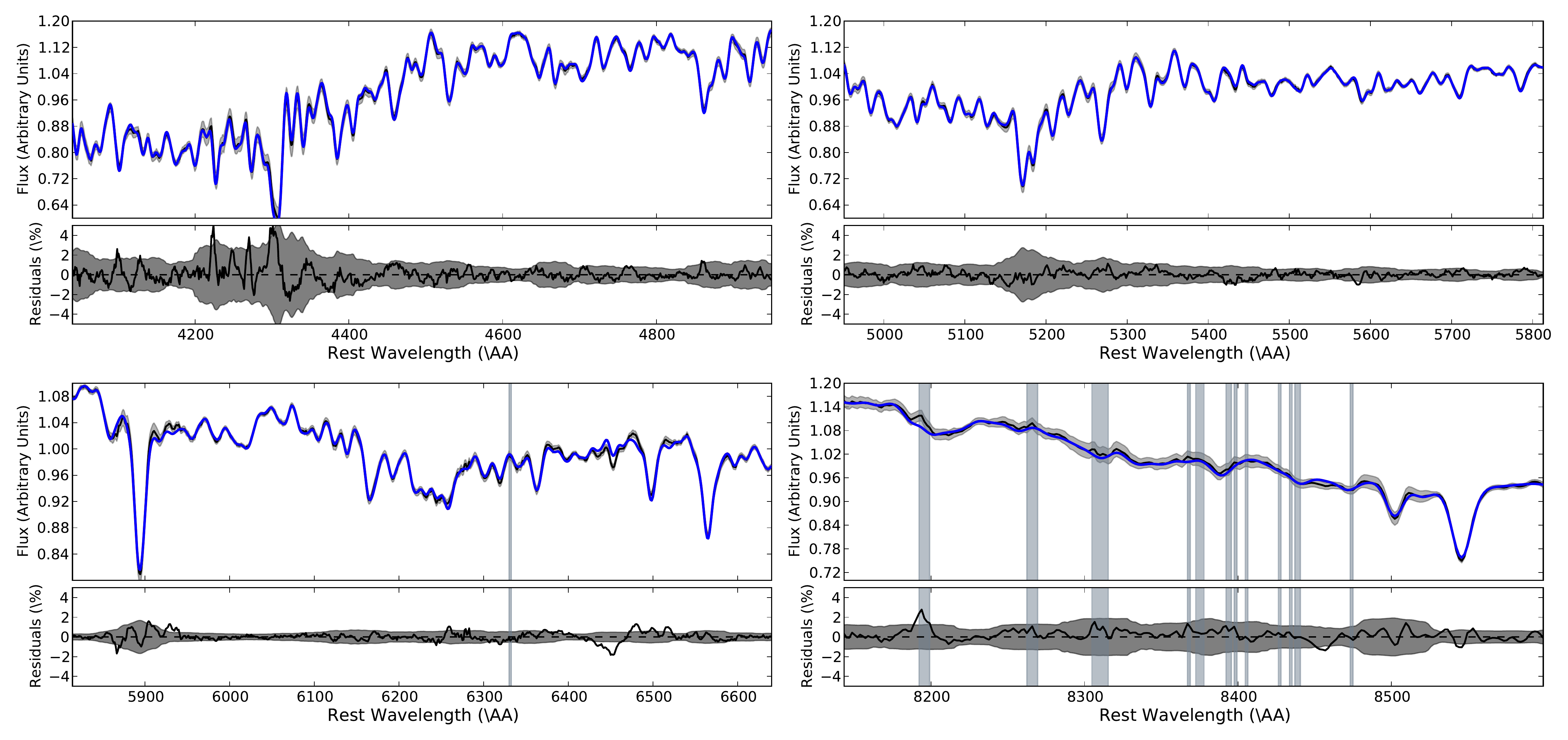}{0.75\textwidth}{(a) \pystaff\space  fit of \obfour, $r$=0.15\,\re}}
\gridline{\fig{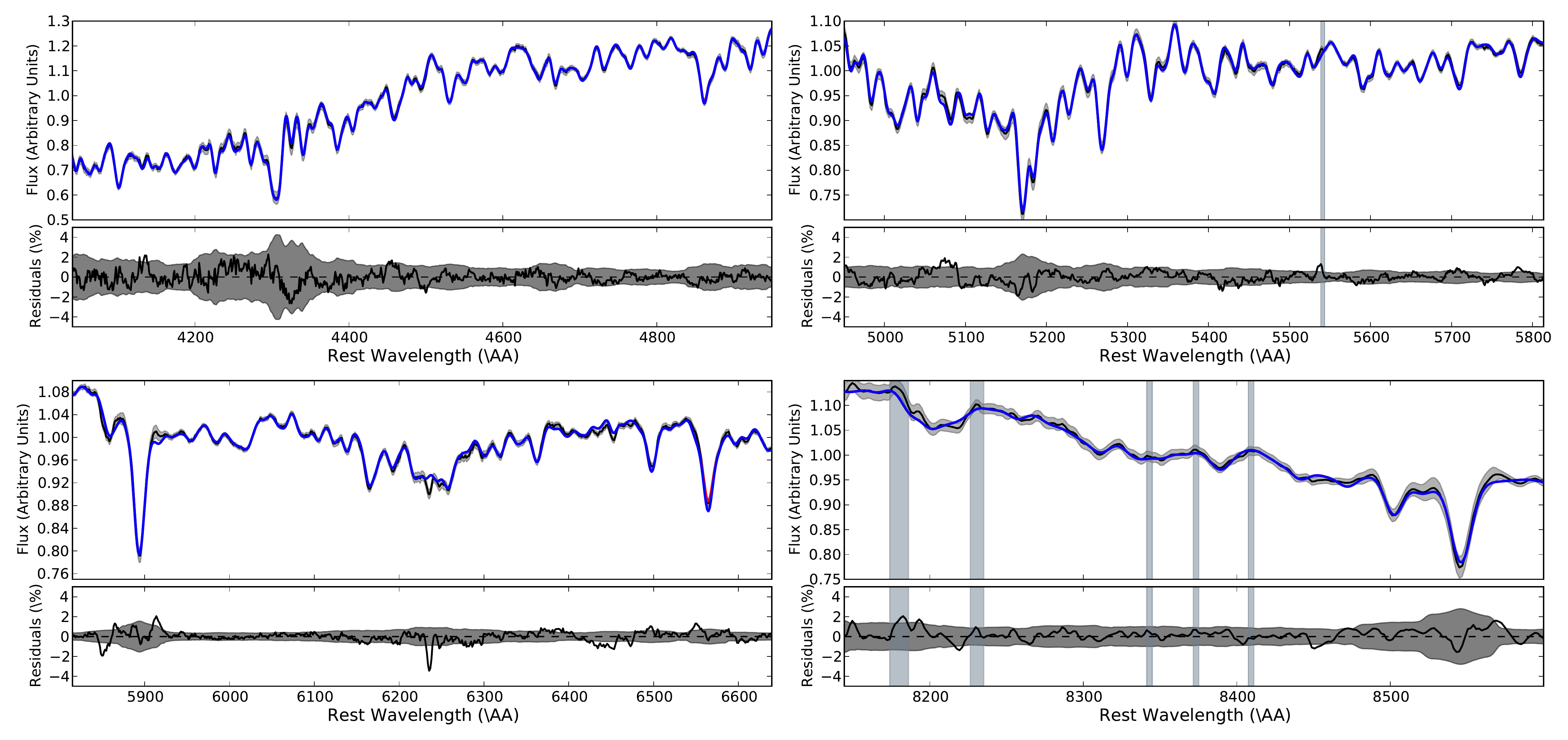}{0.75\textwidth}{(b) \pystaff\space  fit of \obfive, $r$=0.13\,\re}}
\gridline{\fig{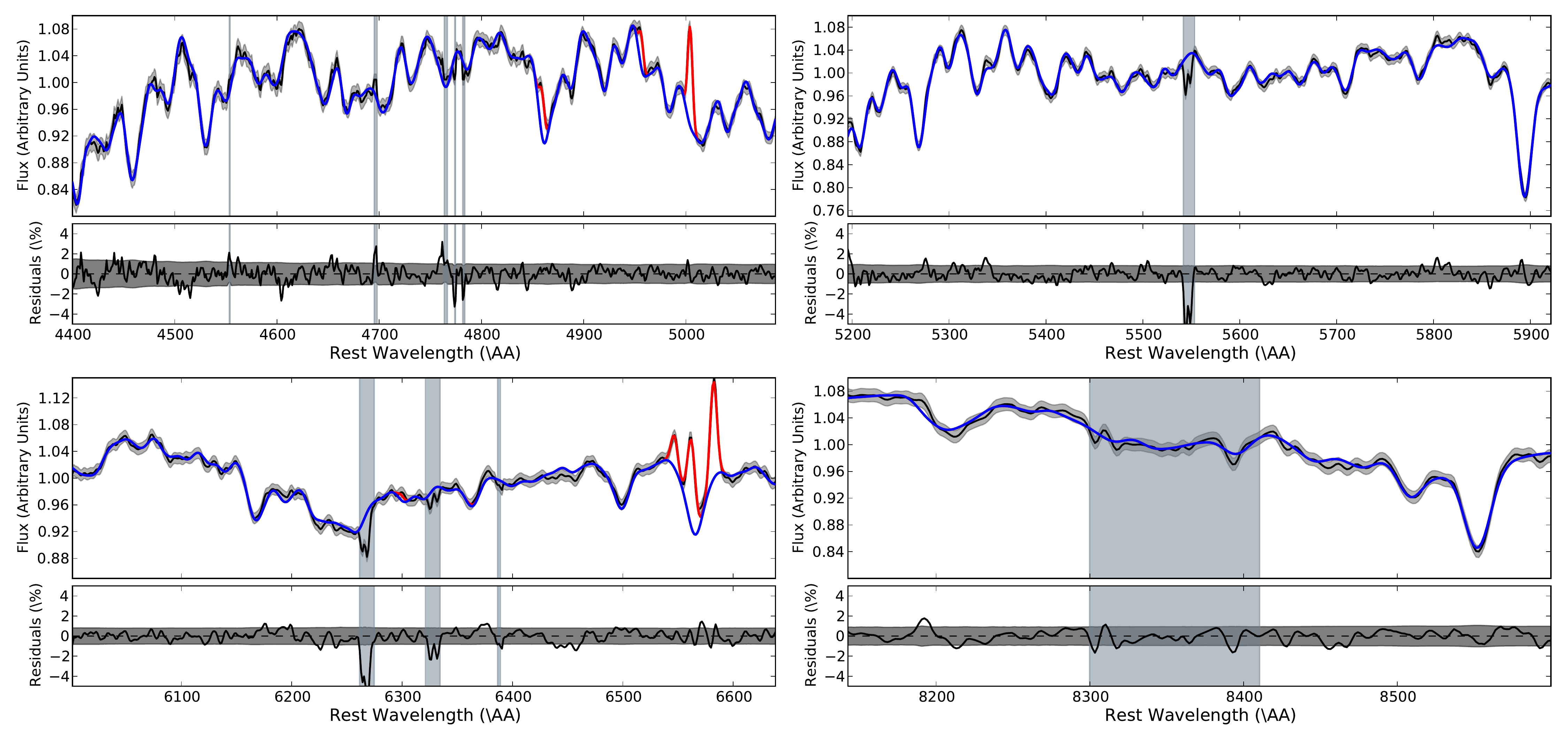}{0.75\textwidth}{(c) \pystaff\space  fit of \obic, $r$=0.1875\,\re\space south}}
 \caption{Examples of spectra (black), the \pystaff\space  best-fit stellar population models (blue),  and gas emission (red). The gray color along the spectrum illustrates the noise level; vertical grey regions are excluded from the fit due to bad pixels,  sky,  telluric, or fringe residuals.  
The plots also show the fit residuals in percent as a function of wavelength. The examples are  \obfour\space at $r$=0.15\,\re, \obfive\space at $r$=0.13\,\re, and \obic\space at $r$=0.1875\,\re\space south and PA=0.
\label{fig:pyfit}}
\end{figure*}

\begin{deluxetable*}{l|l|l}
\tablecaption{Summary of  \pystaff\space Fit Settings\label{tab:pystaff}}
\tablecolumns{3}
\tablewidth{0pt}
\tablehead{
\colhead{Galaxy} &
\colhead{Approximate Rest-wavelength Fitting Range} &
\colhead{Emission Lines} \\
\nocolhead{}&\colhead{(\AA)}&\nocolhead{}
}
\startdata
 \obm&4230--4950, 4950--5814, 5814--6455, 8143--8600&H$\gamma$, H$\beta$\\
 (M~89)&4040--4950, 4950--5814, 5814--6640, 8143--8600&H$\delta$, H$\gamma$, H$\beta$, H$\alpha$ \\
\obone &  4230--4950, 4950--5814, 5814--6455, 8143--8600&H$\gamma$, H$\beta$\\
&4040--4950, 4950--5814, 5814--6640, 8143--8600 
& H$\delta$, H$\gamma$, H$\beta$, H$\alpha$\\
 \obthree\tablenotemark{b,d}& 4285--4893, 	5144--5813, 5885--6417, 8164--8524& H$\gamma$, H$\beta$ \\
 & 4040--5030, 5090--5814, 5885--6640, 8143--8524&H$\delta$, H$\gamma$, H$\beta$, H$\alpha$  \\
\obfour &  4230--4950, 4950--5814, 5814--6455, 8143--8600&H$\gamma$, H$\beta$\\
&4040--4950, 4950--5814, 5814--6640, 8143--8600
& H$\delta$, H$\gamma$, H$\beta$, H$\alpha$\\
\obfive & 4230--4950, 4950--5814, 5814--6455, 8143--8600 &H$\gamma$, H$\beta$\\
& 4040--4950, 4950--5814, 5814--6640, 8143--8600
& H$\delta$, H$\gamma$, H$\beta$, H$\alpha$\\
\obsev & 4230--4950, 4950--5814, 5814--6455, 8143--8600 &H$\gamma$, H$\beta$\\
& 4040--4950, 4950--5814, 5814--6640, 8143--8600  
& H$\delta$, H$\gamma$, H$\beta$, H$\alpha$\\
\obsevsev\tablenotemark{a} & 4230--4685, 4685--5238, 5238--5802, 5802--6455 & H$\gamma$, H$\beta$ \\
& 4040--4685, 4685--5238, 5238--5802, 5802--6640
& H$\delta$, H$\gamma$, H$\beta$, H$\alpha$\\
 \obic\space (PA=0\degr)\tablenotemark{c,d}& 4400--5088, 5195--5921, 6001--6455, 8143--8600 &H$\beta$, [OI], [OIII], [NII]\\
& 4400--5088, 5195--5921, 6001--6640, 8143--8600 &H$\beta$, H$\alpha$, [O\,I], [O\,III], [N\,II]\\
\obic\space (PA=30\degr)\tablenotemark{b,c,d}& 4276--4992, 5091--5813, 5906--6455, 8144--8580 &H$\gamma$, H$\beta$, [O\,I], [O\,III], [N\,II], [N\,I] \\
& 4040--4992, 5091--5813, 5906--6640, 8144--8580 &H$\delta$, H$\gamma$, H$\beta$, H$\alpha$, [O\,I], [O\,III], [N\,II], [N\,I] \\
 \hline
\enddata
\tablecomments{a: no CaT, b: shorter CaT, c: shorter optical range, d: wavelength gaps}
\end{deluxetable*}
All SSP \textsc{PyStaff} fits included Balmer gas emission lines. We tied the individual Balmer lines to the same kinematics and fixed the relative fluxes of the Balmer lines (H$\delta$=0.259\,H$\beta$, H$\gamma$=0.468\,H$\beta$, H$\alpha$= 2.86\,H$\beta$, \citealt{1989agna.book.....O,1997MNRAS.291..403R}). We constrained the gas kinematics to be close to the stellar kinematics ($\pm$30\,\kms)\footnote{The exception is \obic, as the gas and stellar kinematics  are decoupled for this galaxy \citep{2002ApJ...578..787C}.}, to ensure that we obtain a reasonable emission line flux estimate. An overestimation of the Balmer emission line flux biases our age result to younger ages, and an underestimated gas emission to older ages.

For \obic\space it was necessary to fit further emission line flux doublets: [N\,II] ([N\,II]$_{\lambda{6548.03}}$=0.33 [N\,II]$_{\lambda{6583.41}}$), 
[O\,III] 
([O\,III]$_{\lambda{4958.92}}$ = 0.33 [O\,III]$_{\lambda{5006.84}}$), 
[O\,I] ([O\,I]$_{\lambda{6363.67}}$ = 0.33 [O\,I]$_{\lambda{6300.30}}$), 
and  [N\,I] ([N\,I]$_{\lambda{5200.39}}$ = 0.7 [N\,I]$_{\lambda{5197.90}}$). We fit the Balmer gas kinematics separately from the other emission line doublets. The results between the kinematics of  Balmer  and the other emission lines differ by usually no more than $\sim$50\,\kms.  
 We show three examples of the data and best-fit model spectra in Figure \ref{fig:pyfit}. The gas emission is  prominent in the spectrum of \obic\space  (bottom panel).

We fit each spectrum using two different wavelength regions, as listed in Table~\ref{tab:pystaff}. The wavelength ranges are split into four sections, and in each section we fit  a Legendre polynomial to correct the continuum shape.  The polynomial order is the nearest integer to ($\lambda_{\text{upper}}-\lambda_{\text{lower}}$)/100.  We also list the emission lines that were included for each fit. 
In most cases, we have one set of fits with two Balmer lines, H$\gamma$ and H$\beta$, and a second set with longer wavelengths, that includes in addition H$\delta$ and H$\alpha$. Usually the results from the two fits are in agreement within 1$\sigma$. However, due to the different wavelength regions and Balmer lines, there can be differences in the results for the Balmer line flux, and as a consequence, the stellar population age $t$.  We find that the shorter wavelength fits result in a larger gradient of the Balmer line flux and $t$ compared to the longer wavelength fits that include H$\delta$ and H$\alpha$. For each galaxy spectrum, we computed the error-weighted mean obtained by  the two fits and used it as final results. We used the standard deviation of the two fits as the systematic uncertainty, added in quadrature   to the mean statistical uncertainty of the two fits. The final  results are shown in Figure \ref{fig:radgrad} and listed in Table \ref{tab:pyresult}.

One exception is \obic, for which we have two sets of observations that we fit separately, and for each of them we fit the spectra north and south of the galaxy center out to 0.5\,\re\space separately. The stellar population parameter trends of these sets are mostly similar. There are some differences, which are probably caused by the slightly different wavelength regions of the two sets. For example, the NaD feature lies in a chip gap in   the PA=30\degr\space spectra, leading to larger statistical uncertainties and a higher [Na/H] estimate. We chose a  narrower fitting range of  [Na/H]  for the PA=30\degr\space spectra based on the PA=0\degr\space results. On the other hand, the Mg\textit{b} feature is missing in the PA=0\degr\space spectra. As this causes large  differences for   [Mg/H] and influences some other elemental abundances, we narrowed the  fitting range  of [Mg/H]  for the PA=0\degr\space fits to values close to the PA=30\degr\space results. Again, we computed the error-weighted mean of each stellar parameter at a given radius, combining the results obtained from  four different spectra and  two different fit settings. In the outer  bins, where the S/N of the PA=0\degr\space spectra is significantly lower than the PA=30\degr\space spectra due to the smaller slit width, we use only the results obtained from the PA=30\degr\space data.
\subsection{Extrapolation of Elemental Response Functions}
\label{sec:extrap}
We measured  elemental abundances  by applying linear multiplicative  response functions for each individual element \citep{2018MNRAS.475.1073V,2018MNRAS.479.2443V}, using  the response functions of  \cite{2018ApJ...854..139C}. These  are  computed at reference values of [X/H]=$\pm$0.3\,dex for most elements,  [C/H] = $\pm$0.15\,dex, and  [Na/H]=$\pm$0.3, +0.6, and +0.9\,dex.  \pystaff\space  applies a Taylor expansion near the reference value to obtain the response functions at a given value of [X/H]. In most of our spectral fits, the elemental abundance value does not exceed the reference value. However, for [O/H] and [C/H], we often obtain results that exceed the reference value. For [O/H], we reach values as high as  0.44\,dex, and for [C/H] 0.26\,dex. 

\begin{figure*}
\gridline{\fig{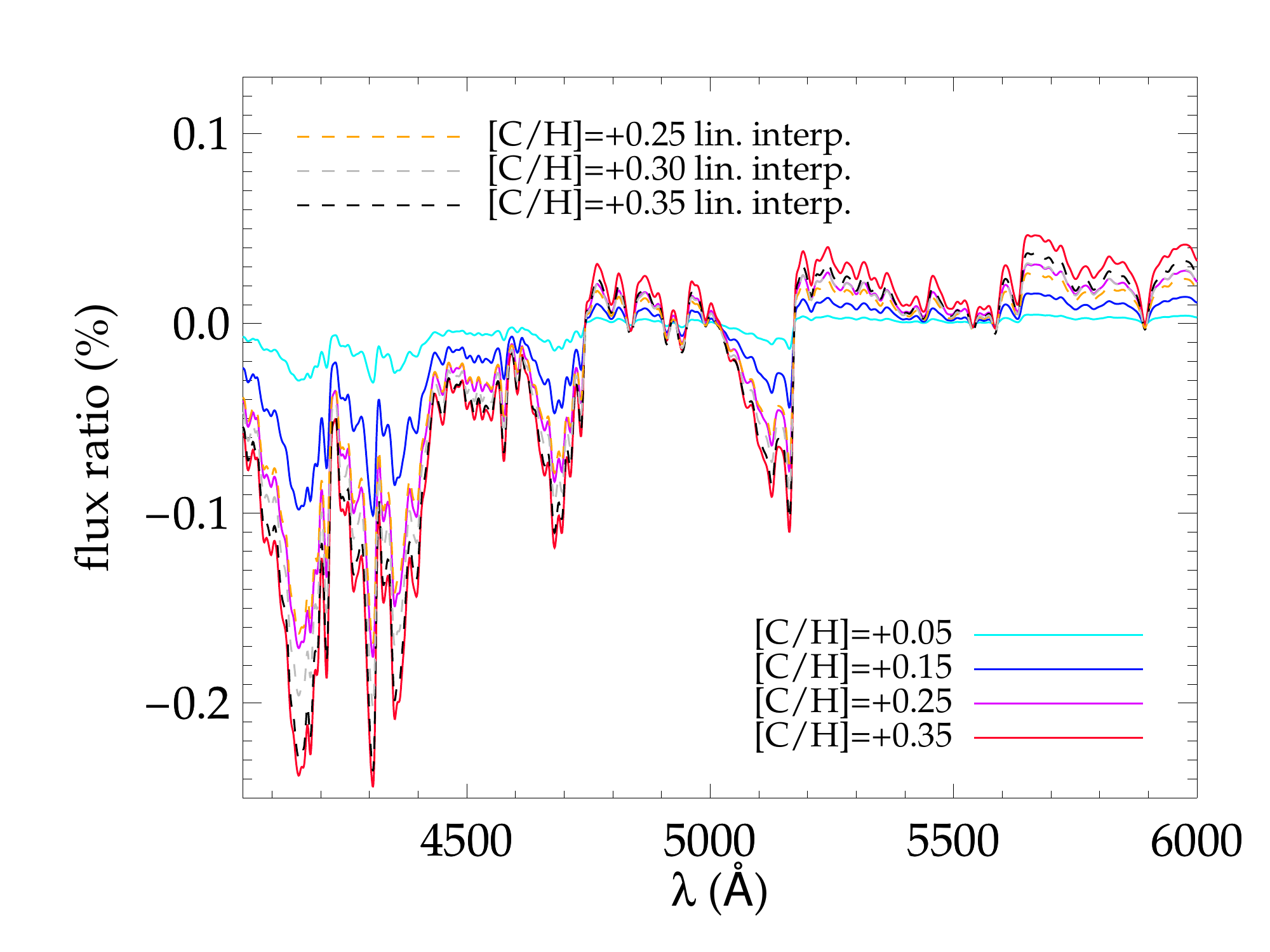}{0.5\textwidth}{}
          \fig{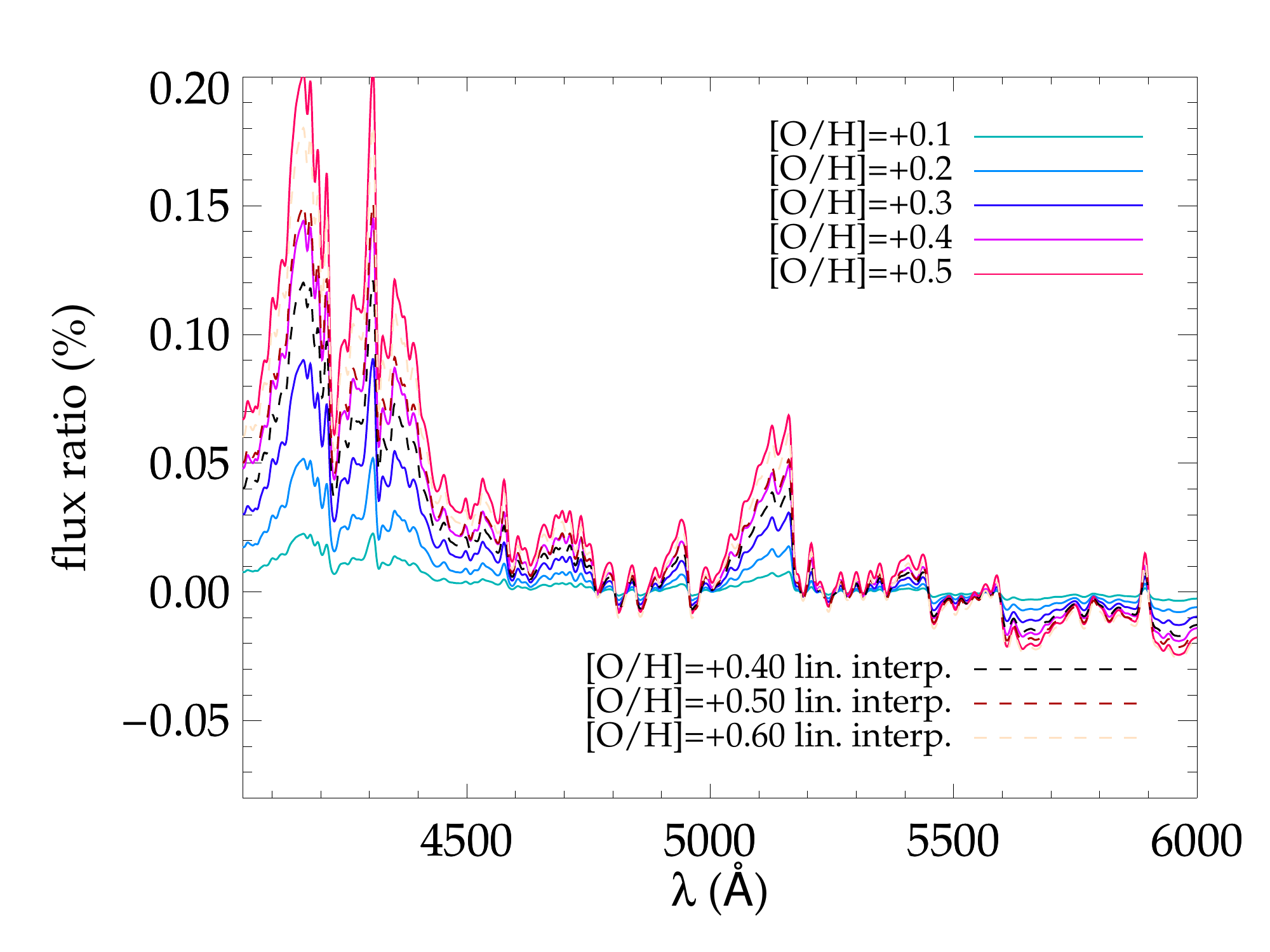}{0.5\textwidth}{}
          }
\caption{Flux ratio of [C/H] enhanced (left) and [O/H] enhanced (right) SSP models with respect to solar models. Solid lines denote the \textsc{PyStaff} Taylor expansion, dashed lines a simple linear extrapolation. 
\label{fig:extrap}}
\end{figure*}

 We compared the Taylor expanson to a linear extrapolation of the response function in   Figure \ref{fig:extrap}. For a SSP with age $t$=10\,Gyr, \met=0.2\,dex, $x_1$=$x_2$=2.3, [Na/H]=0.6\,dex, and otherwise [X/H]=0\,dex, we show the flux ratio of [C/H] (left panel) and [O/H] (right panel) enhanced SSP models to a  model with [C/H]=0\,dex and  [O/H]=0\,dex, respectively. Solid lines denote the \pystaff\space  Taylor expansion at different values, dashed lines denote a simple linear extrapolation from [C/H]=0 and [C/H]=0.15\,dex to the denoted values (0.25, 0.30, and 0.35\,dex). For carbon, we see that the linear extrapolation (orange and black dashed lines) has lower flux ratios than the  respective \pystaff\space  Taylor expansion (pink and red) at the same values. This means that the \pystaff\space extrapolated model deviates more from the solar-scaled model at a given value than a simple linear extrapolation. For a certain observed spectrum with non-solar abundance, the required \pystaff\space abundance value to match the spectrum will be lower. Our comparison with linearly interpolated models in 0.05\,dex steps (Figure \ref{fig:extrap}) suggests  that this effect is \textless0.05\,dex for [C/H].

For [O/H] (right panel) the difference is larger. A \textsc{PyStaff} model with [O/H]=0.5\,dex (solid red) deviates more from the [O/H]=0.0\,dex spectrum than a linearly extrapolated [O/H]=0.6\,dex spectrum (dashed orange). Our \textsc{PyStaff} extrapolated model at [O/H]=0.4\,dex is similar to the linearly interpolated [O/H]=0.5\,dex model. We conclude that our [O/H] measurements that exceed the reference value of [O/H]=0.3\,dex may be biased to lower values by up to 0.1\,dex compared to other methods of extrapolation.

\subsection{Stellar Population Results per Radial Bin and Galaxy}
We list our final SSP fit results in Table \ref{tab:pyresult}. The results were obtained by fitting the spectra in the wavelength ranges and with the emission lines listed in Table \ref{tab:pystaff}. We used two wavelength fitting ranges   for each spectrum and   list the error-weighted mean of the two settings, the standard deviation of the two settings is included in the listed  uncertainties.

 \subsection{Comparison to Other Studies}
 \label{sec:comparison}

\subsubsection{Ages and \met}
Our ranges of age $t$ and \met, and the respective gradient measurements, are within typical values of elliptical galaxies found in other studies, including those by
 \cite{2003A&A...407..423M,2007MNRAS.377..759S,2007A&A...463..455A,2010MNRAS.408...97K,2014A&A...562A..47G,2015MNRAS.449..328W,2020A&A...635A.129K}. 
 
Some studies  derived ages and \met\space for individual galaxies in our sample, though sometimes only for the center. We are in agreement to within the uncertainties
for the ages of \obone\space \citep{2000AJ....120..165T,2014ApJ...783..135A},  \obm\space \citep{2000AJ....119.1645T,2005ApJ...621..673T,2014ApJ...783..135A,2015MNRAS.447.1033M,2015MNRAS.448.3484M,2017ApJ...841...68V},  \obthree\space \citep{2014ApJ...783..135A,2017MNRAS.472.2889C},  \obfour\space \citep{2000AJ....119.1645T,2005ApJ...621..673T,2007A&A...463..455A,2010MNRAS.401L..29S,2015MNRAS.448.3484M},  \obfive\space \citep{2014ApJ...783..135A},  \obsevsev\space \citep{2005ApJ...621..673T,2018Ap&SS.363..131R}, and  \obic\space \citep{2014ApJ...783..135A}. 
However, some studies found younger ages by several Gyr than we did and we are not in agreement to within the uncertainties, in particular for \obthree\space \citep{2005ApJ...621..673T},   \obfour\space \citep{2014ApJ...783..135A},    \obm\space \citep{2007A&A...463..455A} and 
\obic\space \citep{2007A&A...463..455A,2010MNRAS.401L..29S,2019MNRAS.488.1679P}.

We evaluated the accuracy and precision of our \pystaff\space fits of \obthree\space   in \cite{2020ApJ...902...12F} using SSP model spectra to simulate realistic observations by adding noise in several realizations. Indeed, the ages tend to be slightly overestimated ($\sim$0.5\,Gyr). We conclude that our ages  may be biased to older ages, but this is only a  small effect. Our  ages agree with several studies in the literature to within their 1$\sigma$ uncertainties.

Most studies in the literature used different methods and SSP models, 
which are computed using different isochrones and have different approaches to deal with the metallicity  \met\space \citep[see review by][]{2013ARA&A..51..393C}.  One exception is  the work by \cite{2019MNRAS.488.1679P}, who used \pystaff\space and \cite{2018ApJ...854..139C} SSP models, as we did, to analyze MUSE ($\sim$ 4700--6800\,\AA) and KMOS data ($\sim$8100--9000, 9700--10100\,\AA) of \obic. They binned their data in three concentric rings and  found a slightly negative metallicity gradient, though it is consistent with being constant \met=0.15\,dex out to $\sim$0.2\,\re.  We analyzed five spectral bins in the same radial range as their three spectra, and found decreasing \met\space from 0.13 to $-0.09$\,dex (deviating from \citealt{2019MNRAS.488.1679P} by 1.0--3.2 $\sigma$), while our ages are higher by 4\,Gyr in the center and $\sim$2.5\,Gyr in the outer region (deviating by 2.7--1.5 $\sigma$). \cite{2019MNRAS.488.1679P} did not include the emission line doublet [N\,I]$_{\lambda\lambda 5197.90,5200.39}$ to their fit, and their Figure 7 shows residuals in the  spectra at 5200\,\AA, especially in the center. 
We tested fitting our central spectrum and neglecting [N\,I] emission. The  change of  most parameters was within 1 
$\sigma$; the age decreased by 1\,Gyr, \met\space decreased by 0.01\,dex. 
 Since [N\,I] emission decreases with increasing radius,  neglecting it  has the largest effect in the center, and may decrease the age difference there from 2.7 to $\sim$2.1 $\sigma$.  Though,  the effect of  including  [N\,I] may be different  for the  wavelength region   used by \cite{2019MNRAS.488.1679P}. 
The lower \met\space we obtain  at 0.2\,\re\space does not come from [N\,I].    
Using fixed solar \met, \cite{2018MNRAS.473.4698B} showed that ages measured from near-infrared spectra are less precise than from optical spectra. The different wavelength regions  might cause the age discrepancy of 2 $\sigma$, and in combination with the age-metallicity anti-correlation, cause the discrepancy of \met.

\subsubsection{Elemental Abundances}

  Our elemental abundance measurements  are in rough agreement with studies of other galaxies. In particular, the ranges of Fe, Mg, Ca, C and N agree with the results of \cite{2007ApJS..171..146S}, \cite{2007ApJ...671..243G} and \cite{2009MNRAS.398..119S}, and the ranges of Mg, O, C, Ca, and Na with \cite{2017ApJ...841...68V}. 
In addition, our ranges of Fe and Ca agree with   \cite{2014ApJ...780...33C} and \cite{2015ApJ...807...11G}.  Though,    Mg, C, and N of \cite{2015ApJ...807...11G} tend to be higher;  while  Mg,  C, and O of \cite{2014ApJ...780...33C} tend to be lower. Also,   we have a wider range of values for  N, Si, Ti than \cite{2014ApJ...780...33C}. 
Our ranges of N and Na roughly agree with \cite{2014ApJ...783...20W}, but
we have higher values of Mg, Ca, and C, and a narrower range of O, and Si. 

The different results are, to some extent, caused by different approaches and data:  \cite{2007ApJS..171..146S}, \cite{2014ApJ...780...33C}, and \cite{2014ApJ...783...20W} did not measure radial gradients but rather total  galaxy parameters. Most of the aforementioned studies used stacked spectra. Only   \cite{2009MNRAS.398..119S} and \cite{2017ApJ...841...68V} analyzed individual galaxies,  \cite{2014ApJ...783...20W} derived abundances for both stacked spectra and individual galaxies. Their analysis showed that the elemental abundance ranges spanned by  individual galaxies exceeds the abundance trends of stacked spectra binned by velocity dispersion. This  explains why several of our abundance ranges are larger than in studies of stacked spectra in the literature.

\subsubsection{Elemental Abundance Gradients}
There are several papers on abundance gradients of individual galaxies. Often  they do  not measure individual elements  but  [$\alpha$/Fe], as the data quality is too low to measure individual elemental abundances. 
These studies \citep[e.g.][]{2003A&A...407..423M} usually rely on the Mg\textit{b}  feature, and obtain  [$\alpha$/Fe] values and gradients in agreement with our [Mg/Fe].  Further, our central values and $\nabla_r$ of  [Mg/Fe] and [C/Fe] are  within the ranges found by  \cite{2019MNRAS.489..608F}.  

 Few studies measured  elemental abundance gradients for a larger number of elements.  \cite{2013ApJ...776...64G,2015ApJ...807...11G} and \cite{2019MNRAS.483.3420P} used spectral indices of stacked galaxy spectra. 
 Our values for  $\nabla_r$[Mg/Fe], $\nabla_r$[Ca/Fe], $\nabla_r$[N/Fe], $\nabla_r$[Na/Fe], and $\nabla_r$[Ti/Fe] are similar to the results of  \cite{2019MNRAS.483.3420P} for most galaxies, though we obtain  a steeper decreasing \met. 
As \cite{2013ApJ...776...64G,2015ApJ...807...11G}, we obtain rather flat  [Mg/Fe] and mild age gradients. However,  we obtain relatively flat profiles for  [Fe/H] and  [C/Fe], while  \cite{2013ApJ...776...64G,2015ApJ...807...11G} find a steeper decrease  with radius.
  One major difference with these studies is that they have no  fitting parameter \met, which  has a steep radial gradient  in our analysis. The different  treatment of metallicity in the \cite{2007ApJS..171..146S} compared to the \cite{2018ApJ...854..139C} SSP models that we use may  cause their large $\nabla_r$[Fe/H]  as opposed to our flat $\nabla_r$[Fe/H]. Further, \cite{2015ApJ...807...11G} found radially varying [C/Fe] and flat [N/Fe], and we find the opposite. 
  \cite{2013ApJ...776...64G,2015ApJ...807...11G}  measured   [C/Fe] using the C$_\text{2}$4668 index, and used this result to infer [N/Fe] from the CN$_1$ index. The predictions of the  \cite{2007ApJS..171..146S}  and   \cite{2018ApJ...854..139C} SSP models for these Lick indices differ at ages \textgreater6\,Gyr, which may cause our discrepant results.

\cite{2017ApJ...841...68V} measured elemental abundances as a function of radius for six galaxies, including \obm, using full spectral fitting. Our general trends for [Mg/Fe], [Ca/Fe], [Na/Fe], [C/Fe] are similar, but \cite{2017ApJ...841...68V} find  steeper increasing [O/Fe],   steeper decreasing  [Fe/H], and overall older ages. They do not show their results for  \met, Si, Ti, and N, so we cannot compare these. \cite{2017ApJ...841...68V} fit more parameters than we do, in particular they include several other elemental abundances (K, V, Cr, Mn, Co, Ni, Sr, Ba), and a second, younger population in their fit. This likely causes the older age of their dominant stellar population compared to our SSP fit. It is not clear how other measurements may be affected, and if including  a second stellar population can account for the different results.

\section{Star Formation History and Time-Scale}
\subsection{Star Formation History}
\label{sec:sfh}

In Section \ref{sec:pystaff}  we fit a  best-fit single stellar population (SSP) spectrum. Early-type galaxies are often dominated by an old population, which makes this approach reasonable for these cases. However,  early-type galaxies can also have  an extended star formation history or several bursts of star formation \citep{2005ApJ...621..673T,2010MNRAS.404.1775T}. In particular, we obtained a  young SSP age for \obfive\space (3--5.5\,Gyr). Here, we investigate if this galaxy has an unusual star formation history compared to the other galaxies. 

\begin{figure}
\plotone{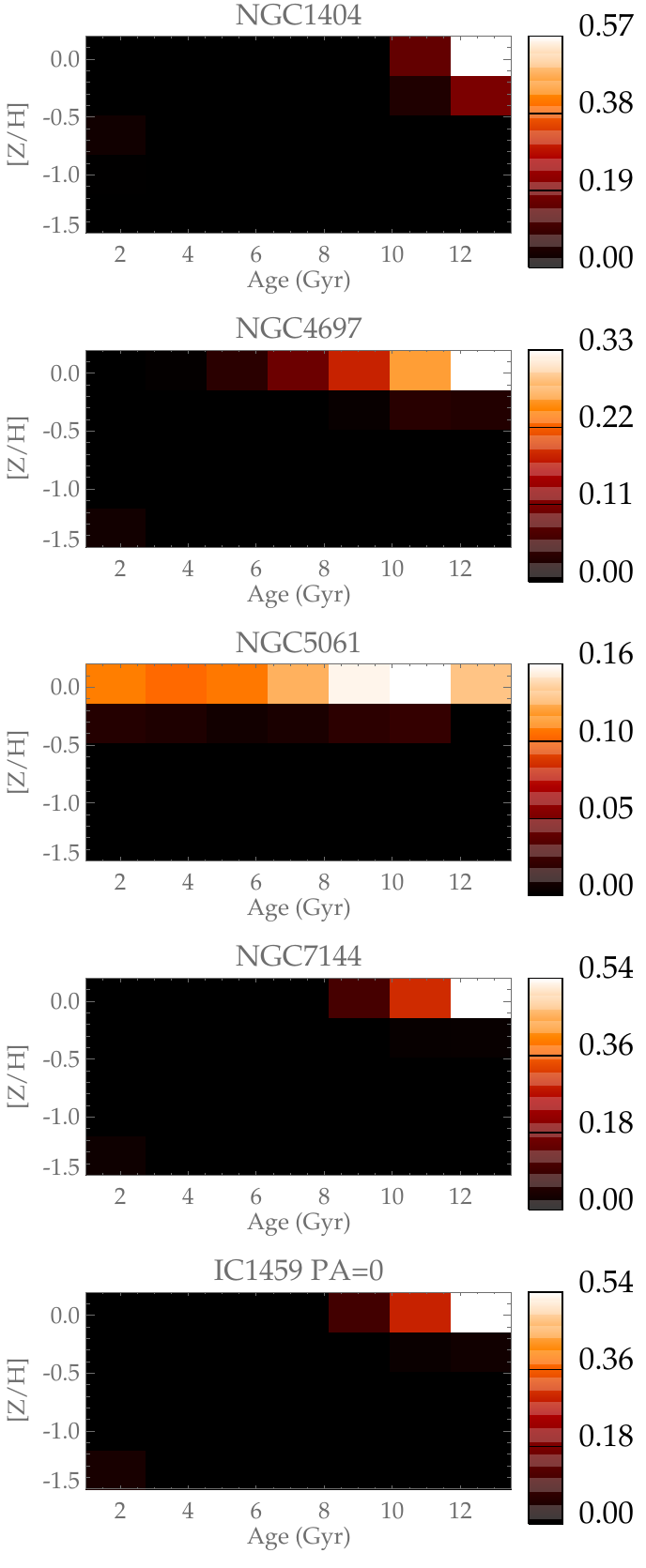}
\caption{Star formation histories of several galaxies in the most central bin, assuming a Kroupa IMF. The colors denote mass weights. All galaxies have solar to supersolar \met, and are dominated by old stellar populations \textgreater 10\,Gyr, including the galaxies which are not shown  for brevity (\obthree, \obm, \obsevsev).  Only \obfive\space has weight contributions \textgreater0.12 at all age bins.}
\label{fig:sfhplots}
\end{figure}

We derived the star formation history using the program \textsc{pPXF} \citep{2004PASP..116..138C,2012ascl.soft10002C,2017MNRAS.466..798C} and the \cite{2018ApJ...854..139C} models with solar abundances as templates.  We applied regularization as described by \cite{2017MNRAS.466..798C}.
The code \textsc{pPXF} obtains a best-fit model by  assigning weights to the template  spectra, which span a grid in age and \met. The best-fit spectrum is a linear combination of the template spectra, and we obtain the weights assigned to  each individual template. We show the weight distributions obtained for the central spectra of five galaxies in Figure \ref{fig:sfhplots}, assuming a Kroupa IMF slope.  Applying the weights, we   derived mass-weighted age and metallicity distributions, with a set of models with either \cite{2001MNRAS.322..231K}-like IMF slopes  (i.e. $x_1$=1.3, $x_2$=2.3), or a bottom-heavy IMF ($x_1$=3.1, $x_2$=3.3). The bottom-heavy IMF resulted in slightly lower ages, but the overall age and metallicity  distributions are similar between the two sets of models. 

The age distribution of \obfive\space is more extended than the age distribution of  the other galaxies in our sample. Most galaxies are dominated by old stars,  formed $\sim$9--13\,Gyr ago, and   have only a small weight contribution of younger ages. Some galaxies have contributions of lower \met\space at larger radii, indicating accreted  material.  However, the    best fit for \obfive\space  gives non-zero contributions of model spectra with ages over the entire age range of the input models  (1-13.5\,Gyr). The mass-weighted \obfive\space ages obtained with \textsc{pPXF} are higher (6--10\,Gyr) than the light-weighted age obtained with \pystaff\space (3--5.5\,Gyr). Young stars are brighter than old stars, and therefore  they  contribute more to light-weighted SSP age measurements than to mass-weighted SFH age \citep[see also][]{2007MNRAS.374..769S,2015MNRAS.448.3484M}.

\cite{2005MNRAS.364.1239R} estimated the flux contributed by stars with ages of  3\,Myr, 100\,Myr, 1\,Gyr, and 10\,Gyr in \obfive, \obone, and \obthree. Their analysis suggests comparable contributions of stars with 10\,Gyr between the galaxies, but a relatively larger contribution of 100\,Myr stars in \obfive\space  (6--12\%) compared to the other two galaxies (3-6\%). We cannot comment on  this result, as we use SSP models  in the age range of 1-13.5\,Gyr. However, this work confirms that the contribution of young stars in \obfive\space is larger than in other galaxies of our sample. 
We conclude that the  \pystaff\space approach to find the best-fit single stellar population model is less appropriate for \obfive, as it has a higher fraction of young stars that bias the results compared to other galaxies in our sample. 

\subsection{Star Formation Time-scale Estimate}
\label{sec:sfts}

\begin{figure}
\plotone{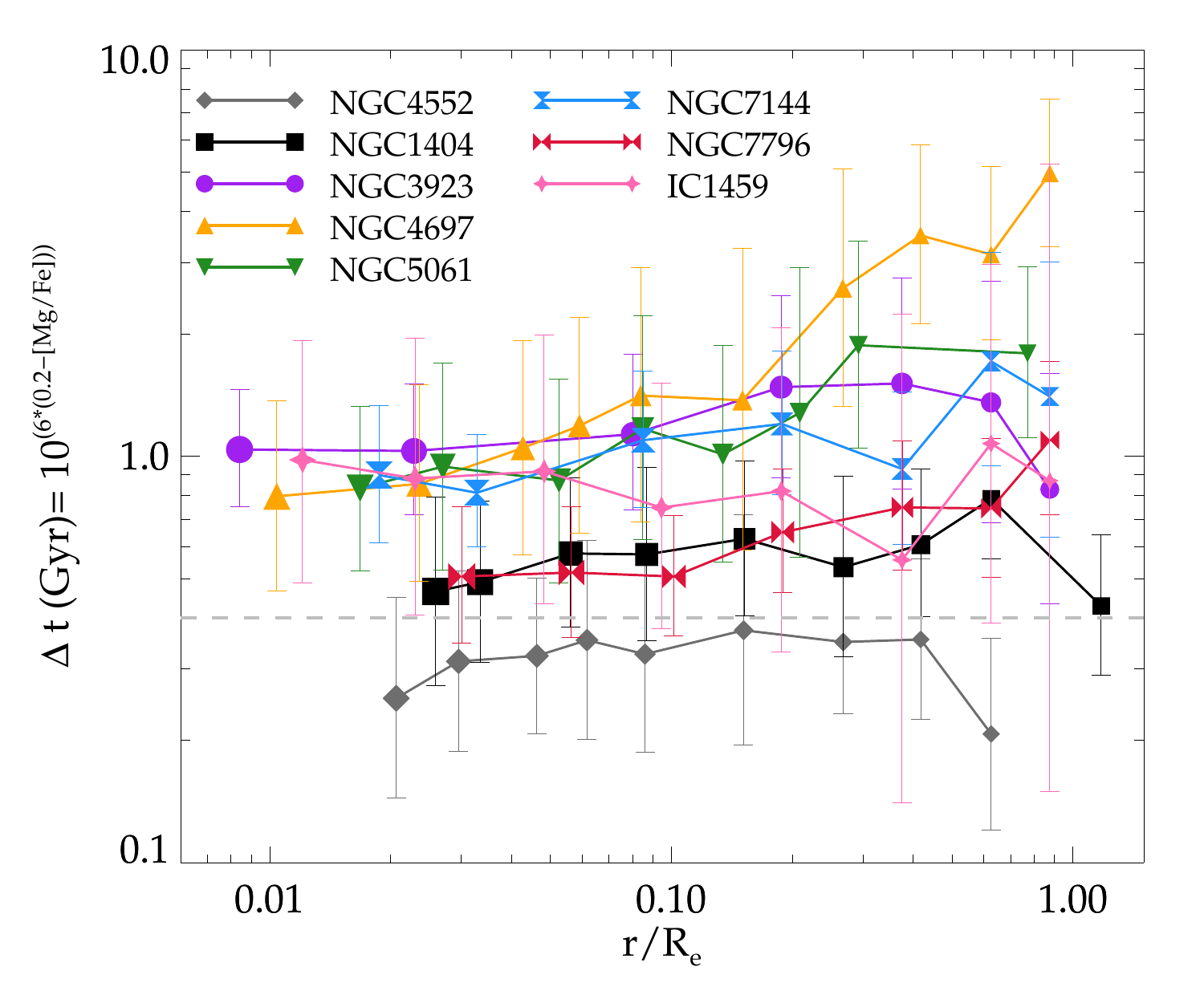}
\plotone{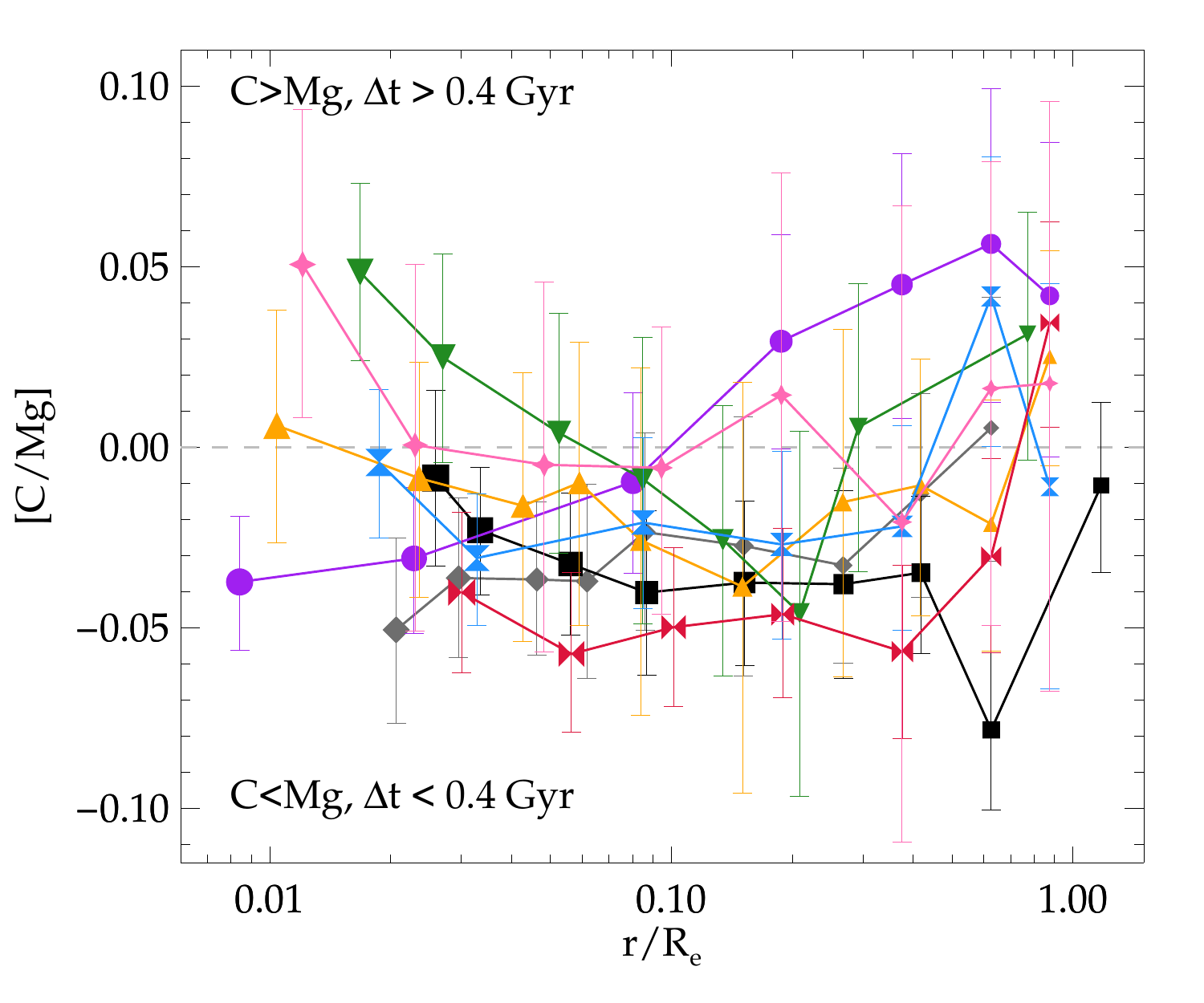}
\caption{Star formation time-scale $\Delta t$ for various galaxies, using the relation for [$\alpha$/Fe] derived by \cite{2005ApJ...621..673T}. In the  upper panel we show $\Delta t$, in the lower panel  [C/Mg] as function of radius. A value  of [C/Mg]$\lesssim$0 indicates $\Delta t$ \textless0.4\,Gyr, [C/Mg]\textgreater 0  longer time-scales. Different colors and symbols denote different galaxies, decreasing symbol s  larger $r$/\re. }
\label{fig:timescale}
\end{figure}

Various elements formed at different time-scales. In particular, $\alpha$-elements (O, Mg, Ca, Si, Ti) are formed in SN II explosions of massive stars, whereas Fe-peak elements are predominantly generated in SN Ia explosions, which happen later. The $\alpha$-element oxygen is, in addition, produced in low-to intermediate mass stars \citep[0.8--3.5\,$M_\sun$, e.g.][]{2001A&A...370..194M}, while nitrogen and carbon are formed in intermediate-mass stars (3-8\,\msun).  
This means that the ratios of different elemental abundances depend on the star formation time-scale $\Delta t$.

\cite{2005ApJ...621..673T}  derived a  relation for $\Delta t$  and   [$\alpha$/Fe],  using simulations of chemical evolution for toy galaxies with various extended star formation histories. 
Applying their relation to [Mg/Fe], we  calculated the star  formation time-scales  as a function of galaxy radius (see Figure \ref{fig:timescale}, top panel). For most galaxies, $\Delta t$ is consistent with being constant with radius within 1\,$\sigma$. Other galaxies show an increasing $\Delta t$ with radius (\obfour, \obfive, and \obsevsev).  \obm, \obone, and \obsevsev\space have  the shortest star formation time-scales. These galaxies are rather old  and  have a high velocity dispersion, which is in  agreement with \cite{2005ApJ...621..673T}, who found   that massive galaxies tend to have shorter $\Delta t$ and therefore  have older SSP ages than  less massive galaxies \citep[see also][]{2005MNRAS.361..897L,2006MNRAS.366..499D}.

We checked the consistency of our   star formation time-scale estimate   by comparing  C and Mg.  \cite{2012MNRAS.421.1908J} argue that  [C/Mg]=0  sets a limit on $\Delta t$, because a significant contribution of C comes from intermediate-mass stars. The lifetime of 3\,$M_\sun$ stars is $\sim$0.4\,Gyr \citep{1998A&A...334..505P},  and this is the star formation time-scale required to reach [C/Mg]=0, or [C/Fe]=[Mg/Fe]. 
We show [C/Mg] as a function of  radius  in the lower panel of Figure \ref{fig:timescale}.  Here we have  [C/Mg]\textless 0\,dex (indicating $\Delta t \lesssim$ 0.4\,Gyr) for a larger number of galaxies compared to $\Delta t$  from  Mg and the \cite{2005ApJ...621..673T} relation.  \obm, \obone, and \obsevsev\space have low  values of [C/Mg], indicating again the short $\Delta t $. 
 Some galaxies  (e.g.  \obfour, \obfive) have different radial trends in the two panels. These galaxies also have radially varying IMF slopes. 
 
The time-scale estimate has several caveats, which may  account for the differences: We assumed that   [$\alpha$/Fe] used by \cite{2005ApJ...621..673T} equals our  [Mg/Fe], which is not necessarily true \citep[see also][]{2014ApJ...780...33C}. 
When estimating the time-scale using [C/Mg],  only stellar masses from 3\,\Msun\space upwards are considered, whereas  [Mg/Fe] considers  lower masses. 
Both time-scale indicators  are based on the assumption of a constant IMF in all galaxies, which we have not found for the low-mass IMF in our data.   \cite{2018MNRAS.475.3700M} argue that the star formation time-scale  alone cannot explain the observed [Mg/Fe] patterns of elliptical galaxies.  Another caveat is that \met\space is not considered, altough both  Mg and C yields are \met\space dependent
\citep{2019A&A...632A.110Y}, and    all galaxies have a significant \met\space gradient (\nablar\met=--0.13 to --0.44\,dex/\logten($r$/\re).

\section{Notes on Individual Galaxies}
\label{sec:indiv}

Here, we summarise the properties of individual galaxies compared to others in the sample. Overall, we find a  wide range of elemental abundance gradients, which reflects the different formation histories for individual galaxies in our sample. 

\textbf{\obic}\space is old in the center, with a decreasing age. Remarkable are the steeply decreasing element abundance gradients (e.g. of Si, Ti, N, and Na), with  high values  in the center,  and low values at 1\,\re\space compared to other galaxies. This galaxy has an AGN and strong  gas emission lines, which results in relatively large measurement uncertainties in the center.   The ionised gas emission, counter-rotating core, shell-like remnants \citep{1995AJ....109.1576F} all are compatible with a gas-rich major merger, though the accretion of two counter-rotating cold gas streams in the first stage of \obic's formation cannot be  ruled out \citep{2019MNRAS.488.1679P}. The rather young age and low elemental abundances at large radii suggest that a significant portion of younger, possibly accreted, stellar populations contributes to the total light.  \obic\space is the central galaxy in a rather small group.

\textbf{\obthree}\space has an extended  shell system, and may have  experienced a major merger event \citep{2017MNRAS.472.2889C}. Like \obic, this galaxy is the central galaxy in a group. However, we found significant differences in the stellar populations. \obthree\space is $\sim$10\,Gyr old, which makes it the youngest among the massive ($\sigma$\textgreater 200\,\kms) galaxies. \cite{2017MNRAS.472.2889C}   found indications for a young sub-population in this galaxy. 
Also remarkable, and possibly connected to the younger age,  are the relatively low values of [O/Fe], [Mg/Fe], [Ca/Fe], [Ti/Fe], [C/Fe] and high values of [Fe/H] in the center, suggesting a  longer and thus apparently later episode of initial star formation. Some abundances are increasing with radius (C, N, O), whereas most $\alpha$ elements, [Fe/H], and  [Na/Fe] are radially constant or slightly decreasing, as for other galaxies.  However, the [Na/Fe] profile is relatively shallow.  As a result, \obthree\space has relatively high C, N, O, and Na at large radii compared to other galaxies. As these elements are produced, at least partially, in intermediate-mass stars, this  may indicate that the IMF was initially favoring intermediate-mass over high-mass stars, thus enhancing C, N, O, and Na.

\textbf{\obone}\space is the second brightest early-type galaxy  of the Fornax cluster, though it is probably part of an infalling subgroup \citep{2001ApJ...548L.139D,2005ApJ...621..663M}. The radial shape of the velocity dispersion is non-monotonic, possibly indicating overlapping structures. \cite{2019A&A...627A.136I} see indications of a kinematically decoupled component in the central $\sim$9\arcsec, with a double-peaked $\sigma$. This  region corresponds to the inner six bins of our data, and explains our irregular $\sigma$ profile. Not only the kinematics, but also the stellar populations in the inner region have non-monotonic radial profiles, e.g. age, IMF slope, C, N, O.

\textbf{\obm}\space is in the  Virgo cluster. It has a high, radially constant age and  a steeply decreasing \met\space  \citep{2021arXiv210702335L}.
The  IMF slope is  slightly  decreasing with radius. Some peculiar properties of this galaxy are the non-monotonic [Na/Fe] profile, with higher values both in the center and at the outermost bin,  and a  radially increasing [C/Fe], [N/Fe] and [O/Fe] in the outer region, similar to \obthree. 
Both \obone\space and \obm\space have rather high values of [Mg/Fe] and [C/Mg], which indicate short star formation time-scales.
\cite{2015MNRAS.448.3484M}
found that Virgo cluster galaxies have systematically higher [$\alpha$/Fe] than non-Virgo galaxies, and suggest a suppression of extended star formation in Virgo galaxies.

\textbf{\obsevsev}\space is dominated by a very old stellar population at all radii, while \met\space is rather low compared to other galaxies, and it has a more shallow gradient. The IMF slope is consistent with being constant and Salpeter-like. For this galaxy we do not have CaT data, which is helpful to constrain the IMF \citep{2020ApJ...902...12F}. Concerning elemental abundances, we found unusually flat radial profiles for O, Si, Ca, Ti,  C, and Fe compared to other galaxies in our sample. [Ca/Fe], [Si/Fe], and  [Ti/Fe] in the outer regions are higher than for any other galaxy. \obsevsev\space is a field galaxy, which suggests a relatively low contribution of galaxy mergers \citep{2010ApJ...718.1158L}, and hence  flatter stellar population gradients. [Mg/Fe]  is relatively high, and slightly decreasing as a function of radius.   Our study confirms that this galaxy formed stars very fast and efficiently \citep{2018Ap&SS.363..131R}.

\textbf{\obfour}\space is younger than other galaxies of our sample (\textless 9\,Gyr), and has on average lower elemental abundances, in particular [O/Fe], [Mg/Fe], [Si/Fe], [C/Fe], [N/Fe],  [Na/Fe]. This suggests a longer timescale for initial star formation, with enough time to enrich the gas with [Fe/H]. [Fe/H] appears to be radially increasing, and several other elements therefore radially decreasing.   The velocity dispersion is, together with \obsev, the lowest in our sample; thus the galaxy is less massive. This is in agreement with the model suggested by \cite{2010MNRAS.404.1775T}, that stars in a galaxy with high mass formed faster and earlier than in a less massive galaxy. \obfour\space is also the only galaxy in our sample  with a radially increasing $\sigma$. The IMF slope is non-monotonic, but overall slightly radially decreasing.

\textbf{\obsev}, part of a sparse group,  has a radially decreasing age, but constant, Salpeter-like IMF slope. \met\space and several elemental abundances (Si, Ca, N, Na) are rather low.  The radial gradients of Si, Ca, Ti are rather steeply decreasing, but flat for C and N.

\textbf{\obfive}\space  has the youngest SSP age, and an extended star formation history (see Appendix \ref{sec:sfh}). Thus, the SSP values we measure are possibly inaccurate. The \met\space and IMF slope of \obfive\space have the highest values of all galaxies in the center, while [O/Fe] is lowest. 
Also the [Ti/Fe] values in the center are higher than for the other galaxies. [Na/Fe]  is decreasing steeply.

\section{Analysis of the giant low-surface brightness spiral galaxy \obu}
\label{sec:ugc}
We observed \obu\space on September 20, 2014, believing it is an elliptical galaxy. 
However, \cite{2016ApJ...826..210H} obtained deep photometry and discovered a disk and spiral arms surrounding \obu. They classified \obu\space as giant low-surface brightness galaxy (LSBG). 
Here we describe the analysis and results for this galaxy. Due to its classification, we did not include this galaxy in our analysis of early-type galaxies. 

We observed \obu\space with a slit width of 1\farcs5, an exposure time of 50 min, and  at a PA of 90\degr\space with the 600$\ell$/9\fdg78 grating during a cloudy night. This causes the rather low S/N of the optical spectra compared to the data of other galaxies, and made the CaT data, and spectra at \textless4390\,\AA, unusable. Due to the classification as spiral galaxy,   we did not repeat the observation to increase the S/N. Our optical spectra cover the bulge (\re=3\farcs4), which dominates the  spectra of the inner four radial bins, and the inner disk out to 1\,\re\space  \citep[16\arcsec,][]{2016ApJ...826..210H}. We reduced the data in a similar way as described in Section \ref{sec:dr}. The S/N was so low that we did not extract the northern and southern spectra separately, but simply summed the spectra without correcting rotation.  Still, the \obu\space  spectra have the lowest S/N, \textless 50 at all radii and wavelengths, see Figure \ref{fig:snr}.  This is caused by the poor observing conditions at the time of the observations.

\begin{figure}
          \plotone{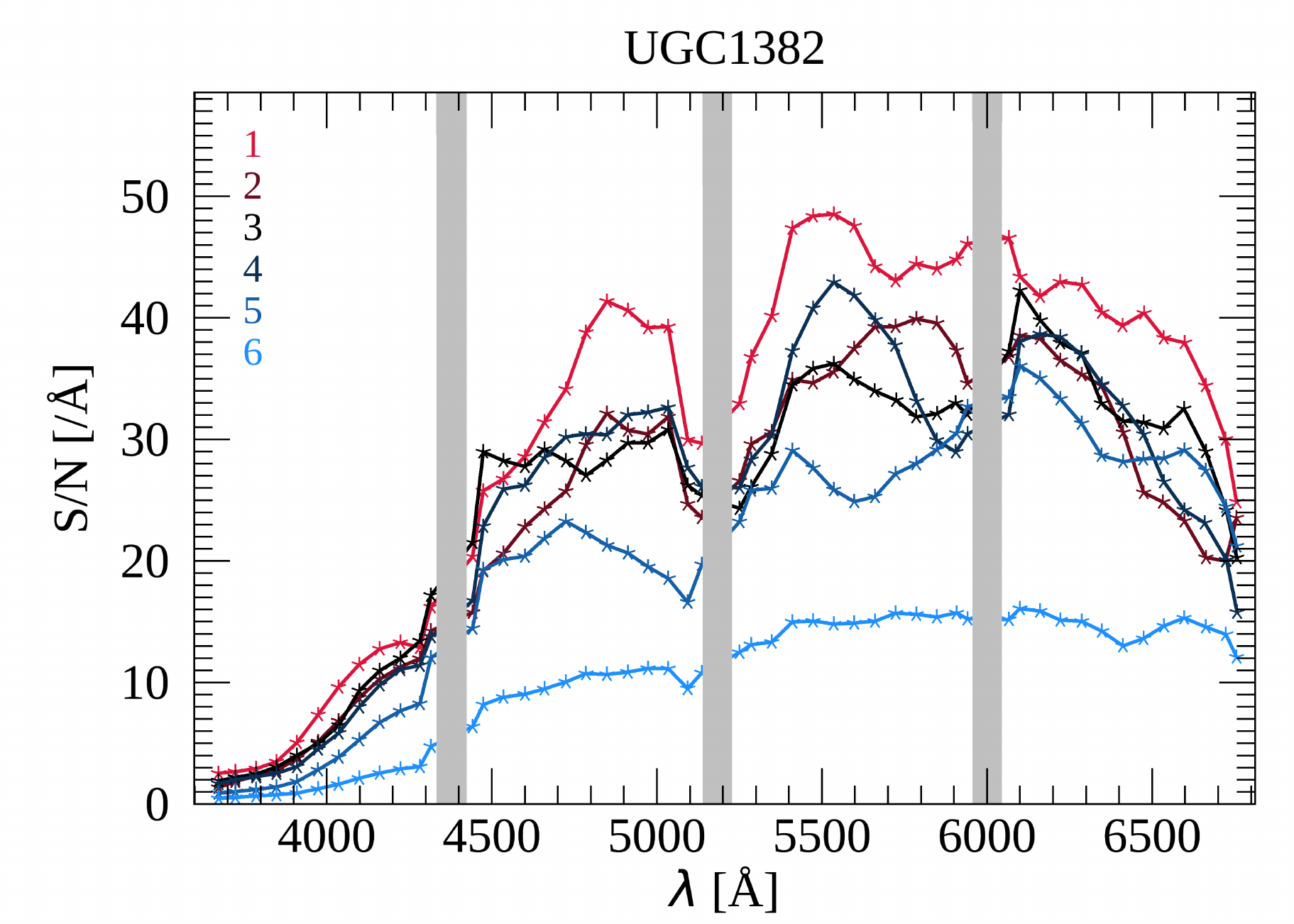}
\caption{Signal-to-noise ratio as function of wavelength for \obu, colors and symbols as in  Figure \ref{fig:snr}. 
\label{fig:snr2}}
\end{figure}

The low S/N required some changes in the fitting procedure compared to the other galaxies. In the \textsc{pystaff} SSP fit, we included the emission lines H$\beta$, H$\alpha$, [OI], [OIII], [NII], similar to \obic, but we tied the gas kinematics of all emission lines together. We used the wavelength ranges 4340--4659, 4659--5053, 5122--5857, 5942--6635\,\AA, which is shorter than for the other galaxies. 
We used only one wavelength setting for this galaxy, as shortening the wavelength region even further may lead to less accurate results. Thus, we do not have systematic uncertainties from varying the wavelength region. Due to the low S/N, the fits already have rather large statistical errors. 

Compared to the other galaxies, we obtain higher [Si/Fe] ($\sim$0.2 to 0.4\,dex), [Ca/Fe] ($\sim$0.0 to 0.15\,dex), [Ti/Fe] ($\sim$0.15 to 0.35\,dex) values, and lower [N/Fe] ($\sim-$0.35 to 0.1\,dex) and  [Na/Fe] ($\sim-$0.3 to 0.3\,dex). Several of our stellar population parameters have a large scatter over the radial range (e.g. IMF, [O/Fe], [C/Fe], [Ti/Fe], see also Table \ref{tab:pyresult}).  For this reason, we are not sure how robust the elemental abundance and IMF measurements are. Also the radial gradient fits have large uncertainties. Rather robust  measurements are the flat age ($\sim$10\,Gyr), and the shallow \met\space gradient  ($-0.12$\,dex/\logten($r$/\re)), with overall low values of \met\space (-0.05 to -0.2\,dex), similar to the also isolated, but slightly older \obsevsev.  The behavior of \met, shared by \obu\space and \obsevsev\space may be connected to their isolated locations.

\begin{longrotatetable}
\begin{deluxetable*}{llrrrrrrrrrrrrr}
\tablecaption{Stellar population results, as function of galaxy radius $r/$\re, where we take the slit width into account,  and distance from the galaxy center along the slit  $x/$\re. IMF denotes the low-mass IMF slope (0.08-1.0\,\Msun). Uncertainties are quadratic sum of statistical and systematic uncertainties. \label{tab:pyresult}}
\tablewidth{700pt}
\tabletypesize{\scriptsize}
\tablehead{
\colhead{$r/$\re} & \colhead{$x/$\re} & \colhead{$\sigma$} & 
\colhead{$t$} & \colhead{\met} & 
\colhead{IMF} & 
\colhead{[O/Fe]} & \colhead{[Mg/Fe]}  & \colhead{[Si/Fe]} & \colhead{[Ca/Fe]} & 
\colhead{[Ti/Fe]} & \colhead{[Fe/H]} &  \colhead{[Na/Fe]} &   \colhead{[C/Fe]} &
\colhead{[N/Fe]}  \\ 
\nocolhead{} &\nocolhead{}& \colhead{(\kms)} & \colhead{(Gyr)} & \colhead{(dex)} & 
\nocolhead{} & \colhead{(dex)} & \colhead{(dex)} & \colhead{(dex)}& \colhead{(dex)}&
\colhead{(dex)} & \colhead{(dex)} & \colhead{(dex)} & \colhead{(dex)}& \colhead{(dex)}
} 
\startdata
\hline
\multicolumn{2}{l}{\obm} \\
\input{all_combine_fecx2d.tex}
\enddata
\end{deluxetable*}
\end{longrotatetable}

\bibliography{bib}{}
\bibliographystyle{aasjournal}

\end{document}

%% file: all_combine_fecx2d.tex
M89 \\
$ 0.021 $&$  0.005$&$ 282^{+ 2}_{- 2}$&$ 11.2^{+ 1.1}_{- 1.0}$&$ 0.16^{+0.02}_{-0.02}$&$  2.7^{+ 0.3}_{- 0.3}$&$ 0.39^{+0.06}_{-0.06}$&$ 0.30^{+0.04}_{-0.04}$&$ 0.18^{+0.04}_{-0.04}$&$ 0.01^{+0.05}_{-0.05}$&$ 0.11^{+0.05}_{-0.05}$&$-0.04^{+0.04}_{-0.03}$&$ 0.67^{+0.07}_{-0.04}$&$ 0.25^{+0.04}_{-0.04}$&$ 0.21^{+0.06}_{-0.06}$\\
$ 0.030 $&$  0.022$&$ 262^{+ 2}_{- 2}$&$ 11.5^{+ 0.7}_{- 0.7}$&$ 0.16^{+0.01}_{-0.01}$&$  2.8^{+ 0.1}_{- 0.2}$&$ 0.38^{+0.04}_{-0.04}$&$ 0.28^{+0.04}_{-0.04}$&$ 0.17^{+0.04}_{-0.04}$&$ 0.02^{+0.05}_{-0.05}$&$ 0.07^{+0.04}_{-0.04}$&$-0.04^{+0.03}_{-0.03}$&$ 0.64^{+0.04}_{-0.04}$&$ 0.25^{+0.03}_{-0.03}$&$ 0.20^{+0.05}_{-0.05}$\\
$ 0.046 $&$  0.042$&$ 255^{+ 2}_{- 2}$&$ 12.5^{+ 0.5}_{- 0.6}$&$ 0.12^{+0.01}_{-0.01}$&$  2.7^{+ 0.1}_{- 0.1}$&$ 0.38^{+0.04}_{-0.04}$&$ 0.28^{+0.03}_{-0.03}$&$ 0.17^{+0.04}_{-0.04}$&$-0.01^{+0.04}_{-0.04}$&$ 0.09^{+0.04}_{-0.04}$&$-0.03^{+0.03}_{-0.03}$&$ 0.62^{+0.03}_{-0.03}$&$ 0.25^{+0.03}_{-0.03}$&$ 0.17^{+0.05}_{-0.05}$\\
$ 0.062 $&$  0.058$&$ 252^{+ 2}_{- 2}$&$ 12.9^{+ 0.6}_{- 0.7}$&$ 0.08^{+0.02}_{-0.02}$&$  2.5^{+ 0.2}_{- 0.2}$&$ 0.38^{+0.04}_{-0.04}$&$ 0.28^{+0.04}_{-0.04}$&$ 0.14^{+0.04}_{-0.04}$&$ 0.00^{+0.05}_{-0.05}$&$ 0.06^{+0.05}_{-0.05}$&$-0.02^{+0.03}_{-0.03}$&$ 0.59^{+0.04}_{-0.04}$&$ 0.24^{+0.04}_{-0.04}$&$ 0.19^{+0.05}_{-0.05}$\\
$ 0.086 $&$  0.084$&$ 248^{+ 2}_{- 2}$&$ 12.5^{+ 0.7}_{- 0.8}$&$ 0.05^{+0.02}_{-0.02}$&$  2.7^{+ 0.2}_{- 0.2}$&$ 0.39^{+0.05}_{-0.05}$&$ 0.28^{+0.04}_{-0.04}$&$ 0.16^{+0.04}_{-0.05}$&$ 0.02^{+0.05}_{-0.05}$&$ 0.07^{+0.05}_{-0.05}$&$-0.04^{+0.03}_{-0.03}$&$ 0.59^{+0.05}_{-0.05}$&$ 0.26^{+0.04}_{-0.04}$&$ 0.21^{+0.06}_{-0.06}$\\
$ 0.151 $&$  0.150$&$ 241^{+ 3}_{- 3}$&$ 11.7^{+ 1.6}_{- 1.6}$&$ 0.01^{+0.05}_{-0.05}$&$  2.8^{+ 0.3}_{- 0.3}$&$ 0.36^{+0.07}_{-0.07}$&$ 0.27^{+0.05}_{-0.05}$&$ 0.14^{+0.05}_{-0.05}$&$ 0.01^{+0.05}_{-0.05}$&$ 0.06^{+0.05}_{-0.06}$&$-0.03^{+0.04}_{-0.04}$&$ 0.52^{+0.09}_{-0.09}$&$ 0.24^{+0.05}_{-0.05}$&$ 0.17^{+0.07}_{-0.07}$\\
$ 0.268 $&$  0.267$&$ 224^{+ 3}_{- 3}$&$ 11.0^{+ 1.3}_{- 1.3}$&$-0.09^{+0.04}_{-0.04}$&$  2.8^{+ 0.4}_{- 0.4}$&$ 0.37^{+0.06}_{-0.06}$&$ 0.28^{+0.03}_{-0.03}$&$ 0.10^{+0.04}_{-0.04}$&$-0.00^{+0.03}_{-0.03}$&$ 0.04^{+0.04}_{-0.05}$&$-0.03^{+0.02}_{-0.02}$&$ 0.50^{+0.07}_{-0.07}$&$ 0.24^{+0.03}_{-0.03}$&$ 0.19^{+0.06}_{-0.06}$\\
$ 0.417 $&$  0.417$&$ 213^{+ 3}_{- 3}$&$ 12.3^{+ 0.9}_{- 0.9}$&$-0.28^{+0.02}_{-0.02}$&$  2.4^{+ 0.2}_{- 0.2}$&$ 0.45^{+0.03}_{-0.03}$&$ 0.28^{+0.03}_{-0.03}$&$ 0.13^{+0.04}_{-0.04}$&$-0.04^{+0.04}_{-0.04}$&$-0.07^{+0.04}_{-0.04}$&$-0.03^{+0.02}_{-0.02}$&$ 0.49^{+0.07}_{-0.07}$&$ 0.26^{+0.02}_{-0.02}$&$ 0.29^{+0.04}_{-0.04}$\\
$ 0.625 $&$  0.625$&$ 216^{+ 5}_{- 5}$&$ 10.9^{+ 1.8}_{- 1.8}$&$-0.39^{+0.03}_{-0.03}$&$  2.3^{+ 0.1}_{- 0.2}$&$ 0.55^{+0.02}_{-0.02}$&$ 0.31^{+0.04}_{-0.04}$&$ 0.15^{+0.06}_{-0.06}$&$-0.02^{+0.04}_{-0.04}$&$-0.03^{+0.06}_{-0.06}$&$-0.11^{+0.02}_{-0.02}$&$ 0.65^{+0.05}_{-0.06}$&$ 0.32^{+0.02}_{-0.02}$&$ 0.41^{+0.05}_{-0.06}$\\
\hline 
NGC1404 \\
$ 0.026 $&$  0.005$&$ 238^{+ 2}_{- 2}$&$ 12.7^{+ 0.6}_{- 0.6}$&$ 0.12^{+0.02}_{-0.02}$&$  2.0^{+ 0.3}_{- 0.3}$&$ 0.41^{+0.04}_{-0.04}$&$ 0.26^{+0.04}_{-0.04}$&$ 0.16^{+0.04}_{-0.04}$&$ 0.04^{+0.04}_{-0.04}$&$ 0.11^{+0.04}_{-0.04}$&$-0.04^{+0.03}_{-0.03}$&$ 0.53^{+0.07}_{-0.07}$&$ 0.25^{+0.04}_{-0.03}$&$ 0.17^{+0.06}_{-0.06}$\\
$ 0.033 $&$  0.022$&$ 226^{+ 2}_{- 2}$&$ 12.2^{+ 0.4}_{- 0.5}$&$ 0.13^{+0.01}_{-0.01}$&$  2.4^{+ 0.2}_{- 0.2}$&$ 0.36^{+0.04}_{-0.04}$&$ 0.25^{+0.03}_{-0.03}$&$ 0.18^{+0.04}_{-0.04}$&$ 0.05^{+0.03}_{-0.04}$&$ 0.11^{+0.04}_{-0.04}$&$-0.04^{+0.03}_{-0.03}$&$ 0.50^{+0.05}_{-0.06}$&$ 0.23^{+0.03}_{-0.03}$&$ 0.15^{+0.08}_{-0.08}$\\
$ 0.056 $&$  0.050$&$ 232^{+ 2}_{- 2}$&$ 11.9^{+ 0.7}_{- 0.8}$&$ 0.09^{+0.02}_{-0.02}$&$  2.5^{+ 0.2}_{- 0.2}$&$ 0.33^{+0.03}_{-0.04}$&$ 0.24^{+0.03}_{-0.03}$&$ 0.13^{+0.03}_{-0.03}$&$ 0.04^{+0.03}_{-0.03}$&$ 0.11^{+0.04}_{-0.04}$&$-0.02^{+0.03}_{-0.03}$&$ 0.41^{+0.06}_{-0.06}$&$ 0.21^{+0.03}_{-0.03}$&$ 0.13^{+0.07}_{-0.07}$\\
$ 0.087 $&$  0.083$&$ 242^{+ 2}_{- 2}$&$ 10.5^{+ 0.8}_{- 0.8}$&$ 0.10^{+0.02}_{-0.02}$&$  2.8^{+ 0.3}_{- 0.3}$&$ 0.30^{+0.05}_{-0.05}$&$ 0.24^{+0.04}_{-0.04}$&$ 0.12^{+0.04}_{-0.04}$&$ 0.05^{+0.03}_{-0.03}$&$ 0.10^{+0.04}_{-0.04}$&$-0.03^{+0.03}_{-0.03}$&$ 0.36^{+0.04}_{-0.04}$&$ 0.20^{+0.03}_{-0.03}$&$ 0.12^{+0.08}_{-0.08}$\\
$ 0.152 $&$  0.150$&$ 242^{+ 2}_{- 2}$&$ 10.0^{+ 0.9}_{- 0.8}$&$ 0.07^{+0.02}_{-0.02}$&$  2.8^{+ 0.2}_{- 0.3}$&$ 0.29^{+0.04}_{-0.04}$&$ 0.23^{+0.03}_{-0.03}$&$ 0.10^{+0.04}_{-0.04}$&$ 0.00^{+0.03}_{-0.03}$&$ 0.07^{+0.04}_{-0.04}$&$-0.02^{+0.03}_{-0.03}$&$ 0.33^{+0.04}_{-0.04}$&$ 0.20^{+0.03}_{-0.03}$&$ 0.02^{+0.11}_{-0.11}$\\
$ 0.268 $&$  0.267$&$ 236^{+ 2}_{- 2}$&$ 10.6^{+ 1.7}_{- 1.7}$&$ 0.02^{+0.05}_{-0.05}$&$  2.5^{+ 0.4}_{- 0.4}$&$ 0.34^{+0.04}_{-0.05}$&$ 0.25^{+0.04}_{-0.04}$&$ 0.08^{+0.05}_{-0.05}$&$ 0.01^{+0.04}_{-0.04}$&$ 0.09^{+0.04}_{-0.04}$&$-0.04^{+0.03}_{-0.03}$&$ 0.35^{+0.06}_{-0.06}$&$ 0.21^{+0.03}_{-0.03}$&$ 0.08^{+0.12}_{-0.12}$\\
$ 0.418 $&$  0.417$&$ 218^{+ 2}_{- 2}$&$ 11.8^{+ 0.9}_{- 1.0}$&$-0.07^{+0.02}_{-0.02}$&$  2.2^{+ 0.2}_{- 0.2}$&$ 0.35^{+0.03}_{-0.04}$&$ 0.24^{+0.03}_{-0.03}$&$ 0.11^{+0.04}_{-0.04}$&$ 0.01^{+0.03}_{-0.04}$&$ 0.06^{+0.04}_{-0.04}$&$-0.01^{+0.02}_{-0.02}$&$ 0.36^{+0.03}_{-0.03}$&$ 0.20^{+0.03}_{-0.03}$&$ 0.06^{+0.06}_{-0.07}$\\
$ 0.626 $&$  0.625$&$ 208^{+ 2}_{- 2}$&$ 10.5^{+ 0.7}_{- 0.7}$&$-0.16^{+0.02}_{-0.02}$&$  2.1^{+ 0.2}_{- 0.2}$&$ 0.28^{+0.04}_{-0.05}$&$ 0.22^{+0.02}_{-0.02}$&$ 0.09^{+0.03}_{-0.03}$&$-0.03^{+0.03}_{-0.03}$&$-0.03^{+0.05}_{-0.05}$&$ 0.03^{+0.02}_{-0.02}$&$ 0.29^{+0.03}_{-0.03}$&$ 0.14^{+0.02}_{-0.02}$&$ 0.11^{+0.05}_{-0.05}$\\
$ 1.175 $&$  1.175$&$ 196^{+ 4}_{- 4}$&$  9.4^{+ 1.0}_{- 1.0}$&$-0.31^{+0.02}_{-0.02}$&$  2.4^{+ 0.1}_{- 0.2}$&$ 0.49^{+0.02}_{-0.02}$&$ 0.26^{+0.03}_{-0.03}$&$ 0.11^{+0.03}_{-0.03}$&$-0.03^{+0.04}_{-0.04}$&$-0.09^{+0.05}_{-0.05}$&$-0.05^{+0.02}_{-0.02}$&$ 0.31^{+0.03}_{-0.03}$&$ 0.25^{+0.02}_{-0.02}$&$ 0.06^{+0.14}_{-0.14}$\\
\hline 
NGC3923 \\
$ 0.008 $&$  0.004$&$ 257^{+ 2}_{- 3}$&$  9.9^{+ 1.3}_{- 1.3}$&$ 0.17^{+0.03}_{-0.03}$&$  2.5^{+ 0.3}_{- 0.3}$&$ 0.36^{+0.03}_{-0.03}$&$ 0.20^{+0.02}_{-0.02}$&$ 0.16^{+0.04}_{-0.04}$&$-0.02^{+0.04}_{-0.04}$&$ 0.03^{+0.05}_{-0.05}$&$ 0.02^{+0.02}_{-0.02}$&$ 0.53^{+0.03}_{-0.03}$&$ 0.16^{+0.02}_{-0.02}$&$ 0.19^{+0.04}_{-0.04}$\\
$ 0.023 $&$  0.022$&$ 254^{+ 3}_{- 3}$&$ 10.1^{+ 1.3}_{- 1.4}$&$ 0.14^{+0.03}_{-0.03}$&$  2.5^{+ 0.3}_{- 0.3}$&$ 0.36^{+0.03}_{-0.03}$&$ 0.20^{+0.03}_{-0.03}$&$ 0.15^{+0.04}_{-0.04}$&$-0.01^{+0.04}_{-0.04}$&$ 0.03^{+0.05}_{-0.05}$&$ 0.01^{+0.02}_{-0.02}$&$ 0.53^{+0.03}_{-0.03}$&$ 0.17^{+0.02}_{-0.03}$&$ 0.19^{+0.04}_{-0.04}$\\
$ 0.080 $&$  0.080$&$ 240^{+ 3}_{- 3}$&$  9.8^{+ 1.4}_{- 1.5}$&$ 0.05^{+0.05}_{-0.04}$&$  2.4^{+ 0.3}_{- 0.3}$&$ 0.38^{+0.03}_{-0.04}$&$ 0.19^{+0.03}_{-0.03}$&$ 0.14^{+0.05}_{-0.05}$&$ 0.01^{+0.05}_{-0.05}$&$-0.03^{+0.06}_{-0.07}$&$-0.00^{+0.02}_{-0.03}$&$ 0.51^{+0.04}_{-0.05}$&$ 0.18^{+0.03}_{-0.03}$&$ 0.21^{+0.05}_{-0.05}$\\
$ 0.188 $&$  0.188$&$ 216^{+ 2}_{- 2}$&$ 10.0^{+ 0.6}_{- 0.6}$&$-0.10^{+0.02}_{-0.02}$&$  2.2^{+ 0.2}_{- 0.3}$&$ 0.43^{+0.03}_{-0.04}$&$ 0.17^{+0.04}_{-0.04}$&$ 0.15^{+0.04}_{-0.05}$&$ 0.00^{+0.05}_{-0.05}$&$-0.04^{+0.06}_{-0.06}$&$-0.02^{+0.03}_{-0.03}$&$ 0.39^{+0.07}_{-0.08}$&$ 0.20^{+0.03}_{-0.03}$&$ 0.24^{+0.06}_{-0.06}$\\
$ 0.375 $&$  0.375$&$ 212^{+ 4}_{- 4}$&$  9.2^{+ 1.1}_{- 1.1}$&$-0.22^{+0.02}_{-0.02}$&$  2.0^{+ 0.2}_{- 0.3}$&$ 0.47^{+0.03}_{-0.03}$&$ 0.17^{+0.04}_{-0.04}$&$ 0.11^{+0.05}_{-0.06}$&$-0.02^{+0.05}_{-0.05}$&$-0.07^{+0.06}_{-0.06}$&$-0.04^{+0.03}_{-0.03}$&$ 0.37^{+0.04}_{-0.04}$&$ 0.22^{+0.03}_{-0.03}$&$ 0.23^{+0.08}_{-0.09}$\\
$ 0.625 $&$  0.625$&$ 205^{+ 5}_{- 5}$&$  8.1^{+ 1.6}_{- 1.6}$&$-0.29^{+0.03}_{-0.04}$&$  2.1^{+ 0.2}_{- 0.2}$&$ 0.50^{+0.03}_{-0.04}$&$ 0.18^{+0.05}_{-0.05}$&$ 0.06^{+0.07}_{-0.07}$&$-0.02^{+0.05}_{-0.05}$&$-0.08^{+0.06}_{-0.06}$&$-0.07^{+0.03}_{-0.03}$&$ 0.36^{+0.04}_{-0.04}$&$ 0.23^{+0.03}_{-0.03}$&$ 0.30^{+0.10}_{-0.10}$\\
$ 0.875 $&$  0.875$&$ 197^{+ 5}_{- 5}$&$  7.8^{+ 1.8}_{- 1.8}$&$-0.35^{+0.04}_{-0.04}$&$  1.8^{+ 0.3}_{- 0.3}$&$ 0.53^{+0.03}_{-0.03}$&$ 0.21^{+0.05}_{-0.05}$&$ 0.04^{+0.07}_{-0.07}$&$-0.02^{+0.05}_{-0.05}$&$-0.11^{+0.05}_{-0.05}$&$-0.10^{+0.02}_{-0.02}$&$ 0.35^{+0.04}_{-0.04}$&$ 0.26^{+0.03}_{-0.03}$&$ 0.34^{+0.17}_{-0.17}$\\
\hline 
NGC4697 \\
$ 0.010 $&$  0.005$&$ 165^{+ 2}_{- 2}$&$  9.2^{+ 0.7}_{- 0.7}$&$ 0.18^{+0.02}_{-0.02}$&$  2.1^{+ 0.3}_{- 0.3}$&$ 0.36^{+0.04}_{-0.04}$&$ 0.22^{+0.04}_{-0.04}$&$ 0.09^{+0.04}_{-0.04}$&$ 0.06^{+0.03}_{-0.03}$&$ 0.07^{+0.04}_{-0.04}$&$-0.05^{+0.03}_{-0.03}$&$ 0.36^{+0.04}_{-0.04}$&$ 0.22^{+0.03}_{-0.03}$&$ 0.11^{+0.06}_{-0.06}$\\
$ 0.024 $&$  0.022$&$ 146^{+ 2}_{- 2}$&$  8.3^{+ 0.6}_{- 0.5}$&$ 0.17^{+0.02}_{-0.02}$&$  2.4^{+ 0.2}_{- 0.3}$&$ 0.31^{+0.04}_{-0.04}$&$ 0.21^{+0.04}_{-0.04}$&$ 0.07^{+0.04}_{-0.04}$&$ 0.05^{+0.03}_{-0.03}$&$ 0.07^{+0.04}_{-0.04}$&$-0.04^{+0.03}_{-0.03}$&$ 0.29^{+0.04}_{-0.04}$&$ 0.20^{+0.03}_{-0.03}$&$ 0.09^{+0.05}_{-0.06}$\\
$ 0.043 $&$  0.042$&$ 150^{+ 2}_{- 2}$&$  7.9^{+ 0.5}_{- 0.5}$&$ 0.14^{+0.02}_{-0.02}$&$  2.6^{+ 0.2}_{- 0.2}$&$ 0.28^{+0.04}_{-0.05}$&$ 0.20^{+0.04}_{-0.04}$&$ 0.05^{+0.04}_{-0.04}$&$ 0.03^{+0.04}_{-0.04}$&$ 0.06^{+0.04}_{-0.04}$&$-0.04^{+0.03}_{-0.03}$&$ 0.23^{+0.04}_{-0.04}$&$ 0.18^{+0.03}_{-0.03}$&$ 0.01^{+0.07}_{-0.07}$\\
$ 0.059 $&$  0.058$&$ 149^{+ 2}_{- 2}$&$  7.5^{+ 0.5}_{- 0.5}$&$ 0.13^{+0.02}_{-0.02}$&$  2.7^{+ 0.2}_{- 0.2}$&$ 0.26^{+0.05}_{-0.05}$&$ 0.19^{+0.04}_{-0.04}$&$ 0.03^{+0.04}_{-0.04}$&$ 0.03^{+0.04}_{-0.04}$&$ 0.05^{+0.04}_{-0.04}$&$-0.04^{+0.03}_{-0.03}$&$ 0.20^{+0.05}_{-0.05}$&$ 0.18^{+0.03}_{-0.03}$&$ 0.01^{+0.06}_{-0.06}$\\
$ 0.084 $&$  0.083$&$ 152^{+ 2}_{- 2}$&$  6.8^{+ 0.5}_{- 0.5}$&$ 0.10^{+0.03}_{-0.03}$&$  3.0^{+ 0.3}_{- 0.3}$&$ 0.18^{+0.06}_{-0.07}$&$ 0.18^{+0.05}_{-0.05}$&$ 0.02^{+0.04}_{-0.05}$&$ 0.02^{+0.04}_{-0.04}$&$ 0.04^{+0.05}_{-0.05}$&$-0.03^{+0.03}_{-0.03}$&$ 0.16^{+0.06}_{-0.06}$&$ 0.15^{+0.03}_{-0.03}$&$-0.06^{+0.07}_{-0.06}$\\
$ 0.150 $&$  0.150$&$ 156^{+ 3}_{- 3}$&$  6.7^{+ 0.8}_{- 0.9}$&$ 0.04^{+0.05}_{-0.05}$&$  2.8^{+ 0.3}_{- 0.3}$&$ 0.20^{+0.07}_{-0.07}$&$ 0.18^{+0.06}_{-0.06}$&$-0.00^{+0.05}_{-0.06}$&$ 0.01^{+0.04}_{-0.04}$&$ 0.03^{+0.06}_{-0.06}$&$-0.02^{+0.03}_{-0.04}$&$ 0.09^{+0.09}_{-0.09}$&$ 0.14^{+0.04}_{-0.04}$&$-0.08^{+0.08}_{-0.08}$\\
$ 0.267 $&$  0.267$&$ 169^{+ 3}_{- 3}$&$  7.2^{+ 0.6}_{- 0.6}$&$-0.10^{+0.03}_{-0.03}$&$  2.8^{+ 0.3}_{- 0.3}$&$ 0.18^{+0.06}_{-0.06}$&$ 0.13^{+0.05}_{-0.05}$&$-0.01^{+0.05}_{-0.05}$&$-0.00^{+0.03}_{-0.03}$&$-0.01^{+0.06}_{-0.06}$&$ 0.00^{+0.02}_{-0.02}$&$-0.00^{+0.07}_{-0.07}$&$ 0.12^{+0.03}_{-0.03}$&$-0.20^{+0.07}_{-0.06}$\\
$ 0.417 $&$  0.417$&$ 168^{+ 3}_{- 3}$&$  7.3^{+ 0.5}_{- 0.5}$&$-0.14^{+0.02}_{-0.02}$&$  2.5^{+ 0.2}_{- 0.2}$&$ 0.24^{+0.04}_{-0.04}$&$ 0.11^{+0.04}_{-0.04}$&$-0.04^{+0.04}_{-0.04}$&$-0.06^{+0.03}_{-0.03}$&$-0.11^{+0.05}_{-0.05}$&$ 0.03^{+0.02}_{-0.02}$&$-0.09^{+0.05}_{-0.06}$&$ 0.10^{+0.02}_{-0.02}$&$-0.15^{+0.08}_{-0.07}$\\
$ 0.625 $&$  0.625$&$ 163^{+ 3}_{- 3}$&$  7.9^{+ 0.9}_{- 0.9}$&$-0.25^{+0.02}_{-0.02}$&$  1.8^{+ 0.3}_{- 0.3}$&$ 0.27^{+0.04}_{-0.05}$&$ 0.12^{+0.03}_{-0.04}$&$-0.09^{+0.04}_{-0.04}$&$-0.09^{+0.03}_{-0.03}$&$-0.13^{+0.05}_{-0.04}$&$ 0.01^{+0.02}_{-0.02}$&$ 0.04^{+0.04}_{-0.05}$&$ 0.10^{+0.02}_{-0.02}$&$-0.10^{+0.16}_{-0.16}$\\
$ 0.875 $&$  0.875$&$ 156^{+ 5}_{- 5}$&$  8.2^{+ 0.8}_{- 0.8}$&$-0.33^{+0.02}_{-0.02}$&$  2.2^{+ 0.2}_{- 0.2}$&$ 0.27^{+0.05}_{-0.05}$&$ 0.08^{+0.03}_{-0.03}$&$-0.05^{+0.05}_{-0.05}$&$-0.04^{+0.03}_{-0.03}$&$-0.19^{+0.04}_{-0.04}$&$-0.02^{+0.02}_{-0.02}$&$-0.01^{+0.05}_{-0.05}$&$ 0.11^{+0.02}_{-0.02}$&$ 0.03^{+0.15}_{-0.15}$\\
\hline 
NGC5061 \\
$ 0.017 $&$  0.005$&$ 206^{+ 2}_{- 2}$&$  3.0^{+ 0.2}_{- 0.2}$&$ 0.28^{+0.02}_{-0.02}$&$  3.3^{+ 0.1}_{- 0.2}$&$ 0.28^{+0.04}_{-0.04}$&$ 0.21^{+0.03}_{-0.03}$&$ 0.15^{+0.04}_{-0.04}$&$-0.01^{+0.03}_{-0.03}$&$ 0.28^{+0.04}_{-0.04}$&$-0.08^{+0.03}_{-0.03}$&$ 0.65^{+0.03}_{-0.03}$&$ 0.26^{+0.03}_{-0.03}$&$ 0.31^{+0.04}_{-0.04}$\\
$ 0.027 $&$  0.022$&$ 184^{+ 2}_{- 2}$&$  3.0^{+ 0.2}_{- 0.2}$&$ 0.27^{+0.02}_{-0.02}$&$  3.3^{+ 0.1}_{- 0.1}$&$ 0.21^{+0.05}_{-0.05}$&$ 0.20^{+0.04}_{-0.04}$&$ 0.11^{+0.04}_{-0.04}$&$-0.02^{+0.04}_{-0.04}$&$ 0.24^{+0.04}_{-0.04}$&$-0.06^{+0.03}_{-0.03}$&$ 0.60^{+0.04}_{-0.04}$&$ 0.23^{+0.04}_{-0.04}$&$ 0.23^{+0.05}_{-0.05}$\\
$ 0.053 $&$  0.050$&$ 179^{+ 2}_{- 2}$&$  3.0^{+ 0.2}_{- 0.2}$&$ 0.22^{+0.02}_{-0.02}$&$  3.4^{+ 0.1}_{- 0.1}$&$ 0.19^{+0.05}_{-0.06}$&$ 0.21^{+0.04}_{-0.04}$&$ 0.09^{+0.04}_{-0.04}$&$-0.02^{+0.04}_{-0.04}$&$ 0.19^{+0.04}_{-0.04}$&$-0.07^{+0.03}_{-0.03}$&$ 0.38^{+0.04}_{-0.04}$&$ 0.21^{+0.03}_{-0.03}$&$ 0.09^{+0.07}_{-0.07}$\\
$ 0.085 $&$  0.083$&$ 181^{+ 2}_{- 2}$&$  3.6^{+ 0.2}_{- 0.2}$&$ 0.16^{+0.02}_{-0.02}$&$  3.4^{+ 0.1}_{- 0.1}$&$ 0.15^{+0.06}_{-0.06}$&$ 0.19^{+0.05}_{-0.05}$&$ 0.07^{+0.04}_{-0.04}$&$-0.04^{+0.04}_{-0.04}$&$ 0.15^{+0.05}_{-0.05}$&$-0.05^{+0.03}_{-0.03}$&$ 0.25^{+0.04}_{-0.04}$&$ 0.18^{+0.03}_{-0.03}$&$-0.02^{+0.07}_{-0.07}$\\
$ 0.134 $&$  0.133$&$ 174^{+ 2}_{- 2}$&$  3.7^{+ 0.2}_{- 0.2}$&$ 0.11^{+0.02}_{-0.02}$&$  3.4^{+ 0.1}_{- 0.1}$&$ 0.13^{+0.06}_{-0.05}$&$ 0.20^{+0.04}_{-0.04}$&$ 0.06^{+0.04}_{-0.05}$&$-0.02^{+0.04}_{-0.04}$&$ 0.15^{+0.05}_{-0.05}$&$-0.06^{+0.03}_{-0.03}$&$ 0.20^{+0.05}_{-0.05}$&$ 0.17^{+0.03}_{-0.03}$&$-0.09^{+0.08}_{-0.08}$\\
$ 0.209 $&$  0.208$&$ 176^{+ 3}_{- 3}$&$  4.0^{+ 0.3}_{- 0.4}$&$ 0.04^{+0.04}_{-0.04}$&$  3.3^{+ 0.2}_{- 0.3}$&$ 0.11^{+0.08}_{-0.08}$&$ 0.18^{+0.06}_{-0.06}$&$ 0.03^{+0.07}_{-0.07}$&$-0.06^{+0.05}_{-0.05}$&$ 0.12^{+0.07}_{-0.07}$&$-0.02^{+0.04}_{-0.04}$&$ 0.11^{+0.08}_{-0.08}$&$ 0.14^{+0.05}_{-0.05}$&$-0.09^{+0.11}_{-0.12}$\\
$ 0.292 $&$  0.292$&$ 168^{+ 3}_{- 3}$&$  4.6^{+ 0.2}_{- 0.2}$&$-0.11^{+0.02}_{-0.02}$&$  2.9^{+ 0.2}_{- 0.2}$&$ 0.27^{+0.06}_{-0.06}$&$ 0.15^{+0.04}_{-0.04}$&$-0.10^{+0.05}_{-0.05}$&$-0.04^{+0.04}_{-0.04}$&$ 0.10^{+0.05}_{-0.06}$&$-0.03^{+0.03}_{-0.03}$&$ 0.10^{+0.05}_{-0.05}$&$ 0.16^{+0.03}_{-0.03}$&$-0.10^{+0.08}_{-0.07}$\\
$ 0.770 $&$  0.770$&$ 167^{+ 5}_{- 5}$&$  5.5^{+ 0.9}_{- 0.9}$&$-0.33^{+0.03}_{-0.03}$&$  1.9^{+ 0.3}_{- 0.4}$&$ 0.42^{+0.05}_{-0.05}$&$ 0.16^{+0.03}_{-0.04}$&$-0.09^{+0.08}_{-0.08}$&$-0.11^{+0.03}_{-0.03}$&$ 0.03^{+0.07}_{-0.07}$&$-0.07^{+0.02}_{-0.02}$&$ 0.12^{+0.04}_{-0.05}$&$ 0.19^{+0.02}_{-0.02}$&$ 0.10^{+0.09}_{-0.09}$\\
\hline 
NGC7144 \\
$ 0.019 $&$  0.007$&$ 158^{+ 1}_{- 1}$&$ 11.6^{+ 0.7}_{- 0.7}$&$ 0.14^{+0.02}_{-0.02}$&$  2.2^{+ 0.2}_{- 0.2}$&$ 0.31^{+0.04}_{-0.04}$&$ 0.21^{+0.03}_{-0.03}$&$ 0.10^{+0.03}_{-0.04}$&$ 0.02^{+0.03}_{-0.03}$&$ 0.06^{+0.04}_{-0.04}$&$-0.01^{+0.02}_{-0.02}$&$ 0.32^{+0.03}_{-0.03}$&$ 0.20^{+0.03}_{-0.03}$&$ 0.04^{+0.08}_{-0.08}$\\
$ 0.033 $&$  0.028$&$ 154^{+ 2}_{- 2}$&$ 10.2^{+ 0.4}_{- 0.4}$&$ 0.08^{+0.01}_{-0.01}$&$  2.4^{+ 0.1}_{- 0.1}$&$ 0.30^{+0.03}_{-0.03}$&$ 0.21^{+0.02}_{-0.02}$&$ 0.02^{+0.08}_{-0.08}$&$ 0.02^{+0.03}_{-0.03}$&$ 0.04^{+0.03}_{-0.04}$&$-0.01^{+0.02}_{-0.02}$&$ 0.26^{+0.03}_{-0.03}$&$ 0.18^{+0.02}_{-0.02}$&$-0.02^{+0.12}_{-0.12}$\\
$ 0.085 $&$  0.083$&$ 147^{+ 1}_{- 1}$&$ 11.0^{+ 0.8}_{- 0.7}$&$-0.10^{+0.02}_{-0.02}$&$  2.1^{+ 0.2}_{- 0.3}$&$ 0.32^{+0.04}_{-0.04}$&$ 0.19^{+0.03}_{-0.03}$&$ 0.09^{+0.03}_{-0.04}$&$-0.04^{+0.03}_{-0.03}$&$ 0.00^{+0.04}_{-0.04}$&$ 0.00^{+0.02}_{-0.02}$&$ 0.17^{+0.04}_{-0.04}$&$ 0.17^{+0.03}_{-0.03}$&$-0.16^{+0.15}_{-0.15}$\\
$ 0.188 $&$  0.188$&$ 138^{+ 2}_{- 2}$&$ 10.0^{+ 0.5}_{- 0.5}$&$-0.17^{+0.02}_{-0.02}$&$  2.3^{+ 0.2}_{- 0.2}$&$ 0.31^{+0.04}_{-0.04}$&$ 0.19^{+0.03}_{-0.03}$&$ 0.08^{+0.04}_{-0.04}$&$-0.07^{+0.03}_{-0.03}$&$ 0.00^{+0.05}_{-0.05}$&$-0.00^{+0.02}_{-0.02}$&$ 0.09^{+0.04}_{-0.04}$&$ 0.16^{+0.03}_{-0.03}$&$-0.20^{+0.15}_{-0.15}$\\
$ 0.375 $&$  0.375$&$ 129^{+ 2}_{- 2}$&$  8.6^{+ 0.8}_{- 0.8}$&$-0.25^{+0.02}_{-0.02}$&$  2.4^{+ 0.1}_{- 0.1}$&$ 0.38^{+0.04}_{-0.04}$&$ 0.21^{+0.03}_{-0.03}$&$ 0.07^{+0.04}_{-0.04}$&$-0.11^{+0.03}_{-0.03}$&$ 0.01^{+0.05}_{-0.05}$&$-0.03^{+0.02}_{-0.02}$&$ 0.09^{+0.04}_{-0.04}$&$ 0.18^{+0.03}_{-0.03}$&$-0.09^{+0.16}_{-0.16}$\\
$ 0.625 $&$  0.625$&$ 139^{+ 3}_{- 3}$&$  6.8^{+ 1.0}_{- 1.0}$&$-0.30^{+0.04}_{-0.04}$&$  2.4^{+ 0.2}_{- 0.2}$&$ 0.45^{+0.05}_{-0.05}$&$ 0.16^{+0.04}_{-0.04}$&$-0.02^{+0.06}_{-0.06}$&$-0.13^{+0.03}_{-0.03}$&$-0.15^{+0.05}_{-0.05}$&$-0.05^{+0.03}_{-0.03}$&$ 0.09^{+0.05}_{-0.05}$&$ 0.20^{+0.04}_{-0.04}$&$-0.17^{+0.16}_{-0.15}$\\
$ 0.875 $&$  0.875$&$ 142^{+ 4}_{- 4}$&$  6.6^{+ 1.1}_{- 1.0}$&$-0.34^{+0.05}_{-0.05}$&$  2.2^{+ 0.2}_{- 0.3}$&$ 0.32^{+0.10}_{-0.11}$&$ 0.18^{+0.06}_{-0.06}$&$-0.07^{+0.08}_{-0.07}$&$-0.20^{+0.05}_{-0.04}$&$-0.17^{+0.07}_{-0.07}$&$-0.05^{+0.04}_{-0.04}$&$ 0.10^{+0.07}_{-0.08}$&$ 0.16^{+0.06}_{-0.05}$&$-0.09^{+0.16}_{-0.13}$\\
\hline 
NGC7796 \\
$ 0.030 $&$  0.015$&$ 246^{+ 2}_{- 2}$&$ 12.9^{+ 0.6}_{- 0.7}$&$ 0.01^{+0.02}_{-0.02}$&$  2.3^{+ 0.2}_{- 0.3}$&$ 0.35^{+0.04}_{-0.04}$&$ 0.25^{+0.03}_{-0.03}$&$ 0.12^{+0.04}_{-0.04}$&$ 0.05^{+0.04}_{-0.05}$&$ 0.09^{+0.04}_{-0.04}$&$-0.00^{+0.02}_{-0.02}$&$ 0.55^{+0.03}_{-0.03}$&$ 0.21^{+0.03}_{-0.03}$&$ 0.25^{+0.04}_{-0.05}$\\
$ 0.056 $&$  0.050$&$ 241^{+ 2}_{- 2}$&$ 12.7^{+ 0.8}_{- 0.8}$&$-0.01^{+0.02}_{-0.02}$&$  2.4^{+ 0.2}_{- 0.2}$&$ 0.31^{+0.04}_{-0.05}$&$ 0.25^{+0.03}_{-0.03}$&$ 0.12^{+0.04}_{-0.04}$&$ 0.06^{+0.04}_{-0.04}$&$ 0.10^{+0.04}_{-0.04}$&$-0.00^{+0.02}_{-0.02}$&$ 0.50^{+0.04}_{-0.05}$&$ 0.19^{+0.03}_{-0.03}$&$ 0.23^{+0.05}_{-0.06}$\\
$ 0.101 $&$  0.098$&$ 236^{+ 2}_{- 2}$&$ 12.9^{+ 0.6}_{- 0.7}$&$-0.05^{+0.02}_{-0.02}$&$  2.4^{+ 0.2}_{- 0.2}$&$ 0.34^{+0.04}_{-0.04}$&$ 0.25^{+0.02}_{-0.03}$&$ 0.12^{+0.03}_{-0.04}$&$ 0.07^{+0.04}_{-0.04}$&$ 0.10^{+0.04}_{-0.04}$&$ 0.00^{+0.02}_{-0.02}$&$ 0.44^{+0.05}_{-0.04}$&$ 0.20^{+0.02}_{-0.02}$&$ 0.21^{+0.05}_{-0.06}$\\
$ 0.189 $&$  0.188$&$ 234^{+ 2}_{- 2}$&$ 12.9^{+ 0.6}_{- 0.8}$&$-0.08^{+0.02}_{-0.02}$&$  2.5^{+ 0.2}_{- 0.3}$&$ 0.30^{+0.04}_{-0.05}$&$ 0.23^{+0.02}_{-0.03}$&$ 0.12^{+0.04}_{-0.04}$&$ 0.02^{+0.04}_{-0.04}$&$ 0.10^{+0.04}_{-0.04}$&$ 0.02^{+0.02}_{-0.02}$&$ 0.35^{+0.03}_{-0.03}$&$ 0.18^{+0.03}_{-0.03}$&$ 0.16^{+0.05}_{-0.06}$\\
$ 0.376 $&$  0.375$&$ 228^{+ 2}_{- 2}$&$ 13.0^{+ 0.6}_{- 0.7}$&$-0.13^{+0.02}_{-0.02}$&$  2.3^{+ 0.3}_{- 0.3}$&$ 0.27^{+0.05}_{-0.05}$&$ 0.22^{+0.03}_{-0.03}$&$ 0.09^{+0.04}_{-0.04}$&$ 0.01^{+0.05}_{-0.05}$&$ 0.11^{+0.04}_{-0.04}$&$ 0.01^{+0.02}_{-0.02}$&$ 0.32^{+0.03}_{-0.03}$&$ 0.16^{+0.03}_{-0.03}$&$ 0.08^{+0.06}_{-0.08}$\\
$ 0.626 $&$  0.625$&$ 214^{+ 3}_{- 3}$&$ 13.4^{+ 0.4}_{- 0.6}$&$-0.20^{+0.02}_{-0.02}$&$  2.2^{+ 0.2}_{- 0.3}$&$ 0.33^{+0.05}_{-0.05}$&$ 0.22^{+0.03}_{-0.03}$&$ 0.10^{+0.04}_{-0.04}$&$ 0.02^{+0.05}_{-0.05}$&$ 0.11^{+0.04}_{-0.05}$&$ 0.00^{+0.02}_{-0.02}$&$ 0.32^{+0.03}_{-0.03}$&$ 0.19^{+0.03}_{-0.03}$&$-0.00^{+0.11}_{-0.11}$\\
$ 0.875 $&$  0.875$&$ 215^{+ 3}_{- 3}$&$ 13.5^{+ 0.5}_{- 0.7}$&$-0.23^{+0.02}_{-0.02}$&$  2.2^{+ 0.3}_{- 0.3}$&$ 0.39^{+0.05}_{-0.05}$&$ 0.19^{+0.03}_{-0.03}$&$ 0.17^{+0.04}_{-0.04}$&$ 0.05^{+0.07}_{-0.07}$&$ 0.12^{+0.05}_{-0.05}$&$-0.03^{+0.02}_{-0.02}$&$ 0.32^{+0.03}_{-0.03}$&$ 0.23^{+0.03}_{-0.03}$&$ 0.16^{+0.12}_{-0.13}$\\
\hline 
UGC1382 \\
$ 0.028 $&$  0.016$&$ 182^{+ 3}_{- 3}$&$  9.5^{+ 0.9}_{- 0.9}$&$-0.05^{+0.02}_{-0.02}$&$  3.1^{+ 0.2}_{- 0.2}$&$ 0.24^{+0.08}_{-0.10}$&$ 0.29^{+0.04}_{-0.04}$&$ 0.22^{+0.06}_{-0.06}$&$ 0.15^{+0.04}_{-0.05}$&$ 0.19^{+0.05}_{-0.05}$&$-0.02^{+0.03}_{-0.03}$&$ 0.06^{+0.16}_{-0.17}$&$ 0.23^{+0.03}_{-0.04}$&$-0.36^{+0.09}_{-0.06}$\\
$ 0.052 $&$  0.047$&$ 179^{+ 3}_{- 4}$&$ 10.9^{+ 1.4}_{- 1.4}$&$-0.05^{+0.03}_{-0.03}$&$  2.4^{+ 0.4}_{- 0.6}$&$ 0.51^{+0.04}_{-0.04}$&$ 0.26^{+0.05}_{-0.04}$&$ 0.31^{+0.06}_{-0.07}$&$ 0.17^{+0.05}_{-0.05}$&$ 0.36^{+0.05}_{-0.05}$&$-0.08^{+0.03}_{-0.03}$&$ 0.23^{+0.15}_{-0.19}$&$ 0.35^{+0.04}_{-0.03}$&$-0.32^{+0.07}_{-0.05}$\\
$ 0.097 $&$  0.094$&$ 173^{+ 3}_{- 3}$&$  9.7^{+ 1.0}_{- 0.9}$&$-0.05^{+0.03}_{-0.03}$&$  3.2^{+ 0.1}_{- 0.2}$&$ 0.27^{+0.09}_{-0.11}$&$ 0.26^{+0.05}_{-0.05}$&$ 0.21^{+0.07}_{-0.07}$&$ 0.00^{+0.06}_{-0.06}$&$ 0.22^{+0.05}_{-0.06}$&$-0.04^{+0.03}_{-0.03}$&$ 0.33^{+0.15}_{-0.16}$&$ 0.25^{+0.04}_{-0.04}$&$-0.34^{+0.09}_{-0.06}$\\
$ 0.189 $&$  0.188$&$ 157^{+ 3}_{- 3}$&$ 10.5^{+ 1.3}_{- 1.1}$&$-0.07^{+0.03}_{-0.03}$&$  2.7^{+ 0.3}_{- 0.3}$&$ 0.19^{+0.08}_{-0.06}$&$ 0.17^{+0.05}_{-0.05}$&$ 0.33^{+0.06}_{-0.07}$&$ 0.15^{+0.06}_{-0.06}$&$ 0.17^{+0.06}_{-0.07}$&$-0.11^{+0.03}_{-0.03}$&$ 0.26^{+0.18}_{-0.21}$&$ 0.18^{+0.04}_{-0.04}$&$-0.13^{+0.16}_{-0.12}$\\
$ 0.376 $&$  0.375$&$ 139^{+ 4}_{- 4}$&$ 11.2^{+ 1.4}_{- 1.4}$&$-0.12^{+0.04}_{-0.04}$&$  2.7^{+ 0.3}_{- 0.4}$&$ 0.45^{+0.08}_{-0.10}$&$ 0.27^{+0.05}_{-0.05}$&$ 0.37^{+0.06}_{-0.08}$&$ 0.11^{+0.07}_{-0.07}$&$ 0.14^{+0.08}_{-0.08}$&$-0.13^{+0.03}_{-0.03}$&$ 0.08^{+0.20}_{-0.19}$&$ 0.27^{+0.05}_{-0.04}$&$-0.07^{+0.18}_{-0.14}$\\
$ 0.625 $&$  0.625$&$ 142^{+ 8}_{- 7}$&$  8.7^{+ 1.4}_{- 1.1}$&$-0.23^{+0.05}_{-0.06}$&$  3.3^{+ 0.1}_{- 0.2}$&$ 0.45^{+0.07}_{-0.08}$&$ 0.32^{+0.09}_{-0.08}$&$ 0.40^{+0.09}_{-0.11}$&$-0.05^{+0.14}_{-0.13}$&$ 0.33^{+0.10}_{-0.11}$&$-0.04^{+0.07}_{-0.06}$&$-0.27^{+0.23}_{-0.12}$&$ 0.29^{+0.08}_{-0.07}$&$ 0.07^{+0.23}_{-0.27}$\\
\hline 
IC1459 \\
$ 0.012 $&$  0.004$&$ 307^{+ 8}_{- 8}$&$ 13.1^{+ 0.9}_{- 1.0}$&$ 0.13^{+0.02}_{-0.02}$&$  2.2^{+ 0.4}_{- 0.5}$&$ 0.44^{+0.07}_{-0.07}$&$ 0.20^{+0.05}_{-0.05}$&$ 0.20^{+0.14}_{-0.14}$&$ 0.04^{+0.07}_{-0.06}$&$ 0.07^{+0.06}_{-0.06}$&$-0.07^{+0.03}_{-0.03}$&$ 0.77^{+0.06}_{-0.05}$&$ 0.25^{+0.03}_{-0.03}$&$ 0.38^{+0.09}_{-0.09}$\\
$ 0.023 $&$  0.020$&$ 295^{+11}_{-11}$&$ 13.5^{+ 0.7}_{- 0.8}$&$ 0.09^{+0.05}_{-0.05}$&$  2.3^{+ 0.4}_{- 0.5}$&$ 0.37^{+0.05}_{-0.05}$&$ 0.21^{+0.06}_{-0.06}$&$ 0.16^{+0.08}_{-0.09}$&$ 0.02^{+0.06}_{-0.06}$&$ 0.08^{+0.08}_{-0.08}$&$-0.01^{+0.03}_{-0.03}$&$ 0.69^{+0.07}_{-0.07}$&$ 0.21^{+0.04}_{-0.04}$&$ 0.28^{+0.11}_{-0.11}$\\
$ 0.048 $&$  0.047$&$ 290^{+ 8}_{- 8}$&$ 12.9^{+ 0.7}_{- 0.9}$&$ 0.06^{+0.04}_{-0.04}$&$  2.4^{+ 0.5}_{- 0.5}$&$ 0.35^{+0.04}_{-0.05}$&$ 0.21^{+0.05}_{-0.06}$&$ 0.15^{+0.11}_{-0.11}$&$ 0.03^{+0.05}_{-0.05}$&$ 0.10^{+0.08}_{-0.08}$&$-0.01^{+0.03}_{-0.03}$&$ 0.60^{+0.06}_{-0.06}$&$ 0.20^{+0.03}_{-0.03}$&$ 0.21^{+0.07}_{-0.07}$\\
$ 0.094 $&$  0.094$&$ 284^{+ 9}_{- 9}$&$ 12.4^{+ 1.1}_{- 1.2}$&$ 0.02^{+0.06}_{-0.06}$&$  2.2^{+ 0.5}_{- 0.5}$&$ 0.37^{+0.05}_{-0.05}$&$ 0.22^{+0.05}_{-0.05}$&$ 0.16^{+0.10}_{-0.11}$&$ 0.07^{+0.07}_{-0.07}$&$ 0.08^{+0.10}_{-0.10}$&$-0.03^{+0.04}_{-0.04}$&$ 0.57^{+0.05}_{-0.06}$&$ 0.22^{+0.04}_{-0.04}$&$ 0.16^{+0.08}_{-0.09}$\\
$ 0.188 $&$  0.188$&$ 272^{+13}_{-13}$&$ 11.6^{+ 1.7}_{- 1.7}$&$-0.09^{+0.05}_{-0.04}$&$  1.9^{+ 0.5}_{- 0.5}$&$ 0.42^{+0.05}_{-0.05}$&$ 0.21^{+0.07}_{-0.07}$&$ 0.15^{+0.09}_{-0.10}$&$ 0.05^{+0.05}_{-0.05}$&$-0.06^{+0.09}_{-0.09}$&$-0.03^{+0.03}_{-0.03}$&$ 0.45^{+0.09}_{-0.10}$&$ 0.23^{+0.03}_{-0.03}$&$ 0.17^{+0.11}_{-0.11}$\\
$ 0.375 $&$  0.375$&$ 252^{+14}_{-14}$&$  8.7^{+ 1.7}_{- 1.7}$&$-0.24^{+0.06}_{-0.06}$&$  1.8^{+ 0.4}_{- 0.5}$&$ 0.45^{+0.07}_{-0.07}$&$ 0.24^{+0.10}_{-0.10}$&$ 0.09^{+0.12}_{-0.12}$&$-0.02^{+0.09}_{-0.09}$&$-0.14^{+0.11}_{-0.11}$&$-0.04^{+0.06}_{-0.06}$&$ 0.37^{+0.07}_{-0.08}$&$ 0.22^{+0.06}_{-0.06}$&$ 0.03^{+0.20}_{-0.21}$\\
$ 0.625 $&$  0.625$&$ 256^{+13}_{-13}$&$  7.8^{+ 3.0}_{- 3.0}$&$-0.31^{+0.18}_{-0.18}$&$  2.1^{+ 0.4}_{- 0.6}$&$ 0.43^{+0.07}_{-0.09}$&$ 0.19^{+0.07}_{-0.07}$&$-0.16^{+0.13}_{-0.12}$&$-0.17^{+0.07}_{-0.07}$&$-0.27^{+0.09}_{-0.08}$&$-0.04^{+0.05}_{-0.04}$&$ 0.38^{+0.08}_{-0.12}$&$ 0.21^{+0.05}_{-0.05}$&$-0.04^{+0.21}_{-0.20}$\\
$ 0.875 $&$  0.875$&$ 251^{+17}_{-17}$&$  6.5^{+ 2.6}_{- 2.5}$&$-0.32^{+0.21}_{-0.21}$&$  2.4^{+ 0.5}_{- 0.7}$&$ 0.45^{+0.13}_{-0.14}$&$ 0.21^{+0.13}_{-0.13}$&$-0.13^{+0.19}_{-0.19}$&$-0.09^{+0.14}_{-0.13}$&$-0.31^{+0.12}_{-0.12}$&$-0.09^{+0.11}_{-0.11}$&$ 0.38^{+0.14}_{-0.19}$&$ 0.23^{+0.12}_{-0.12}$&$-0.10^{+0.24}_{-0.22}$\\
\hline 